\documentclass[11pt,a4paper,preprint]{article}
\pdfoutput=1
\usepackage{jheppub} 
\usepackage{graphicx}
\usepackage{graphics}
\usepackage[mathscr]{eucal}
\usepackage{amssymb}
\usepackage{amsmath}
\usepackage{xspace}
\usepackage{listings}
\usepackage{ulem}
\usepackage{hyperref}
\usepackage[capitalise, english]{cleveref}
\usepackage{multirow}
\usepackage{booktabs}
\usepackage[exponent-product = \times]{siunitx}
\usepackage{bbold}
\usepackage{soul}
\usepackage{color}
\usepackage{centernot}
\usepackage{enumerate}

\definecolor{dkgreen}{rgb}{0,0.6,0}
\definecolor{gray}{rgb}{0.5,0.5,0.5}
\definecolor{mauve}{rgb}{0.58,0,0.82}

\newcommand{\SARAH}[0]{{\tt SARAH}\xspace}

\newcommand{\GeV}[0]{\text{GeV}\xspace}

\newcommand{\MUP}{{\ensuremath{M_{\rm up-type}}}\xspace}
\newcommand{\MUPCKM}{\ensuremath{V_{\rm CKM}^\dagger \MUP V_{\rm CKM}}\xspace}

\def\gsim{\raise0.3ex\hbox{$\;>$\kern-0.75em\raise-1.1ex\hbox{$\sim\;$}}}
\def\lsim{\raise0.3ex\hbox{$\;<$\kern-0.75em\raise-1.1ex\hbox{$\sim\;$}}}
\definecolor{mkgreen}{rgb}{0.2,.70,.3}

\definecolor{tobycolour}{rgb}{.5,.0,.5}

\newcommand{\Umns}{\ensuremath{U_{\rm PMNS}}}
\newcommand{\dCP}{\ensuremath{\delta_{\rm CP}}}
\newcommand{\mnul}{\ensuremath{m_{\nu}^{\rm light}}}
\newcommand{\BDi}{\ensuremath{B_D^{(i,i)}}}
\newcommand{\ADi}{\ensuremath{A_D^{(i,i)}}}

\begin{document}

\title{Perspectives for Detecting Lepton Flavour Violation in Left-Right Symmetric Models}

\author[a]{Cesar Bonilla,}
\author[b]{Manuel\ E.\ Krauss,} 
\author[b]{Toby\ Opferkuch,} 
\author[c]{Werner\ Porod}

\affiliation[b]{
Bethe Center for Theoretical Physics \& Physikalisches Institut der
 Universit\"at Bonn, \\
Nussallee 12, 53115 Bonn, Germany }

\affiliation[c]{
Institut f\"ur Theoretische Physik und Astronomie,
Universit\"at W\"urzburg\\
Am Hubland,
97074 W\"urzburg}

\affiliation[a]{
   AHEP Group, Instituto de F\'{\i}sica Corpuscular --
    C.S.I.C./Universitat de Val{\`e}ncia \\
    Edificio de Institutos de Paterna,
 C/Catedratico Jos\'e Beltr\'an, 2 E-46980 Paterna (Val\`{e}ncia) - Spain}
 
\emailAdd{cesar.bonilla@ific.uv.es}
\emailAdd{mkrauss@th.physik.uni-bonn.de}
\emailAdd{toby@th.physik.uni-bonn.de}
\emailAdd{porod@physik.uni-wuerzburg.de}

\preprint{Bonn-TH-2016-08}

\abstract{
We investigate lepton flavour violation in a class of minimal left-right symmetric models where the left-right symmetry is broken by triplet scalars.
In this context we present a method to consistently calculate the triplet-Yukawa
couplings which takes into account the experimental data while simultaneously respecting the underlying
symmetries. 
Analysing various scenarios, we then calculate the full set of tree-level and one-loop contributions to all radiative and three-body flavour-violating fully leptonic decays as well as $\mu-e$ conversion in nuclei. 
Our method illustrates how these processes depend on the underlying parameters of the theory.
To that end we observe that, for many choices of the model parameters, there is a strong complementarity between the different observables.  For instance, in a large part of the parameter space, lepton flavour violating $\tau$-decays have a large enough branching ratio to be measured in upcoming experiments. Our results further show that 
experiments coming online in the immediate future, like Mu3e and BELLE~II, or longer-term, such as PRISM/PRIME, will probe significant portions of the currently allowed parameter space.
}

\maketitle

\section{Introduction}

Left-right (LR) symmetric extensions of the  Standard Model automatically contain the 
correct ingredients to explain the observed
neutrino masses and mixings. The right-handed neutrino field $\nu_R$ is necessarily part
of the theory and breaking the LR symmetry by $SU(2)_R$ triplets generates a Majorana mass term for the $\nu_R$ and thus a seesaw mechanism 
\cite{Minkowski:1977sc,GellMann:1980vs,Mohapatra:1979ia,Schechter:1980gr,Cheng:1980qt}. 
In an LR symmetric model one typically expects a combination of seesaw type I and
type II. These models are also interesting from the point of view of  
grand unified theories (GUT)  based on $SO(10)$ gauge symmetry \cite{Fritzsch:1974nn} where
they form part of its maximal subgroup, the Pati-Salam group \cite{Pati:1973uk}.
A further attractive feature of LR models is parity restoration which occurs for example
together with charge conjugation symmetry in an $SO(10)$ GUT context~\cite{Aulakh:1997fq}.

In a LR model one would also expect that the Higgs sector respects the LR symmetry 
particle-wise, e.g.\ that there is a one-to-one correspondence between the Higgs bosons
charged under $SU(2)_L$ and $SU(2)_R$, respectively. 
In a minimal model with Majorana mass terms
for neutrinos 
one requires a bi-doublet charged under both $SU(2)$ factors and in each sector a 
triplet \cite{Deshpande:1990ip}. As a consequence, lepton flavour violating (LFV) decays
are possible at tree-level \cite{Raidal:1997hq} which is already heavily constrained
by existing data, for example BR($\mu \to 3 e) \leq 10^{-12}$ \cite{Bellgardt:1987du} which
 will be further constrained by upcoming experiments like Mu3e \cite{Blondel:2013ia}. If the new scalar 
 particles are at the TeV scale, one can therefore expect measurable rates in the near future. 
 In \cref{tab:sensi} we give an overview of the relevant LFV observables and their current bounds as well as expected future sensitivities.
 
A priori, the scale of LR breaking could be anywhere between the TeV and the GUT scale. When requiring gauge coupling unification, one finds that the Weinberg angle turns out to be too large for a low breaking scale in the minimal LR-symmetric model. It has been shown, however, that this problem can be solved once the discrete LR parity is broken at a higher scale \cite{Chang:1983fu,Chang:1984uy}. In Ref.~\cite{Arbelaez:2013nga}, a class of LR models consistent with $SO(10)$ unification has been developed which can have breaking scales down to $\mathcal O({\rm TeV)}$.\footnote{In many supersymmetric realizations, a TeV-scale LR symmetry is even preferred for different reasons like an intimate connection between the LR- and the supersymmetry-breaking scale \cite{Babu:2008ep, Cvetic:1983su}, vacuum stability considerations \cite{Basso:2015pka}, or gauge coupling unification \cite{Dev:2009aw,Hirsch:2015fvq}.}
 
Left-right symmetric models with a TeV-scale breaking in various variants
have been considered in the past,
investigating lepton flavour and lepton number violation 
\cite{Deshpande:1990ip,Hirsch:1996qw,Awasthi:2013ff,Barry:2013xxa,Awasthi:2015ota,Bambhaniya:2015ipg,Tello:2010am}, CP
violation \cite{Zhang:2007da}, bounds on the heavy additional vector bosons
\cite{Basso:2015pka,Nemevsek:2011hz,Helo:2015ffa,Lindner:2016lpp}, 
 potential Higgs signals at the LHC \cite{Mohapatra:2013cia,Bambhaniya:2013wza,Bambhaniya:2015wna,Dev:2016dja} as well as
 lepton flavour and number violating signals at the LHC 
 \cite{Tello:2010am,Das:2012ii,Chen:2013fna,Lindner:2016lxq,Gluza:2016qqv}.
Beside the constraints due to LFV processes further constraints arise from observables
in the $K$- and $B$-meson sector, see e.g.\ \cite{Blanke:2011ry,Bertolini:2014sua} for recent
updates, and direct searches for new states, in particular LHC searches. The  latter
put e.g.\ a bound of 2.9~TeV on the mass of the $W_R$ \cite{ATLAS:2016lvi,Khachatryan:2015dcf}. 
Note, however, that such bounds are model dependent and can be weaker if additional decay 
channels of the $W_R$ and/or $\nu_R$ are present as discussed e.g.\ in
\cite{Basso:2015pka}.
In addition, the $\rho$ parameter \cite{Baak:2014ora,Olive:2016xmw},
or more generally the oblique parameters~\cite{Peskin:1991sw,Lavoura:1993nq},
 constrain the vacuum expectation value (VEV) of a $SU(2)_L$ scalar triplet 
 to be at most $\mathcal O({\rm GeV})$ \cite{Gunion:1990dt,Arhrib:2011uy,Kanemura:2012rs,Maiezza:2016bzp}.
In the majority of these works only parts of a complete model have
been considered, e.g.\ the lepton sector or the Higgs sector, without checking whether
the other parts are consistently implemented.

In the present paper we will  discuss  a model which particle-wise is manifest
LR symmetric and where the different scales of the $SU(2)_L$ and $SU(2)_R$ breaking
occur dynamically. We will assume true LR-symmetry in the Yukawa sector of the model
where parity restoration is implemented via discrete parity symmetry or charge conjugation.
As a result of these discrete symmetries, it is possible to parametrise the triplet Yukawa couplings as a function of
only the underlying model parameters and the measured neutrino data \cite{Akhmedov:2005np}. 
Here we expand upon this method and show how a simple analytic expression for the solution can be obtained.
Clearly, the existing data on
lepton masses and mixing is not sufficient to uniquely specify these couplings even in
this restricted context. Consequently we will discuss how LFV decays further
constrain these couplings. However, the results depend on the details of the
Higgs sector, in particular on the value of the masses of the heavier Higgs bosons
as well as on $v_L$, the vacuum expectation values of the $SU(2)_L$ triplet $\Delta_L$.

\begin{table}[tb!]
\centering
\begin{tabular}{lll}
\toprule
LFV Process & Present Bound & Future Sensitivity  \\
\midrule
$\mu \rightarrow  e \gamma$ & $\SI{4.2E-13}{}$~\cite{TheMEG:2016wtm}  & $\SI{6E-14}{}$~\cite{Baldini:2013ke} \\
$\tau \to e \gamma$ & $\SI{3.3E-8}{}$~\cite{Aubert:2009ag}& $ \sim\SI{3E-9}{}$~\cite{Aushev:2010bq}\\
$\tau \to \mu \gamma$ & $\SI{4.4E-8}{}$~\cite{Aubert:2009ag}& $ \sim\SI{E-9}{}$~\cite{Aushev:2010bq} \\
$\mu \rightarrow e e e$ &  $\SI{1.0E-12}{}$~\cite{Bellgardt:1987du} &  $\sim\SI{3E-16}{}$~\cite{Blondel:2013ia} \\
$\tau \rightarrow e e e$ & $\SI{2.7E-8}{}$~\cite{Hayasaka:2010np} &  $\sim \SI{5E-10}{}$~\cite{Aushev:2010bq,BelleII2015} \\
$\tau \rightarrow \mu \mu \mu$ & $\SI{2.1E-8}{}$~\cite{Hayasaka:2010np} & $\sim \SI{4E-10}{}$~\cite{Aushev:2010bq,BelleII2015} \\
$\tau^- \rightarrow e^- \mu^+ \mu^-$ &  $\SI{2.7E-8}{}$~\cite{Hayasaka:2010np} & $\sim \SI{5E-10}{}$~\cite{Aushev:2010bq,BelleII2015} \\
$\tau^- \rightarrow \mu^- e^+ e^-$ &  $\SI{1.8E-8}{}$~\cite{Hayasaka:2010np} & $\sim\SI{3E-10}{}$~\cite{Aushev:2010bq,BelleII2015} \\
$\tau^- \rightarrow \mu^+ e^- e^-$ & $\SI{1.5E-8}{}$ \cite{Hayasaka:2010np} & $\sim\SI{3E-10}{}$~\cite{BelleII2015}\\
$\tau^- \rightarrow e^+ \mu^- \mu^-$ & $\SI{1.7E-8}{}$ \cite{Hayasaka:2010np} & $\sim\SI{3E-10}{}$~\cite{BelleII2015}\\
$\mu^- \rightarrow e^-, \mathrm{Ti}$ &  $\SI{4.3E-12}{}$~\cite{Dohmen:1993mp} & $\sim\SI{E-18}{}$~\cite{PRIME,Barlow:2011zza} \\
$\mu^- \rightarrow e^-, \mathrm{Au}$ & $\SI{7E-13}{}$~\cite{Bertl:2006up} & - \\
$\mu^- \rightarrow e^-, \mathrm{Al}$ & - & $\SI{E-16}{}-\SI{3E-17}{}$~\cite{Bartoszek:2014mya,Cui:2009zz,Kurup:2011zza} \\
$\mu^- \rightarrow e^-, \mathrm{SiC}$ & - & $\SI{E-14}{}$~\cite{Aoki:2012zza} \\
\bottomrule
\end{tabular}
\caption{Current experimental bounds and future sensitivities for
  low-energy LFV observables.}
\label{tab:sensi}
\end{table}

This paper is organized as follows: in \cref{sec:model} we present the details of the model.
In \cref{sec:neutrinos} we discuss particularities in the neutrino sector, in particular
our way of parametrising the Yukawa couplings. 
We stress that this section is crucial for understanding the subsequent parts of the paper.
In \cref{sec:results} we present our
numerical results. Here we first discuss in detail the different contributions to different LFV observables and their 
behaviours as a function of the free parameters. 
Our main results are located in \cref{sec:2D}. There
 where we show which regions of parameter space can be probed by which experiments in the near future. Finally we
conclude in \cref{sec:conclusion}. Some more details on the calculation
of the Yukawa couplings, the mass matrices of the Higgs sector as well as the program
implementation via {\tt SARAH} \cite{Staub:2008uz,Staub:2009bi,Staub:2010jh,Staub:2012pb,Staub:2013tta,Staub:2015kfa} are given in the appendices. There we also present for completeness the results for both the degenerate neutrino masses and inverted neutrino mass hierarchy which are not covered in the main text.

\section{The minimal left-right symmetric model}
\label{sec:model}

We consider the minimal phenomenologically acceptable model with left-right (LR) symmetry at the Lagrangian level. This means that, in addition to promoting $SU(2)_L$-singlet fields to $SU(2)_R$ multiplets, there has to be an additional sector which breaks $SU(2)_R\times U(1)_{B-L} \to U(1)_Y$. The most economical choice for the LR breaking which also at the same time leads to neutrino mass generation via a seesaw mechanism is $SU(2)$ triplets.

\subsection{Model definition}

The minimal particle content and the irreducible representations under $SU(3)_c\times SU(2)_L\times SU(2)_R \times U(1)_{B-L}$, are given by:
\begin{subequations}
\begin{align}
\text{Fermions:}\quad& \notag\\
Q_L &= \begin{pmatrix}
u_L \\ d_L
\end{pmatrix} \in (\mathbf{3},\mathbf{2},\mathbf{1},1/3)\,, & Q_R &= \begin{pmatrix}
u_R \\ d_R
\end{pmatrix} \in (\mathbf{3},\mathbf{1},\mathbf{2},1/3)\,, \\
L_L &= \begin{pmatrix}
\nu_L \\ \ell_L
\end{pmatrix} \in (\mathbf{1},\mathbf{2},\mathbf{1},-1)\,, & L_R &= \begin{pmatrix}
\nu_R \\ \ell_R
\end{pmatrix} \in (\mathbf{1},\mathbf{1},\mathbf{2},-1)\,. \\
\text{Scalars:}\quad& \notag\\
\Phi &=\begin{pmatrix}
\phi^{0}_{1}   &   \phi^{+}_{1} \\
\phi^{-}_{2}   &   \phi^{0}_{2}
\end{pmatrix} \in  (\mathbf{1},\mathbf{2},\mathbf{2},0) \,, \\
\Delta_{L} &=\begin{pmatrix}
\frac{\delta^{+}_{L}}{\sqrt{2}}& \delta^{++}_{L} \\
\delta^{0}_{L}                 & -\frac{\delta^{+}_{L}}{\sqrt{2}}
\end{pmatrix}\in (\mathbf{1},\mathbf{3},\mathbf{1},2)\,, 
& \Delta_{R} &=\begin{pmatrix}
\frac{\delta^{+}_{R}}{\sqrt{2}}& \delta^{++}_{R} \\
\delta^{0}_{R}                 & -\frac{\delta^{+}_{R}}{\sqrt{2}}
\end{pmatrix}\in (\mathbf{1},\mathbf{1},\mathbf{3},2)\,.
\end{align}
\end{subequations}
Here we use the convention that the electric charge is given by 
\begin{align}
Q_{\rm em} &= T_{3L} + T_{3R} + \frac{B-L}{2}\,.
\end{align}
The Yukawa interactions can be split into interactions of the quark and lepton fields with the bidoublet, $\mathcal L_Y^{\Phi}$, leading to Dirac-type masses for all fermions after electroweak symmetry breaking, as well as interactions with the triplets, $\mathcal L_Y^\Delta$, leading to Majorana-type mass terms for the neutrinos after LR-symmetry-breaking. 
The respective terms are
\begin{align} \label{eq:model:dirac_masses}
  -\mathcal{L}_Y^\Phi = \overline{Q_L}\left(Y_{Q_1}\Phi +Y_{Q_2} \tilde{\Phi}\right) Q_{R} +
   \overline{L_L}\left(Y_{L_1}\Phi +Y_{L_2} \tilde{\Phi}\right) L_{R} + {\rm h.c.}\,,
\end{align}
where $\tilde{\Phi}\equiv -\sigma_{2}\Phi^{\ast}\sigma_{2}$, and
\begin{align} \label{eq:model:majorana_masses}
-\mathcal L_Y^{\Delta} &= \overline{L_L^C} \,Y_{\Delta_{L}}\,(i\sigma_2) \Delta_L\, L_L 
                             +\overline{L_R^C} \, Y_{\Delta_{R}}\, (i\sigma_2) \Delta_R\, L_R 
                + {\rm h.c.}\,,
\end{align}
where 
\begin{equation}
\overline{\Psi^C} = \Psi^T  C \qquad {\rm and } \qquad C = i \gamma_2 \gamma_0\,.
\end{equation}

\subsection{Discrete symmetries}
There are two possible discrete symmetries, discrete parity \cite{Chang:1983fu,Chang:1984uy}, and charge conjugation symmetry, denoted as $\mathcal P$ and $\mathcal C$ in the following (see also Ref.~\cite{Maiezza:2010ic} and references therein).\\[0.15cm]
{\emph{\bf{Parity symmetry $\mathcal P$:}}}\\
Parity symmetry exchanges $L$ and $R$, hence, the symmetry operation is 
\begin{align}
\label{eq:model:Pparity}
L_L \leftrightarrow L_R\,,\qquad \Delta_L \leftrightarrow \Delta_R\,,\qquad \Phi \leftrightarrow \Phi^\dagger\,.
\end{align}
Requiring invariance of the Lagrangian yields the following constraints on the model parameters:
\begin{align}\label{eq:parityyukawas}
Y_{\alpha_i} &= Y_{\alpha_i}^\dagger \,, \qquad Y_{\Delta_L} = Y_{\Delta_R} \,,  
\end{align}
where $\alpha=Q,L$ and $i=1,2$. \\[0.15cm]
{\emph{\bf{Charge conjugation symmetry $\mathcal C$:}}}\\
Charge conjugation symmetry exchanges 
\begin{align}
\label{eq:model:Cparity}
L_L \leftrightarrow L_R^C\,,\qquad \Delta_L \leftrightarrow \Delta_R^*\,,\qquad \Phi \leftrightarrow \Phi^T\,.
\end{align}
Once again invariance of the Lagrangian yields 
\begin{align}\label{eq:chargeconjyukawas}
Y_{\alpha_i} &= Y_{\alpha_i}^T\,, \qquad Y_{\Delta_L} = Y_{\Delta_R}^*\,.
\end{align}
\subsection{Scalar sector and gauge symmetry breaking}

The most general $\mathcal C$- and $\mathcal P$-conserving renormalizable Higgs potential invariant under the discrete parity and charge conjugation 
symmetries
is given
by \cite{Deshpande:1990ip}
\begin{eqnarray}
\label{VLR}
V_{LR} &&= - \mu_1^2 {\rm Tr} (\Phi^{\dag} \Phi) - \mu_2^2
\left[ {\rm Tr} (\tilde{\Phi} \Phi^{\dag}) + {\rm Tr} (\tilde{\Phi}^{\dag} \Phi) \right]
- \mu_3^2 \left[ {\rm Tr} (\Delta_L \Delta_L^{\dag}) + {\rm Tr} (\Delta_R
\Delta_R^{\dag}) \right] 
\\
&&+ \lambda_1 \left[ {\rm Tr} (\Phi^{\dag} \Phi) \right]^2 + \lambda_2 \left\{ \left[
{\rm Tr} (\tilde{\Phi} \Phi^{\dag}) \right]^2 + \left[ {\rm Tr}
(\tilde{\Phi}^{\dag} \Phi) \right]^2 \right\} \nonumber \\
&&+ \lambda_3 {\rm Tr} (\tilde{\Phi} \Phi^{\dag}) {\rm Tr} (\tilde{\Phi}^{\dag} \Phi) +
\lambda_4 {\rm Tr} (\Phi^{\dag} \Phi) \left[ {\rm Tr} (\tilde{\Phi} \Phi^{\dag}) + {\rm
Tr}
(\tilde{\Phi}^{\dag} \Phi) \right]\nonumber \\
&& + \rho_1 \left\{ \left[ {\rm Tr} (\Delta_L \Delta_L^{\dag}) \right]^2 + \left[ {\rm
Tr} (\Delta_R \Delta_R^{\dag}) \right]^2 \right\} \nonumber \\ && + \rho_2 \left[ {\rm
Tr} (\Delta_L \Delta_L) {\rm Tr} (\Delta_L^{\dag} \Delta_L^{\dag}) + {\rm Tr} (\Delta_R
\Delta_R) {\rm Tr} (\Delta_R^{\dag} \Delta_R^{\dag}) \right] \nonumber
\\
&&+ \rho_3 {\rm Tr} (\Delta_L \Delta_L^{\dag}) {\rm Tr} (\Delta_R \Delta_R^{\dag})+
\rho_4 \left[ {\rm Tr} (\Delta_L \Delta_L) {\rm Tr} (\Delta_R^{\dag} \Delta_R^{\dag}) +
{\rm Tr} (\Delta_L^{\dag} \Delta_L^{\dag}) {\rm Tr} (\Delta_R
\Delta_R) \right]  \nonumber \\
&&+ \alpha_1 {\rm Tr} (\Phi^{\dag} \Phi) \left[ {\rm Tr} (\Delta_L \Delta_L^{\dag}) +
{\rm Tr} (\Delta_R \Delta_R^{\dag})  \right] \nonumber
\\
&&+ \left\{ \alpha_2 e^{i \delta_2} \left[ {\rm Tr} (\tilde{\Phi} \Phi^{\dag}) {\rm Tr}
(\Delta_L \Delta_L^{\dag}) + {\rm Tr} (\tilde{\Phi}^{\dag} \Phi) {\rm Tr} (\Delta_R
\Delta_R^{\dag}) \right] + {\rm h.c.}\right\} \nonumber
\\
&&+ \alpha_3 \left[ {\rm Tr}(\Phi \Phi^{\dag} \Delta_L \Delta_L^{\dag}) + {\rm
Tr}(\Phi^{\dag} \Phi \Delta_R \Delta_R^{\dag}) \right] + \beta_1 \left[ {\rm Tr}(\Phi
\Delta_R \Phi^{\dag} \Delta_L^{\dag}) +
{\rm Tr}(\Phi^{\dag} \Delta_L \Phi \Delta_R^{\dag}) \right] \nonumber \\
&&+ \beta_2 \left[ {\rm Tr}(\tilde{\Phi} \Delta_R \Phi^{\dag} \Delta_L^{\dag}) + {\rm
Tr}(\tilde{\Phi}^{\dag} \Delta_L \Phi \Delta_R^{\dag}) \right] + \beta_3 \left[ {\rm
Tr}(\Phi \Delta_R \tilde{\Phi}^{\dag} \Delta_L^{\dag}) + {\rm Tr}(\Phi^{\dag} \Delta_L
\tilde{\Phi} \Delta_R^{\dag}) \right] \,. \nonumber
\end{eqnarray}
The neutral scalar fields in the above potential can be expressed in terms of their CP-even and -odd components:
\begin{subequations}
\begin{align}
\phi_1^0 &= \frac{1}{\sqrt{2}} \left(v_1 + \sigma_1 + i \varphi_1 \right)\,, &  \delta_L^0 &= \frac{1}{\sqrt{2}} \left(v_L + \sigma_L + i \varphi_L \right)\,,\\
\phi_2^0 &= \frac{1}{\sqrt{2}} \left(v_2 + \sigma_2 + i \varphi_2 \right)\,,  & \delta_R^0 &= \frac{1}{\sqrt{2}} \left(v_R + \sigma_R + i \varphi_R \right)\,,
\end{align}
\end{subequations}
where we use the generic symbols $\sigma$ and $\varphi$ to label the CP-even and -odd states, respectively. 
For the vacuum expectation values, which we assume to be real, we use the following parametrisation:
\begin{align}
v_1 &= v \cos \beta\,,\qquad v_2 = v \sin\beta\,, \qquad t_\beta \equiv \tan\beta=\frac{v_2}{v_1}\,,
\end{align}
where $v_L\ll v \ll v_R$ so that $v$ can be identified as the SM VEV. The masses of the new gauge bosons therefore read
\begin{align}
M_{Z_R} \simeq \sqrt{g_R^2 + g_{BL}^2}\, v_R\,, \qquad\qquad M_{W_R} \simeq \frac{g_R}{\sqrt{2}} \, v_R \,.
\end{align}
Due to LR symmetry, we take the $SU(2)$ gauge coupling to be equal, namely $g_R=g_L$. Solving the four minimisation conditions for the potential we eliminate the following four parameters:

\begin{subequations}
\begin{align}
\mu^2_1 &= v^2\left(  \lambda_1  -2 \lambda_4 \frac{t_\beta}{ t_\beta^2+1}  \right) +v_L v_R   (\beta_1 - 2 \beta_3  t_\beta) \frac{t_\beta}{(t_\beta^2-1) } +\frac{v_L^2 v_R^2}{v^2} (\rho_3- 2 \rho_1 ) \frac{(t_\beta^2+1)}{(t_\beta^2-1) }  \notag\\
&\qquad\qquad +\left(v_L^2 + v_R^2\right)\left( \alpha_1 + \alpha_3 \frac{t_\beta^2}{t_\beta^2-1}\right)\,,\\
\mu_2^2 &= v^2 \left(\frac{\lambda_4}{2}-(2 \lambda_2 + \lambda_3) \frac{t_\beta}{1+t_\beta^2}\right) +\frac{v_L^2}{4} \left(2\alpha_2+\alpha_3\frac{t_\beta}{t_\beta^2-1}\right) +\frac{v_L v_R}{4}\left((\beta_1 - 2\beta_3 t_\beta)\frac{t_\beta^2+1}{t_\beta^2-1}\right) \notag \\ 
&\qquad\qquad -\frac{v_L^2 v_R^2}{2v^2} \left((2\rho_1-\rho_3) \frac{t_\beta(t_\beta^2+1)}{t_\beta^2-1}\right) + \frac{v_R^2}{2} \left(\alpha_2 + \frac{\alpha_3}{2} \frac{t_\beta}{t_\beta^2-1}\right)\,,\\
\mu_3^2 &= \frac{v^2}{2}\left(\alpha_1 +\left(\alpha_3 t_\beta-4\alpha_2\right)\frac{t_\beta}{t_\beta^2+1}\right)+(v_L^2+v_R^2) \rho_1 \,, \\
\beta_2 &= (\beta_1 - \beta_3 t_\beta) t_\beta - \frac{v_L v_R}{v^2} (2 \rho_1 - \rho_3)(1+t_\beta^2)\,.
\end{align}
\end{subequations}
From the last expression above one can derive the VEV seesaw relation as noted in~\cite{Deshpande:1990ip}.
Using the above expressions $\mu^2_i$, where $i=1,2,3$, and $\beta_2$ can be eliminated from 
the potential and the scalar mass matrices of the theory can be derived. These expressions are given 
in full detail in \cref{sec:scalarmasses}. Here we only quote the results after diagonalisation of 
the mass matrices, see \cref{sec:scalarmasses} for details on all assumptions made. Firstly, the bidoublet-like scalar masses:
\begin{subequations}
\begin{align}
m_h^2 &\simeq 2 \lambda_1 v^2-\frac{8 \lambda_4^2 v^4}{\alpha_3 v_R^2}\,, & 
m_H^2 &\simeq 2(2\lambda_2+\lambda_3)v^2 + \frac{\alpha_3}{2} v_R^2\,,  \label{eq:bidoubletHiggs_masses}  \\
m_{A}^2 &\simeq 2 \alpha_3 v_R^2 + 2(\lambda_3 - 2 \lambda_2)  v^2 \,, &
m_{H^{\pm}}^2&\simeq \frac{1}{4}\alpha_3(v^2+2v_R^2)\,.
\end{align}
\end{subequations}
Here, $h$ corresponds to the SM-like Higgs boson; $H, A$ and $H^\pm$ are the bidoublet-like heavier neutral scalar and pseudoscalar states as well as the mostly bidoublet-like charged Higgs. The triplet-scalar sector masses are:
\begin{subequations}
\begin{align}
m_{H_L}^2 &\simeq \frac{1}{2} \left(\rho_3 - 2 \rho_1\right) v_R^2\, &
m_{H_R}^2 &\simeq 2 \rho_1 v_R^2\,,  \\
m_{A_{L}}^2 &\simeq \frac{1}{2}(\rho_3 - 2\rho_1) v_R^2\,, &
m_{H^{\pm}_L}^2&\simeq \frac{1}{2}v_R^2(\rho_3 -2 \rho_1)\,, \\
m_{H^{\pm\pm}_1}^2&\simeq 2 \rho_2 v_R^2 +\frac{1}{2} \alpha_3 v^2\,, & 
m_{H^{\pm\pm}_2}^2&\simeq \frac{1}{2} \left((\rho_3-2\rho_1)v_R^2 + \alpha_3 v^2 \right)\,.
\end{align}
\end{subequations}
Particles with an index $L(R)$ mostly consist of $\Delta_{L(R)}$ components. The doubly-charged Higgses can in general be strongly mixed which is why we only label them as $H^{\pm\pm}_{1/2}$.

\section{Neutrino sector}
\label{sec:neutrinos}
Using information from neutrino oscillation experiments, we can  determine the neutrino mass matrix $m_{\nu}$ which we express as follows
\begin{align}
\mnul&=\Umns^*\text{diag}(m_{\nu_1},m_{\nu_2},m_{\nu_3}) \Umns^\dag\,,
\label{eq:mlight}
\end{align}
where $\Umns=\Umns(\theta_{12},\theta_{13},\theta_{23};\dCP)$ is the lepton mixing matrix and $m_{i}$ are the neutrino masses. Using the standard parametrisation in a basis where the lepton mass matrix is flavour-diagonal, the neutrino mixing matrix is given by
\begin{align}
\label{eq:Upmns}
\Umns =
\begin{pmatrix}
c_{12} c_{13}                          &  c_{13} s_{12}                        & s_{13} e^{i\dCP}\\
-c_{23} s_{12} - c_{12} s_{13} s_{23} e^{i\dCP}  & c_{23} c_{12} - s_{12} s_{13}  s_{23} e^{i\dCP}  & c_{13} s_{23}\\
 s_{23} s_{12}  -c_{12} c_{23} s_{13} e^{i\dCP} & - c_{12} s_{23} -c_{23} s_{12} s_{13}  e^{i\dCP} & c_{13} c_{23}
\end{pmatrix} K\,.
\end{align}
Here $c_{ij} = \cos \theta_{ij}, s_{ij} = \sin \theta_{ij}$, $\dCP$ 
corresponds to the Dirac CP-violating phase and $K$ is a complex diagonal matrix which contains the two Majorana phases. 
From global fits of neutrino oscillation parameters \cite{Forero:2014bxa,Gonzalez-Garcia:2014bfa,Capozzi:2013csa} 
the best fit values and the $3\sigma$ intervals for a normal neutrino mass hierarchy (NH) are:
\begin{subequations}
\begin{align}
\label{eq:nudata}
\sin^2\theta_{13}&= 0.0234^{+0.0060}_{-0.0057}\,, & \Delta m^2_{21}&=7.60^{+0.58}_{-0.49}\times10^{-5}\,\text{eV}\,, \\
 \sin^2\theta_{12}&=0.323^{+0.052}_{-0.045}\,, & \Delta m^2_{31}&=2.48^{+0.17}_{-0.18}\times10^{-3}\,\text{eV}\,, \\
 \sin^2\theta_{23}&=0.567^{+0.175}_{-0.076}\,.
\end{align}
\end{subequations}

\subsection{Neutrino masses}
From \cref{eq:model:dirac_masses} and (\ref{eq:model:majorana_masses}) the neutrino mass matrix follows as
\begin{align}
-\mathcal{L}_Y &\supset \frac{1}{2} \begin{pmatrix}\overline{\nu_L} & \overline{\nu_R^C}\end{pmatrix} \, M_\nu \, \begin{pmatrix}\nu_L^C\\ \nu_R \end{pmatrix} \,+\, {\rm h.c.}\,,
\end{align}
where
\begin{align}
\label{eq:neutrino_mass_matrix}
\mathcal{M}_{\nu} &=\begin{pmatrix}
   M_L^* &   M_D\\
   M_D^T         &   M_R
\end{pmatrix}\,.
\end{align}
In the above expression we have used the following definitions
\begin{align}\label{eq:ML_MR_mD_defs}
 M_{L}=\sqrt{2}\,Y_{\Delta_{L}}v_{L}\,,\ \ 
 M_{R}=\sqrt{2}\,Y_{\Delta_{R}}v_{R}\,,\ \ \text{and}\ \
 M_D=\frac{1}{\sqrt{2}}\left(Y_{L_{1}}v_1 + Y_{L_{2}}v_2\right)\,. 
\end{align}
Note the conjugate of $M_L$ in the (1,1) entry of \cref{eq:neutrino_mass_matrix}. This conjugate is crucial in the case of non-zero phases but is however usually forgotten in the literature.
Since $v_{R}\gg v_L,v_{1,2}\,$, the see-saw approximation can be used to determine the light-neutrino mass eigenstates, yielding
\begin{align} \label{eq:lightnumassmatrix}
\mnul&=\left(M_L^*-M_D\,M_{R}^{-1}\,M_D^T\right)\,.
\end{align}
As shown in \cref{eq:ML_MR_mD_defs}, the Dirac neutrino mass matrix $M_D$ arises as the sum of two different Yukawas multiplied with their respective VEVs. Consequently at loop-level, corrections are proportional to the individual Yukawa coupling values rather than $M_D$. 
Therefore, in regions where $\tan\beta \simeq 1\,$, loop corrections to these two Yukawas spoil the cancellation required for small $M_D$ values if imposed at tree-level. 
As $\tan \beta$ has negligible impact on the lepton flavour-violating operators discussed below, we choose to restrict our analysis to the small $\tan\beta$ scenario in the following numerical studies. In this limit $M_D\propto Y_{L_1} v$, while the charged lepton masses are $M_\ell \propto Y_{L_2} v$. 

\subsection{Parametrisation of the Yukawa matrices}\label{sec:ParameterisationYuks}
Under the discrete symmetries of the theory, namely parity $\mathcal P$ and charge-conjugation $\mathcal C$, the resulting light-neutrino mass matrices can be re-expressed as
\begin{align} 
\mnul &\overset{\mathcal{C}}{=}\left(\frac{v_L}{v_R} \,M_R -M_D\,M_{R}^{-1}\,M_D\right)\,,\label{eq:lightnumassmatrix}\\
 \mnul&\overset{\mathcal{P}}{=}\left(\frac{v_L}{v_R} \,M_R^* -M_D\,M_{R}^{-1}\,M_D^*\right)\,. \label{eq:lightnumassmatrixParity}
\end{align}
Both discrete symmetries exhibit favourable structures, relating $M_L$ to $M_R$. In particular, this enables an elegant parametrisation for fitting the neutrino masses which we will outline in what follows.

The parametrisation, first proposed in Ref.~\cite{Akhmedov:2005np}, allows one to explicitly solve for the triplet-Yukawa couplings $Y_{\Delta_{L}}$ and $Y_{\Delta_{R}}$ given a specific input for $M_D$. The parametrisation relies on solving a quadratic polynomial for each diagonal entry of \cref{eq:lightnumassmatrix} or \cref{eq:lightnumassmatrixParity} after diagonalisation. Here, our method differs slightly to Ref.~\cite{Akhmedov:2005np}. We have exploited the fact that \cref{eq:lightnumassmatrix,eq:lightnumassmatrixParity} can both be manipulated into a form requiring only a single unitary
rotation matrix $R$ to bring both sides into their respective diagonal forms. Full details of the procedure can be found in \cref{sec:TripCoupParameterization}. We can therefore express the right-triplet-Yukawa as
\begin{align}\label{eq:TripletCouplSolution}
Y_\Delta^{(\pm\pm\pm)} \equiv  Y_{\Delta_{L/R}}^{(\pm\pm\pm)} &= \frac{1}{{2\sqrt{2} v_L}} M_D^{(*)1/2}R^* \text{diag}\left(\BDi \pm \sqrt{\left(\BDi\right)^2 + 4 \alpha}\right)
R^\dag M_D^{1/2}\,,
\end{align}
where $B_D = R^{\dagger} M_D^{-1/2} \mnul M_D^{(*)\,-1/2} R^*$ is a diagonal $3\times3$ matrix, $R$ is the aforementioned rotation matrix and $\alpha=v_L/v_R$. Finally $(*)$ is an additional conjugation of $M_D$ required in the case of a parity symmetric neutrino sector. \cref{eq:TripletCouplSolution} is valid for:
\begin{enumerate}
\item[($i$)] both possible discrete left-right symmetries if $\dCP = 0$,
\item[($ii$)] all possible CP phases if the Lagrangian is $\mathcal P$-symmetric.
\end{enumerate}
As before, further details can be found in \cref{sec:TripCoupParameterization}. 

This result forms the basis of our subsequent numerical studies. By choosing a form for $M_D$ and requiring that $m^{\rm light}_\nu$ satisfies \cref{eq:mlight}, one can determine $R$ such that $B_D$ is diagonal. $R$ therefore contains the information from the experimental neutrino data.
From \cref{eq:TripletCouplSolution} we see that there does not exist a unique solution to the triplet-Yukawa. Rather for each diagonal entry there appears a sign choice in front of the square-root. Considering the possible permutations, there are in total eight unique solutions. This parametrisation is therefore advantageous in comparison to the Casas-Ibarra-like parametrisations \cite{Casas:2001sr} as it by construction respects the discrete symmetries of the theory. This is crucial, as the finite number of solutions is a direct consequence of invariance under a discrete left-right symmetry.

\section{Results}
\label{sec:results}
\subsection{Numerical Set-up}
In this section we present a numerical study of the model. In order to do so we have used the {\tt Mathematica} package {\tt SARAH} \cite{Staub:2008uz,Staub:2009bi,Staub:2010jh,Staub:2012pb,Staub:2013tta,Staub:2015kfa} for which we have created the necessary model files, see \cref{sec:app:model_file}. Along with this paper, the respective code is also available on the \href{https://sarah.hepforge.org/trac/wiki}{{\SARAH} model database}. {\tt SARAH} interfaces to the spectrum generator {\tt SPheno} \cite{Porod:2003um,Porod:2011nf} which  enables the computation of the mass spectrum 
and particle decays as well as quark and lepton flavour violating observables via the the link to {\tt FlavorKit} \cite{Porod:2014xia}.

\begin{figure}
\includegraphics{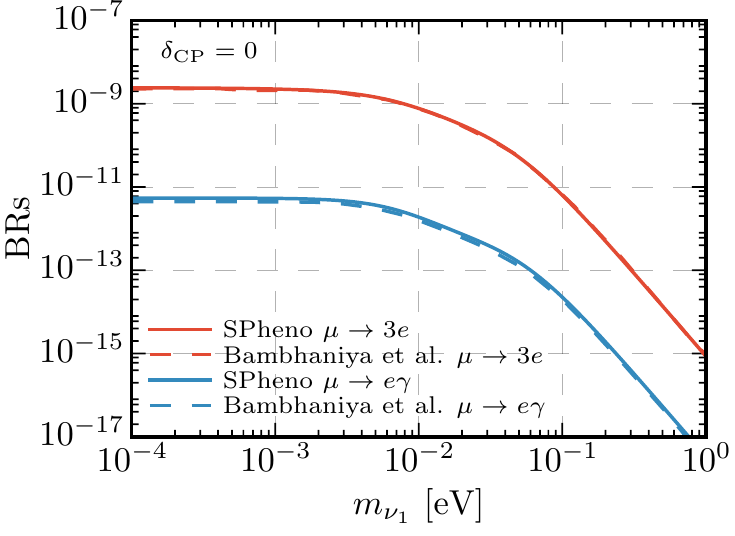}
\includegraphics{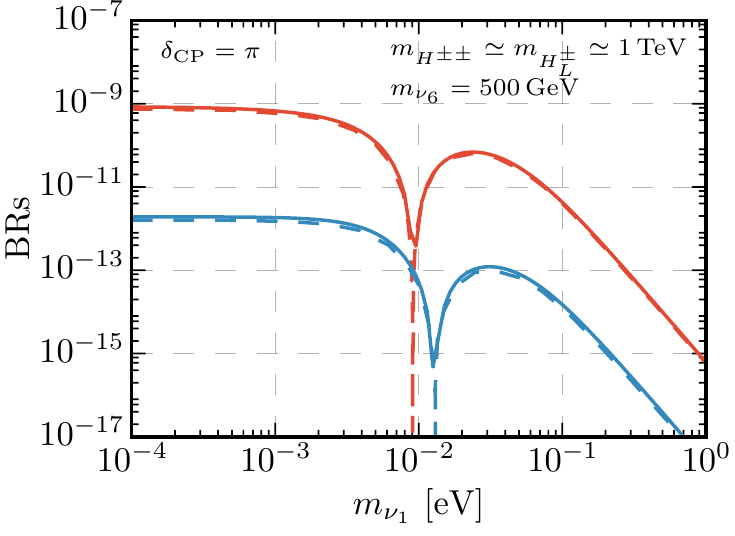}
\caption{Comparison of the {\tt SPheno} code with results in Fig.~3.4 from Ref.~\cite{Bambhaniya:2015ipg}.}
\label{fig:Validation_plots}
\end{figure}

As a first step we have compared the $\mu\to 3e$ and $\mu\to e \gamma$ branching ratios with those from Ref.~\cite{Bambhaniya:2015ipg}. To do so we consider a similar setup where $M_L=0$ and $M_D \propto \mathbb{1}$ leading to a pure type-I seesaw mechanism where the light neutrino masses and mixings are encoded in $Y_{\Delta_R}$ couplings. In addition, Ref.~\cite{Bambhaniya:2015ipg} neglected contributions arising from both neutral scalars and $W_L-W_R$ mixing which is a well justified approximation. Shown in \cref{fig:Validation_plots} are the rates for $\mu \to 3e$ and $\mu \to e \gamma$ from this work (solid lines) and, for comparison, the results from Fig. 3.4 of Ref.~\cite{Bambhaniya:2015ipg} (dashed lines), where the triplet masses are set to \SI{1}{\TeV}. We observe good agreement between the respective results, with only small deviations in the rates for $\mu \to e \gamma$.
The main reason for these small deviations is that our analysis considers a complete model where the scalar masses are a function of the model parameters. This prevents one from varying the scalar masses independently. Therefore the resulting spectrum does not correspond exactly to the mass choices of Ref.~\cite{Bambhaniya:2015ipg}. As both of the observables are highly sensitive functions of the scalar masses,  a 5\% deviation in the mass spectrum leads to the observed small mismatch in the flavour observables.

In the subsequent analysis we study lepton flavour violating rare decays based on the best-fit NH oscillation parameters given in \cref{eq:nudata} choosing the lightest mass to be $m_{\nu_1}=10^{-4}\,$eV. We consider the impact of varying these two choices in \cref{sec:alt_neutrino_params}. Lastly, we choose $\dCP=0$, but consider non-zero choices and $\dCP=3\pi/2$, as suggested by recent global fits \cite{Esteban:2016qun}, in later sections. The model parameters used, unless otherwise stated, are given in \cref{tab:benchmark}. The value chosen for $v_R$ leads to $W_R$ and $Z_R$ masses which are outside of the reach of the LHC. However, in the presence of a low-scale discrete $\mathcal C$  symmetry, the $K$- and $B$-meson constraints only allow the heavy bidoublet Higgs to be as `light' as 20~TeV \cite{Bertolini:2014sua} which, in combination with a perturbativity constraint on $\alpha_3$, dictates a lowest possible $v_R$ value of $\sim 15~$TeV, cf. \cref{eq:bidoubletHiggs_masses}. This can lead to scalar triplet masses of $\mathcal O({\rm 1~TeV})$ and therefore within the LHC reach, it however pushes $M_{W_R,Z_R}$ to $\mathcal O({\rm 10~TeV})$. 

\begin{figure}
\includegraphics[width=0.5\linewidth]{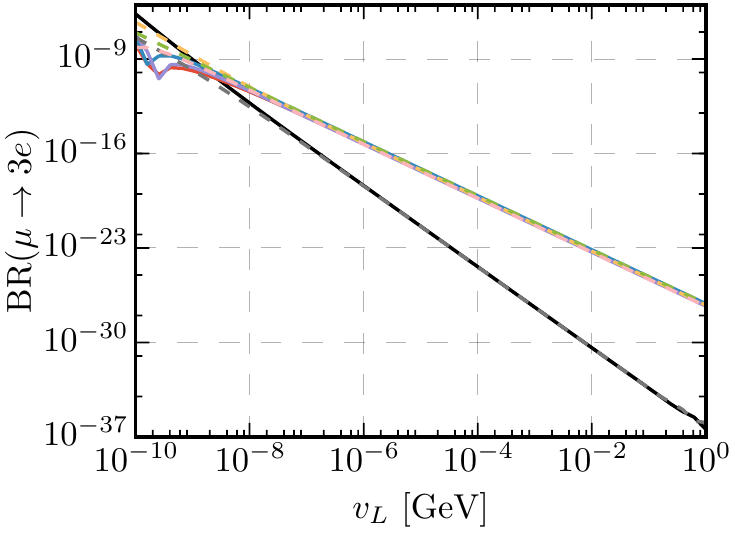}
\includegraphics[width=0.5\linewidth]{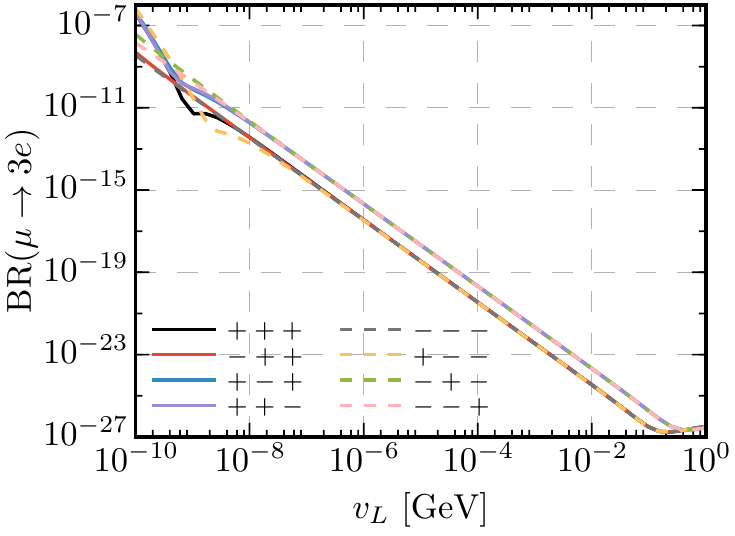}
\caption{Dependence of the observable BR$(\mu\to3e$) on the eightfold degenerate solutions in the cases that $M_D=x \mathbb{1}$ [GeV]  (left-hand panel) and $M_D =x\MUPCKM$ (right-hand panel), where in both cases $x=\SI{E-4}{}$.}
\label{fig:Mu3e_sign_dep} 
\end{figure}

The remaining parameters and choices which we investigate are as follows:
\begin{itemize}
\item $v_L$, which we typically vary between \SI{0.1}{\eV} and \SI{1}{\GeV}.
\item $M_D$, the Dirac neutrino mass matrix which in our parametrisation is an input parameter. We study three different possibilities:
\begin{enumerate}[(i)]
\item $M_D = x\mathbb{1}\,$GeV,
\item $M_D = x \MUP\,$,
\item $M_D = x \MUPCKM\,$,
\end{enumerate}
where $\MUP$ is the diagonal up-type quark-mass matrix. For each choice we have also added the parameter $x$, which we use to vary the overall mass scale of the  matrix $M_D$. 
\item Sign choice of the diagonal $\pm$ signs appearing in \cref{eq:TripletCouplSolution}. In the numerical studies we investigate two different choices of the possible eight, namely $(+++)$ and $(+-+)$. This is well motivated as these eight solutions can be divided into two subgroups, whereby each subgroup leads to similar results. This is demonstrated in \cref{fig:Mu3e_sign_dep}, where we show the branching ratio for $\mu\to 3e$ for all eight sign choices varying $v_L$, with two different extreme examples of $M_D$.  Here we clearly see the grouping of the eight solutions into the two classes (i) {\it same-sign} and (ii) {\it mixed-sign} solutions. 
\end{itemize}

\begin{table}
\begin{center}
\begin{tabular}{lr@{\hspace{5em}}lr}\toprule
\multicolumn{4}{c@{}}{Model Parameters} \\\midrule
$\lambda_1$ & \SI{0.13}{} & $v_L$ & $10^{-10}...$ \SI{1}{\GeV} \\
$\lambda_2$ & \SI{1.0}{} & $v_R$ & \SI{20}{\TeV}\\
$\lambda_3$ & \SI{1.0}{} & $\tan\beta$ & $10^{-4}$\\
$\lambda_4$ & \SI{0}{} & $\alpha_1$ & \SI{0}{} \\
$\rho_1$ & \SI{3.2E-4}{} &$\alpha_2$ & \SI{0}{} \\
$\rho_2$ & \SI{2.5E-4}{} &  $\alpha_3$ & \SI{2.0}{} \\
$\rho_3$ & \SI{1.8E-3}{} & $\beta_1$  & \SI{0}{} \\
$\rho_4$ & \SI{0}{}& $\beta_2$ & \SI{3.83E-4}{} \\
$\mu_1^2$ & \SI{7.87E+3}{\GeV^2} & $\beta_3$ & \SI{0}{} \\
$\mu_2^2$ & \SI{-2.00E+4}{\GeV^2} & $\mu_3^2$ & \SI{1.28E+5}{\GeV^2}\\ \midrule
\multicolumn{4}{c@{}}{Resulting Mass Spectrum} \\ \midrule
$m_h$& \SI{125.5}{\GeV} & $m_H$ & \SI{20}{\TeV}\\
$m_{A}$ & \SI{20}{\TeV} & $m_{H^\pm}$ & \SI{20}{\TeV}\\
$m_{H_L}$& \SI{482}{\GeV} & $m_{H_R}$& \SI{506}{\GeV} \\ 
$m_{A_L}$ & \SI{482}{\GeV} &$m_{H^\pm_L}$& \SI{512}{GeV} \\ 
$m_{H^{\pm\pm}_1}$& \SI{511}{\GeV} & $m_{H^{\pm\pm}_2}$ & \SI{541}{\GeV}\\
$M_{W_R}$ & \SI{9.37}{\TeV} & $M_{Z_R}$ & \SI{15.7}{\TeV} \\
\bottomrule
\end{tabular}
\end{center}
\caption{Benchmark point used in the subsequent LFV study. All parameters and masses are compatible with the constraints derived in Refs.~\cite{Maiezza:2016bzp,Bertolini:2014sua}.}
\label{tab:benchmark}
\end{table}

\subsection{Numerical Results}
\label{sec:numericalresults}

\begin{figure}
\begin{center}
\includegraphics[width=.17\linewidth]{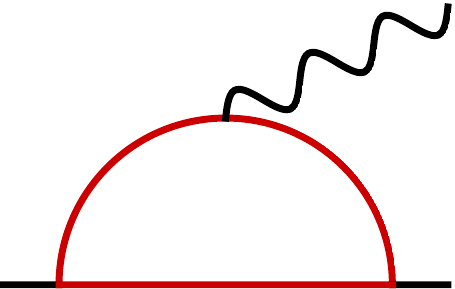}\hspace{.02\linewidth}
\includegraphics[width=.17\linewidth]{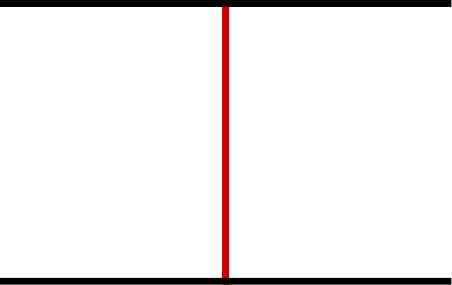}\hspace{.02\linewidth}
\includegraphics[width=.17\linewidth]{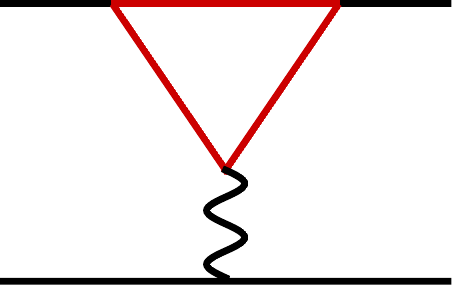}\hspace{.02\linewidth}
\includegraphics[width=.17\linewidth]{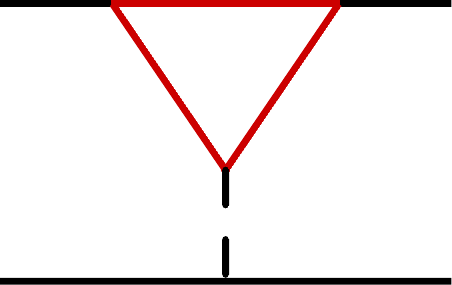}\hspace{.02\linewidth}
\includegraphics[width=.17\linewidth]{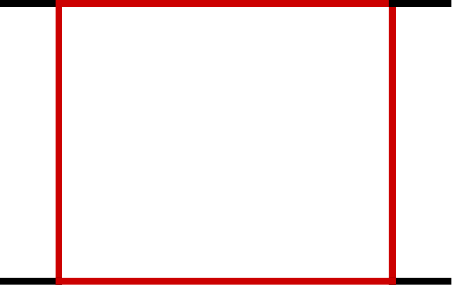}
\end{center}
\caption{Representative lepton flavour violating Feynman diagrams. Here, red solid lines represent particles of all spins. Diagrams of the left-hand type lead to the
radiative lepton decays $\ell_\alpha \to \ell_\beta \gamma$. The other four diagrams induce LFV three-body decays as well as  $\mu-e$ conversion in nuclei. We shall label them {\it ``tree-level scalar''}, {\it ``vector penguin''}, {\it ``scalar penguin''} as well as {\it ``box''} contributions.}
\label{fig:diagrams}
\end{figure}

As pointed out beforehand, the free parameters in our study which determine the neutrino sector are $M_D$, $v_L$ as well as $\delta_{\rm CP}$. As we shall see, they are crucially important for determining which type of diagram dominates the lepton flavour violating process. 
We decompose the relevant diagrams into different categories which are depicted in \cref{fig:diagrams}. 

The radiative decays $\ell_\alpha \to \ell_\beta \gamma$ are described by the first type of diagram, the vector line corresponding to an on-shell photon whereas the particles running in the loop can be (i) $H^{\pm \pm}_i-\ell_\delta^\mp$, (ii) $H^{0}_i-\ell_\delta^\pm$, (iii) $H_i^\pm - \nu_j$, (iv) $W_{L/R}^\pm - \nu_j$ (where $j=1,\dotso,6$).

The three-body decays as well as $\mu-e$ conversion processes receive contributions from both tree-level as well as one-loop diagrams. As the heavy neutral bidoublet-like Higgs $H$ couples to both leptons and quarks generically in a flavour-non-conserving manner, it contributes to both $\mu -e$ conversion as well as $\ell_\alpha \to \ell_\beta \ell_\gamma \ell_\delta$. Depending on the flavour structure of the lepton Dirac Yukawa couplings, this contribution can be both sizeable or small (in case of a flavour-diagonal $M_D$, its contribution is zero). 
The tree-level diagram mediated by the doubly-charged scalars vanishes for the $\mu-e$ conversion processes since the triplet doesn't couple to quarks. In case of the the LFV three-body decays one can expect in large portions of the parameter space a dominance of those tree-level diagrams since $Y_\Delta$ is typically much larger than the Dirac Yukawas. It is interesting to note that the $\tau$ three-body decays with a mixed $e/\mu$ final state, $\tau^\pm \to \ell_\alpha^\mp \ell_\beta^\pm \ell_\beta^\pm $ are much more frequent than $\tau^\pm \to \ell_\alpha^\pm \ell_\beta^\mp \ell_\beta^\pm $ whenever the triplet tree-level diagram is dominating the LFV observables and the flavour-violating $Y_\Delta$ entries are small; this is simply because of the doubly-charged mediator: the process $\tau^\pm \to \ell_\alpha^\pm \ell_\beta^\mp \ell_\beta^\pm $ needs a flavour-violating coupling at each vertex whereas $\tau^\pm \to \ell_\alpha^\mp \ell_\beta^\pm \ell_\beta^\pm $ contains one flavour-violating and one flavour-conserving vertex. This is in contrast to the loop-induced contributions including virtual neutral or singly-charged bosons which, in order for a $\tau^\pm \to \ell_\alpha^\mp \ell_\beta^\pm \ell_\beta^\pm $ decay to happen, require at least two flavour-violating vertices in the dominant contributions \cite{Abada:2014kba}.
\begin{figure}
\includegraphics{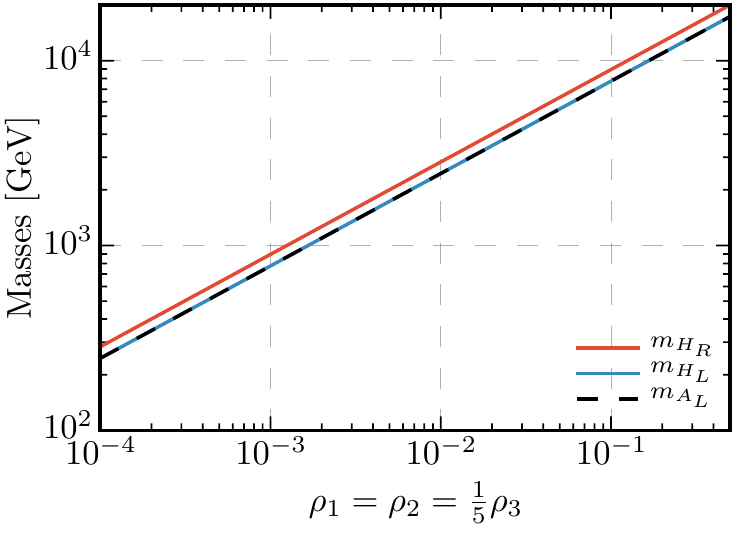}
\includegraphics{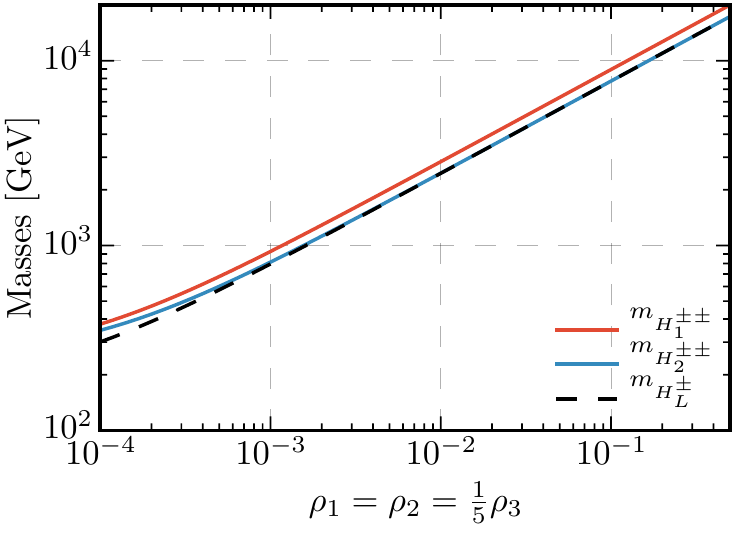}
\caption{Variation of the neutral and charged triplet scalar masses that are used in subsequent figures. Here, the values of the additional parameters not shown in the figure are given in \cref{tab:benchmark}. The light and heavy neutral bi-doublet masses are fixed to \SI{125.5}{\GeV} and \SI{20}{\TeV}, respectively. }
\label{fig:VaryTripletScalarMasses}
\end{figure}

The remaining diagrams are scalar and vector penguins as well as box diagrams. It is known from studies in other models with low-scale seesaw mechanisms that the boxes and vector penguins with $W_L$ bosons and right-handed neutrinos running in the loop can be very important
\cite{Ilakovac:2009jf,Alonso:2012ji,Dinh:2012bp,Ilakovac:2012sh,Abada:2014kba}. In left-right symmetric theories, other very important contributions arise from triplet scalars and neutrinos/leptons  in the loop as well as $W_R - \nu_R$ diagrams. Diagrams including a $W_{L/R}$ and a right-handed neutrino in the loop are expected to be important in the case of small $Y_\Delta$. While penguin diagrams featuring triplet-scalars in the loop are loop-suppressed with respect to the corresponding tree-level diagrams, certain flavour structures of $Y_\Delta$ may suppress the tree-level w.r.t.\ the loop-level diagrams. We shall see examples of this behaviour later on; see, for instance, \cref{subsec:Mu}.

We now start the discussion by looking at the different contributions to the LFV observables as a function 
of the model parameters. In particular, we will vary the masses of the triplet scalars while keeping the bidoublet masses constant. We will do so 
choosing different parametrisations of $M_D$ and values for $v_L$. The reader should be reminded that $v_L$ not only determines the size of the seesaw-II contribution to the neutrino masses, see \cref{eq:neutrino_mass_matrix}, but also feeds into the determination of $Y_\Delta$ for a given $M_D$ following \cref{eq:TripletCouplSolution}.\\

\subsubsection{Case I: $M_D \propto \mathbb 1$}
\label{subsec:diag_MD}
Let us first examine the simplest case where the Dirac neutrino mass is diagonal and flavour-universal. This results in, for the majority of the parameter space, an almost degenerate spectrum of right-handed neutrinos due to almost degenerate diagonal $Y_\Delta^{(i,i)}$ entries. More importantly, all the lepton flavour violation arises through the triplet Yukawas, meaning that the bidoublet states have only lepton flavour-conserving interactions. Quite generically, this also means that the rather uniform structure of neutrino mixing is translated to the triplet Yukawas. Hence, there is no large hierarchy between the Yukawa matrix elements which mix the 1st, 2nd or 3rd generation.\footnote{In this context, `no large hierarchy' means no more than an order of magnitude of difference, therefore small compared to the hierarchy in quark flavour mixing.} 

{\emph{\bf{$(+++)$ Solution.}}}
As a numerical example, choosing $v_L=\SI{2E-7}{\GeV}$, $M_D = \mathbb{1}\SI{}{\MeV}$ \cref{eq:TripletCouplSolution}, yields
\begin{equation}
Y_\Delta^{(+++)} = \begin{pmatrix}
\SI{1.12E-2}{}  & \SI{-1.41E-5}{}   & \SI{2.97E-6}{}  \\
\SI{-1.41E-5}{} & \SI{1.12E-2}{}   & -\SI{3.78E-5}{}   \\
\SI{2.97E-6}{} & -\SI{3.78E-5}{} & \SI{1.12E-2}{}
\end{pmatrix}\,.
\label{eq:Ydelta_diag_ppp}
\end{equation}
\begin{figure}
\includegraphics{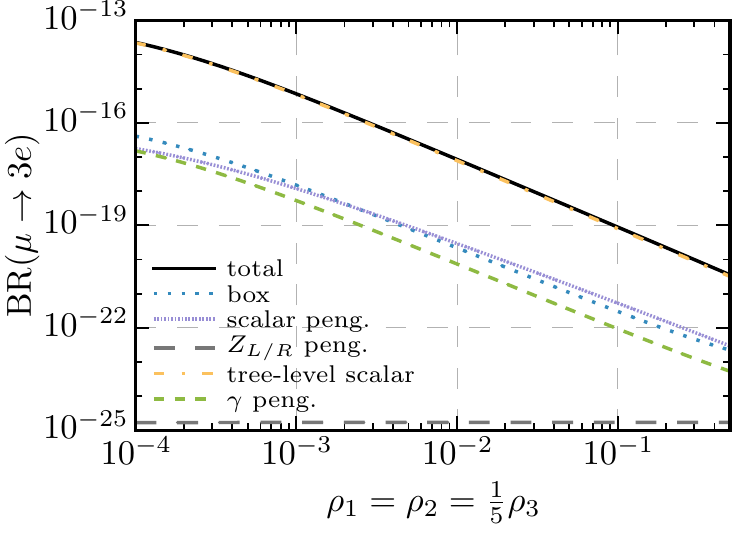}
\includegraphics{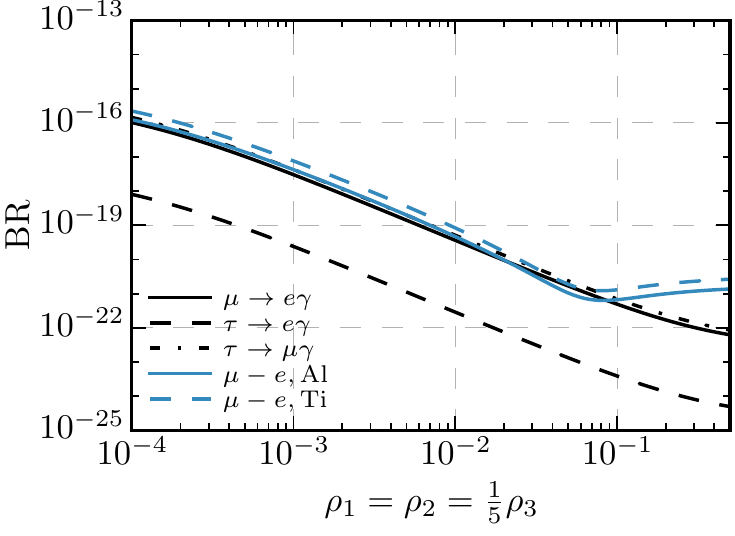}\\
\includegraphics{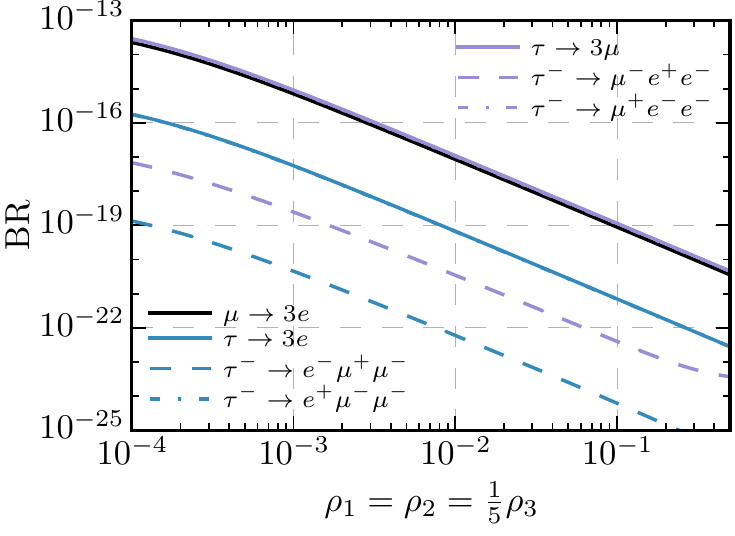}
\includegraphics{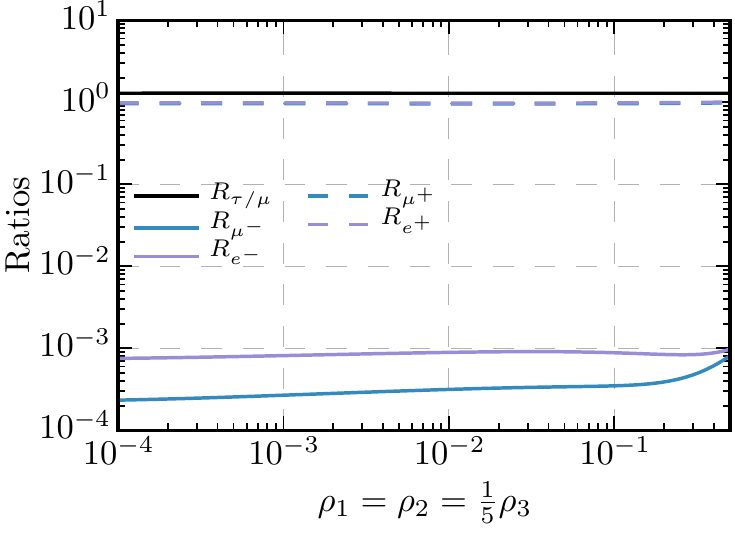}
\caption{A cross-section of different LFV observables for the choice $v_L=\SI{2E-7}{\GeV}$, $M_D = \mathbb{1}\,\SI{}{\MeV}$ and the Yukawa solution sign choice $(+++)$. Top left: Total branching ratio of $\mu \to 3e$ and the different contributing types of diagrams. Top right: $\ell_\alpha \to \ell_\beta\gamma$ and $\mu-e$ conversion in different nuclei. Bottom left: Different 3-body decay channels of muons and taus, note that the channels $\tau^-\to e^+ \mu^-\mu^-$ and $\tau^-\to \mu^+ e^-e^-$ cannot be seen as as they lie very close to the branching ratios $\tau \to 3\mu$ and $\mu \to 3e$, respectively. Bottom right: Ratios of the different 3-body decay modes, see \cref{eq:ratiolabels} for a description of the labels.}
\label{fig:diag_scalar_variation}
\end{figure}
From here we can already draw some conclusions: (i) the doubly-charged Higgs as the tree-level mediator dominates the LFV three-body decays, which means that (ii)
the magnitudes of the $\mu$ and $\tau$ LFV decays are of comparable size (within at most an order of magnitude or two) and that (iii) the three-body decays are much more abundant than the radiative decays $\ell_\alpha \to \ell_\beta \gamma$. A LFV process observed at the Mu3e experiment with no evidence for $\mu \to e \gamma$ would therefore be a smoking gun for these scenarios with LFV triplet scalar interactions. 

We will now move to discussing numerical examples starting with the dependence of various LFV observables on the triplet scalar sector. Unless noted otherwise, all model parameters are chosen as given in \cref{tab:benchmark}. We therefore vary the model parameters $\rho_1$, $\rho_2$ and $\rho_3$, where we show the resulting masses in \cref{fig:VaryTripletScalarMasses}.

In the left upper panel of \cref{fig:diag_scalar_variation}  the magnitude of $\mu \to 3e$ is shown using the parametrisation of \cref{eq:Ydelta_diag_ppp} with the different diagrammatic contributions split according to \cref{fig:diagrams}. As discussed above, the tree-level diagram with a doubly-charged mediator completely dominates over other contributions. In the lower left panel we also show the other LFV three-body decays. The radiative decays, shown on the upper right panel, are smaller by roughly two orders of magnitude which is due to the loop suppression w.r.t.\ the three-body decays. The reason why BR$(\tau \to e\gamma)\ll {\rm BR}(\tau\to \mu\gamma)$, BR$(\mu\to e\gamma)$ as well as BR$(\tau \to 3e)\ll {\rm BR} (\tau\to 3\mu)$, BR$(\mu\to 3 e)$ is simply the order of magnitude difference between $Y_\Delta^{(1,3)}$ and the other two off-diagonal Yukawa entries. For the $\mu-e$ conversion observables we first see a decrease of the conversion rate with an increasing mass scale of the triplet scalar sector.  The reason is that for this choice of parameters, for triplet masses up to \SI{5}{\TeV} the $\gamma$-penguin diagrams with triplets running in the loop are dominating. For higher scalar masses, the $W_{L/R} -\nu_R$-mediated box diagrams which are independent of the scalar sector parameters become more important (as the triplets don't couple to quarks, the most important $\mu-e$ conversion box contribution is always coming from these internal particles). For a heavy scalar sector, we can therefore even have CR$(\mu-e) > {\rm BR}(\mu\to e\gamma)$; this could be interesting for future experiments which have  better prospects for sensitivity in  $\mu-e$ conversion than for the radiative muon decay. For $\mu\to 3e$,  the size of the boxes is determined by the triplets for all of the parameter regions shown. Finally in the lower right panel of \cref{fig:diag_scalar_variation} we show ratios of the three-body branching ratios. The labels in the figure correspond to
\begin{align}\label{eq:ratiolabels}
R_{\tau/\mu} &= \frac{\text{BR}(\tau \to 3\mu)}{\text{BR}(\mu \to 3e)}\,, \quad
R_{e^\mp} = \frac{\text{BR}(\tau^- \to e^\mp \mu^\pm \mu^-)}{\text{BR}(\tau \to 3e)}\,, \quad
R_{\mu^\mp} = \frac{\text{BR}(\tau^- \to \mu^\mp e^\pm e^-)}{\text{BR}(\tau \to 3\mu)}\,.
\end{align}

\begin{figure}
\includegraphics{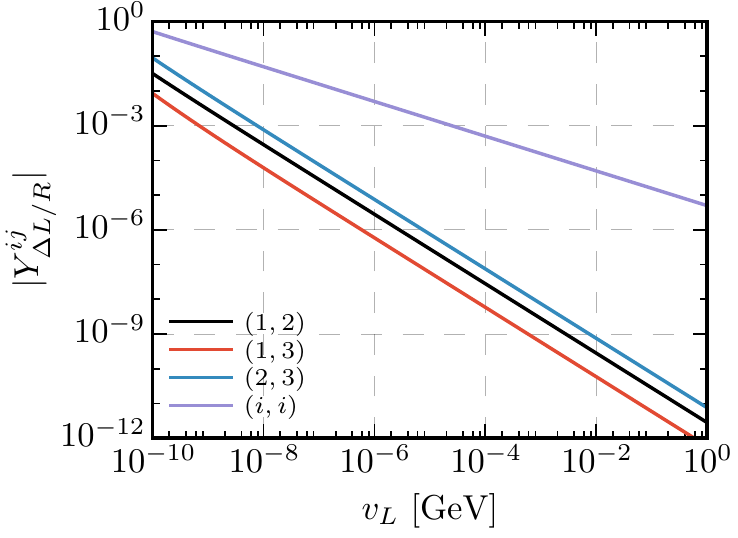}
\includegraphics{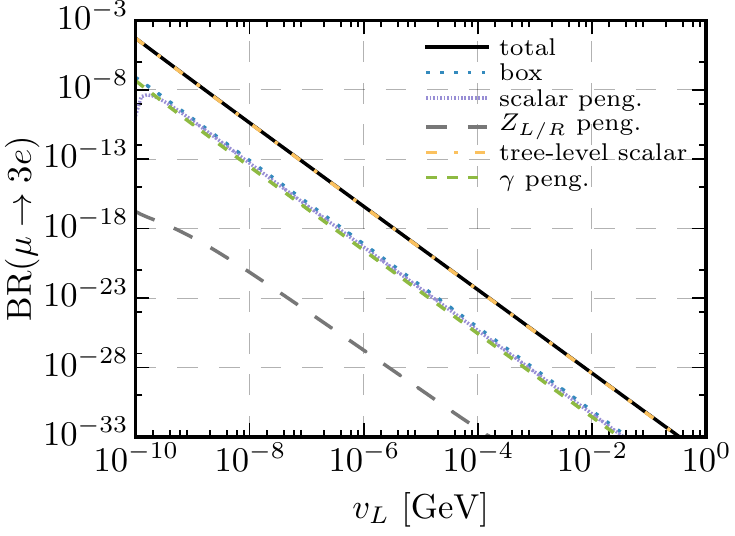}
\caption{Dependence of the triplet Yukawa coupling on left triplet VEV $v_L$ using the $(+++)$ solution of \cref{eq:TripletCouplSolution} (left) and the consequential decoupling of the different contributions to $\mu\to 3e$ (right). }
\label{fig:diag_vL_variation}
\end{figure}

Let us now fix the scalar sector to the benchmark values of \cref{tab:benchmark} and consider the dependence of the LFV rates on the input parameter $v_L$ which we vary from \SI{0.1}{\eV} to \SI{1}{\GeV}. It is important to realize that, for these parameter values, $\sqrt{v_L/v_R} \gg \BDi$, where $B_D = R^{T} M_D^{-1/2} m_\nu^{\rm light} M_D^{-1/2} R$ as used in \cref{eq:TripletCouplSolution}. Therefore the diagonal elements of $Y_\Delta$ approximately scale with $1/\sqrt{v_L}$. The off-diagonal $Y_\Delta$ elements, however, vanish to zeroth order in $\sqrt{v_L/v_R}/\BDi $ for diagonal $M_D$ and both the $(+++)$ or $(---)$ solutions, see \cref{sec:TripCoupParameterization} for further details. Therefore, they are generated by the terms proportional to $B_D$. With the overall $1/v_L$ pre-factor in \cref{eq:TripletCouplSolution}, the off-diagonal $Y_D$ entries decouple like $1/v_L$. This is numerically shown in the left-hand panel of \cref{fig:diag_vL_variation}. On the right-hand panel we show the corresponding decoupling behaviour of the muon three-body decay. The other observables scale accordingly.

\begin{figure}
\includegraphics{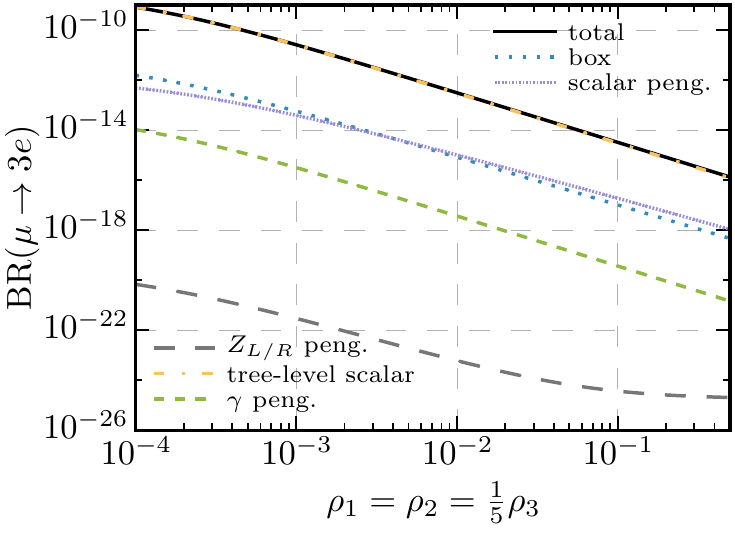}
\includegraphics{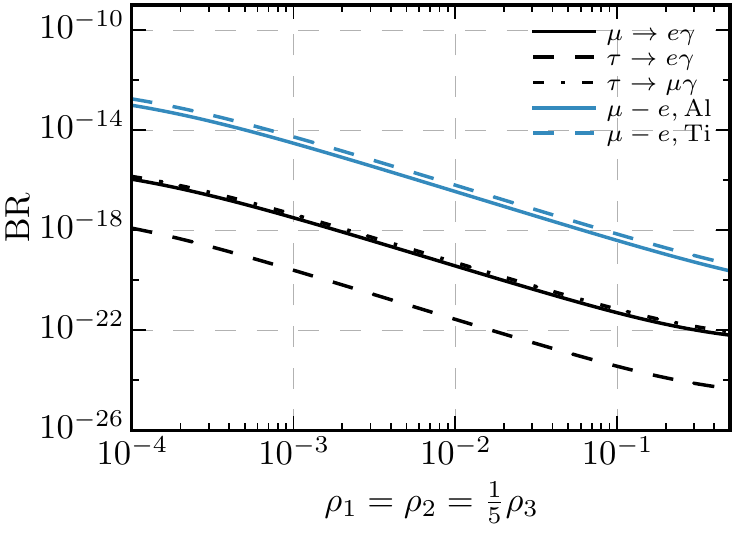}\\
\includegraphics{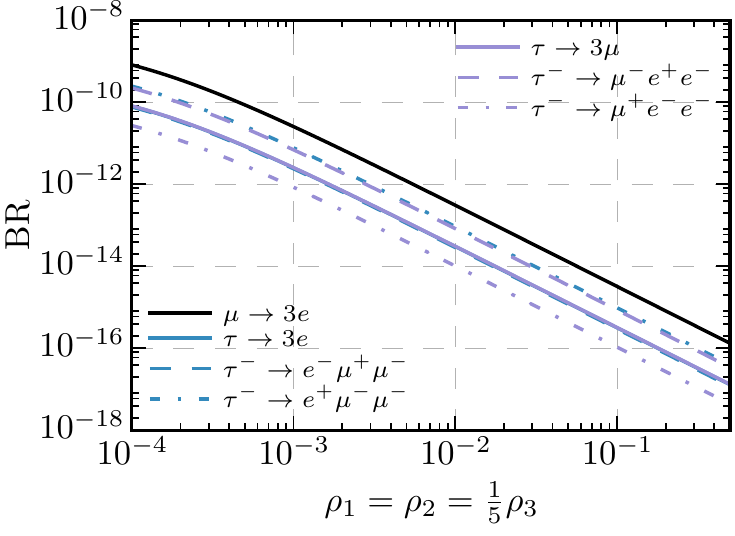}
\includegraphics{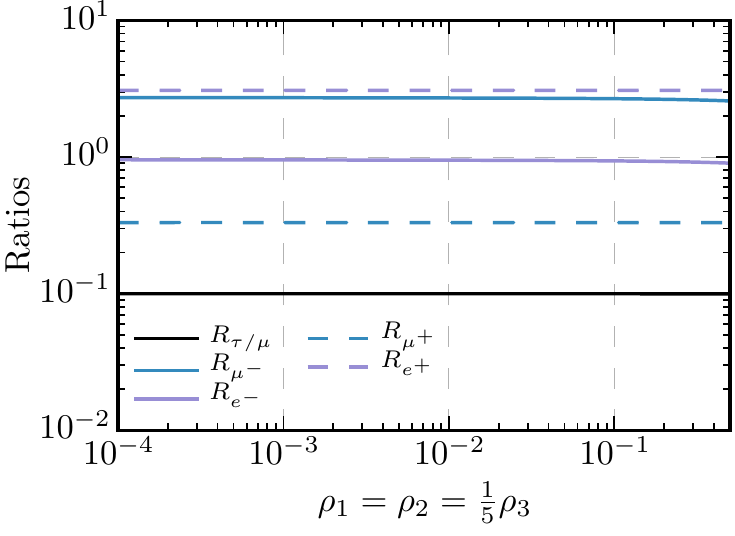}
\caption{A variety of different LFV observables for the choice $v_L=\SI{2E-7}{\GeV}$, $M_D = \mathbb{1}\SI{}{\MeV}$ and the Yukawa solution sign choice $(+-+)$. For an explanation of the four panels see \cref{fig:diag_scalar_variation}. Note, that the cases with $\rho_1 \lesssim  \SI{6E-3}{}$ 
are excluded by the bounds on $\mu\to 3 e$.}
\label{fig:diag_sign_3_scalar_variation}
\end{figure}

{\emph{\bf{$(+-+)$ Solution.}}}
Let us now consider another possibility out of the eight different solutions for $Y_\Delta$ according to \cref{eq:TripletCouplSolution}. As illustrated in detail in \cref{sec:TripCoupParameterization}, the choice of the solution is of particular importance in the case where $M_D$ is diagonal: while the flavour-conserving $Y_\Delta$ elements get reduced by less than an order of magnitude when switching from a $(+++)$ or $(---)$ solution to one with differing sign choices, the 
flavour-violating entries get enhanced sizeably. The reason is that for $Y_\Delta^{(k,l)}$ the entries with $k\neq l$ do not vanish at zeroth order in $\sqrt{v_L/v_R}/\BDi$.  For comparison, using the chosen benchmark point, the Yukawa matrix from the $(+++)$ case in \cref{eq:Ydelta_diag_ppp} reads for the $(+-+)$ case
\begin{equation}
Y_\Delta^{(+-+)} = \begin{pmatrix}
\SI{-3.61E-3}{} & \SI{-8.53E-3}{} & \SI{-6.27E-3}{}  \\
\SI{-8.53E-3}{}  & \SI{6.33E-3}{} & \SI{-3.65E-3}{}   \\
\SI{-6.27E-3}{} & \SI{-3.65E-3}{} & \SI{8.56E-3}{}
\end{pmatrix}\,.
\label{eq:Ydelta_diag_pmp}
\end{equation}
Naturally, this results in a rate enhancement of the LFV observables by many orders of magnitude. In \cref{fig:diag_sign_3_scalar_variation} we show the analogue to \cref{fig:diag_scalar_variation} but this time using the $(+-+)$ solution. We see an interesting effect here: while $\mu \to 3e$ is enhanced by roughly four orders of magnitude, $\mu-e$ conversion observables are only enhanced by three orders. The radiative decays, in turn, are hardly changed at all. The reason for this is as follows. The three-body decays are still dominated by the tree-level $H^{\pm\pm}$ mediation; therefore their amplitude scales with the respective off-diagonal $Y_\Delta$ entry which is enhanced by three orders of magnitude from \cref{eq:Ydelta_diag_ppp} to \cref{eq:Ydelta_diag_pmp}. For the radiative decays, the diagrams with a charged lepton and a doubly-charged Higgs in the loop dominate. For each decay, the internal lepton can be electron, mu or tau flavoured. Taking as an example the decay $\mu \to e \gamma$, the coupling combination entering the amplitude is therefore $Y_\Delta^{(2,1)} Y_\Delta^{(1,1)} c_e + Y_\Delta^{(2,2)} Y_\Delta^{(2,1)} c_\mu + Y_\Delta^{(2,3)} Y_\Delta^{(3,1)} c_\tau $, where $c_i$ denotes the loop function depending on $m_{\ell_i}$ and $m_{H^{\pm\pm}}$. For the photonic dipole loop functions we find that $c_e\simeq c_\mu \simeq c_\tau$. Taking the respective $Y_\Delta^{(k,l)}$ entries from \cref{eq:Ydelta_diag_pmp} we then observe a cancellation between the different terms so that the sum is actually almost as small as the respective combination using the values from the $(+++)$ parametrisation. This leads to an almost unchanged magnitude of the radiative decays from one case to the other. The $\mu-e$ conversion rates are also dominated by the photon penguin; however, what enters here is the monopole contribution. While the aforementioned cancellation also holds for the diagram where the photon couples to the doubly-charged Higgs, the monopole loop functions differ significantly between the lepton flavours for the diagram where the photon couples to the charged lepton in the loop -- therefore spoiling the cancellation. As a result, there is only a partial cancellation and the increase of the conversion rate from the $(+++)$ case to the $(+-+)$ case is only about an order of magnitude smaller than for the three-body decays. This observation generalises to the five other sign choices where one sign is different from the two others.

\begin{figure}
\includegraphics{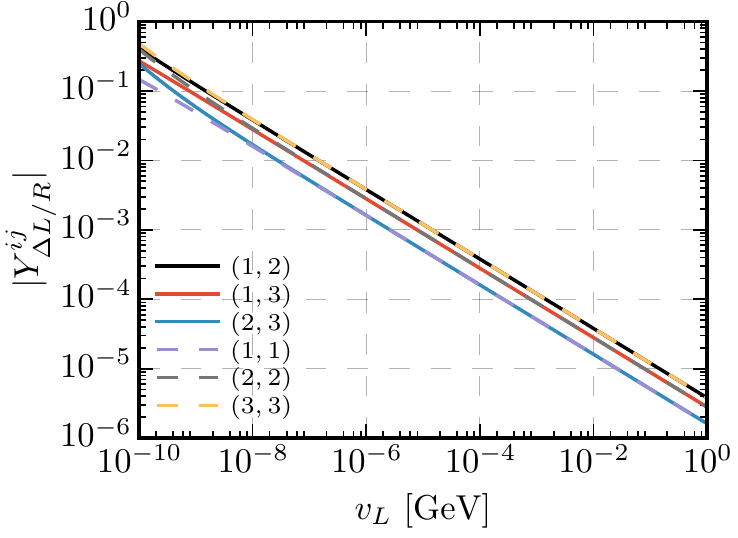}
\includegraphics{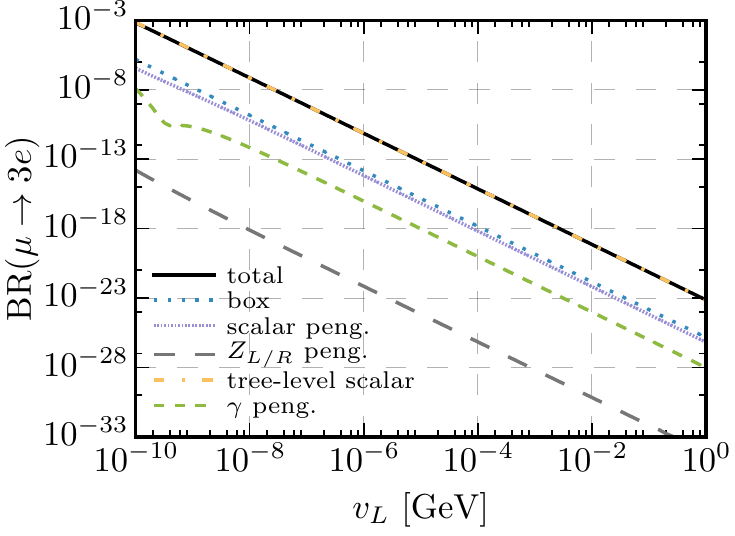}
\caption{Dependence of the triplet Yukawa coupling on left triplet VEV $v_L$ using the $(+-+)$ solution of \cref{eq:TripletCouplSolution} (left) and the consequential decoupling of the different contributions to $\mu\to 3e$ (right). }
\label{fig:diag_sign_3_vL_variation}
\end{figure}

Another consequence of switching to a mixed-sign solution for $Y_\Delta$, besides the size of the off-diagonal elements, is the dependence on $v_L$: while for the same-sign solutions, the off-diagonals vanished to first approximation, leading to a scaling with $1/v_L$, they do not vanish in the mixed-sign case -- leading to the same parametric dependence of $1/\sqrt{v_L}$ as for the diagonal elements. This is depicted in \cref{fig:diag_sign_3_vL_variation} where at the same time we show the decoupling of all contributions to  BR$(\mu\to 3e)$.

\subsubsection{Case II: $M_D \propto \MUP$}
\label{subsec:Mu}

Let us now consider the case where $M_D$ is proportional to the up-type quark matrix.
This choice is motivated from $SO(10)$ unification, where one typically expects unification of the up-and down-type Yukawas. While the individual couplings run differently when evolved from the high to the low scale,\footnote{This of course depends on details  of the intermediate symmetry breaking steps and the scales where this occurs.} let us assume 
for simplicity that the hierarchy in the diagonal Yukawa entries remains approximately unchanged. 
In an $SO(10)$ unification context, one would also expect a non-trivial flavour structure in the up-type Yukawa couplings. We will address this case in the next subsection \ref{subsec:Mu_CKM} and first consider a diagonal $M_D$ here. Obviously, because of the large hierarchy in $\MUP$ any solution to \cref{eq:TripletCouplSolution} also requires a hierarchical structure of $Y_\Delta$, resulting in $m_{\nu_R}^{(e)}/m_u \simeq m_{\nu_R}^{(\mu)}/m_c \simeq m_{\nu_R}^{(\tau)}/m_t$. 

{\emph{\bf{$(+++)$ Solution.}}}
As an explicit example, for $v_L=\SI{5E-5}{}$, $x=10^{-2}$ and this sign choice the triplet Yukawa reads
\begin{equation}
Y_\Delta^{(+++)} = \begin{pmatrix}
\SI{1.77E-5}{} & \SI{-5.63E-8}{} & \SI{1.18E-8}{}  \\
\SI{-5.63E-8}{}  & \SI{8.98E-3}{} & \SI{-1.50E-7}{}   \\
\SI{1.18E-8}{} & \SI{-1.50E-7}{} & \SI{1.23}{}
\end{pmatrix}\,.
\label{eq:Ydelta_Mu_ppp}
\end{equation}
Compared to the case with flavour-universal $M_D$, the resulting off-diagonal structure of $Y_\Delta$ is far less intuitive as the solutions to the respective matrix elements of \cref{eq:TripletCouplSolution} are more involved.\footnote{Note that this Yukawa structure leads to a lightest right-handed neutrino which is lighter than the $\tau$. However, due to the suppression of the corresponding $\tau$ decay by the scale of the $W_R$ boson, the $\tau$ branching ratios will not be changed in an observable way. Similarly, the decays of heavy mesons also do not yet
place any constraints on this scenario.} 
\begin{figure}
\includegraphics{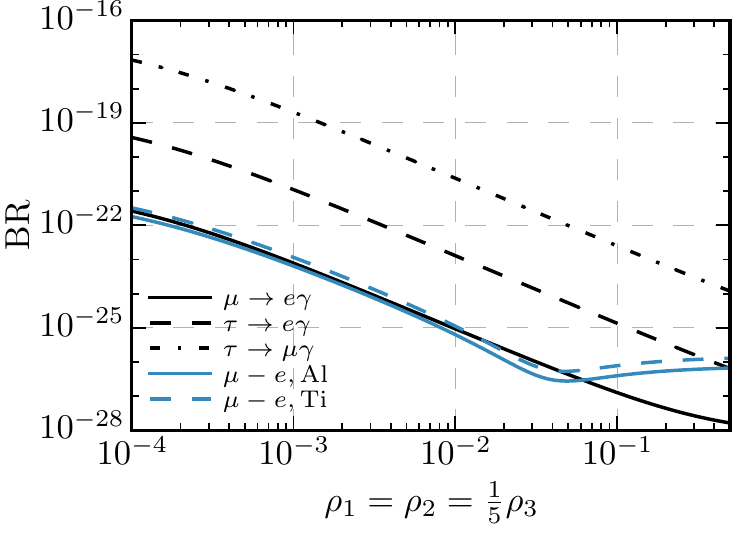}
\includegraphics{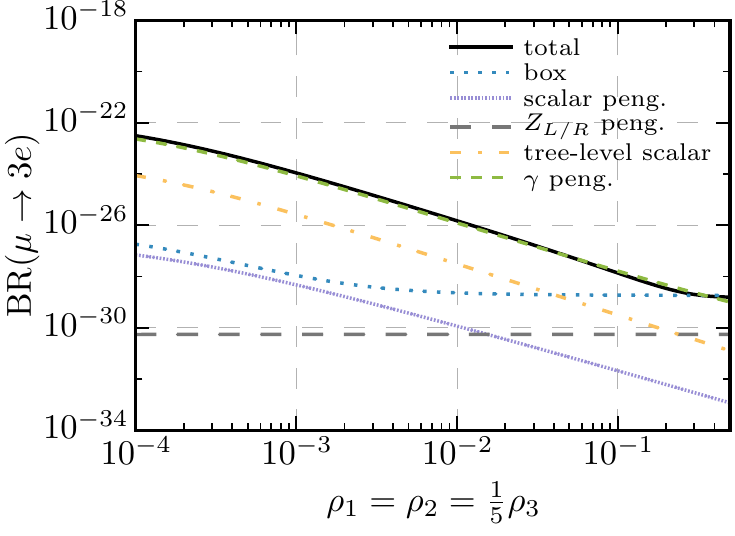}\\
\includegraphics{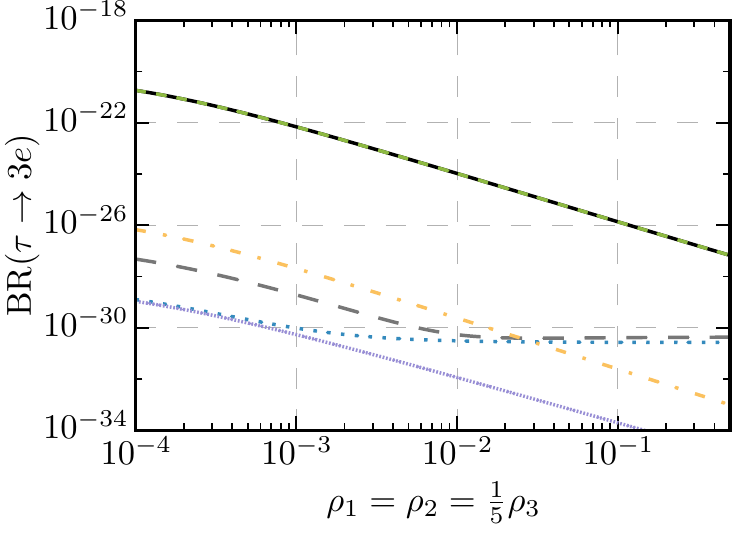}
\includegraphics{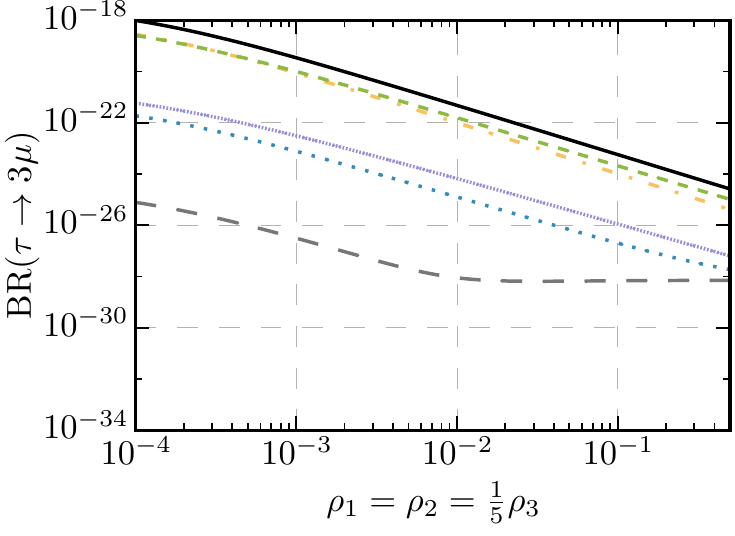}\\
\caption{A cross-section of different LFV observables for the choice $M_D=x \MUP$, with $x=10^{-2}$, $v_L=\SI{5E-5}{\GeV}$ and the Yukawa solution sign choice $(+++)$.}
\label{fig:Mu_Sign_1_scalar_variation}
\end{figure}

What one can already deduce for the relative magnitude of LFV decays is that $\tau \to 3\mu$ will have the largest rates: for this decay, the combination of couplings which enter the tree-level decay mediated by $H^{\pm\pm}$ is $Y_\Delta^{(2,3)} Y_\Delta^{(2,2)}$. For $\tau \to 3e$, in turn, it is $Y_\Delta^{(1,3)} Y_\Delta^{(1,1)}$. As $Y_\Delta^{(1,1)} \simeq m_u/m_c\, Y_\Delta^{(2,2)}$, there is a large hierarchy to be expected between these observables. Furthermore, we can have the case that for three-body decays ending in a $e^+e^-$ pair, loop-induced diagrams dominate over the tree-level mediation for the same reason. Consider again $\tau\to 3e$: the small $Y_\Delta^{(1,3)} Y_\Delta^{(1,1)}$ factor always enters the tree-level amplitude, making it small. In the vector penguins, there is for instance a contribution which involves a $H^{\pm\pm} - e$ loop, scaling with the same combination of matrix entries. In addition, however, there's the $H^{\pm\pm} - \tau$ loop, scaling with $Y_\Delta^{(3,3)} Y_\Delta^{(1,3)}$. The respective amplitude can therefore become even larger than the tree-level contribution despite the loop suppression. 
For the decay $\tau \to 3\mu$, not only is the tree-level contribution correspondingly larger but also the vector penguin as $Y_\Delta^{(3,3)}/Y_\Delta^{(2,2)} \simeq m_t/m_c \simeq \mathcal O(16 \pi^2)$. Therefore the corresponding one-loop amplitude is as important as the tree-level contribution. This is explicitly seen in \cref{fig:Mu_Sign_1_scalar_variation} where we show the dependence of various LFV observables\footnote{ Note that the overall size of the different flavour observables is typically unobservable even with the upcoming projections noted in \cref{tab:sensi}. However, this particular choice of $x$ and $v_L$ serves as a useful benchmark point to highlight the differences when considering both the different $M_D$ choices proportional to $\MUP$ and different sign choices of the Yukawa solutions in the forth-coming sections.} on the mass scale of the scalar sector using the $(+++)$ solution for $Y_\Delta$, in analogy to \cref{fig:diag_scalar_variation}.
Since the LFV $\mu$ decays are suppressed w.r.t.\ the LFV $\tau$ decays due to the smaller Yukawa couplings involved, those diagrams which involve gauge couplings and which are hence independent of the scale of the  scalar sector become relevant much earlier.
This is most prominently seen in the $\mu\to 3e$ as well as $\mu-e$ conversion rates which are dominated by $W_{L/R}-\nu_R$ box diagrams for $\rho_1 \gtrsim 0.3$ and $0.1$, respectively.
Note that the small dip of the $\mu-e$ conversion rates around $\rho_1\simeq \SI{4E-2}{}$ is a result of a destructive interference between the box diagrams and the $\gamma$ penguins. The rates however approach a constant value once the photonic contribution decouples and the box diagrams dominate which is seen at larger $\rho_1$ values.

\begin{figure}[htbp]
\includegraphics{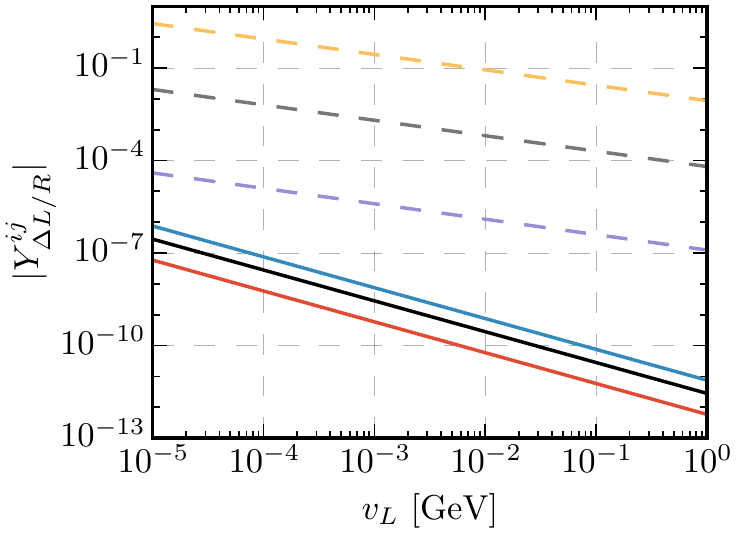}
\includegraphics{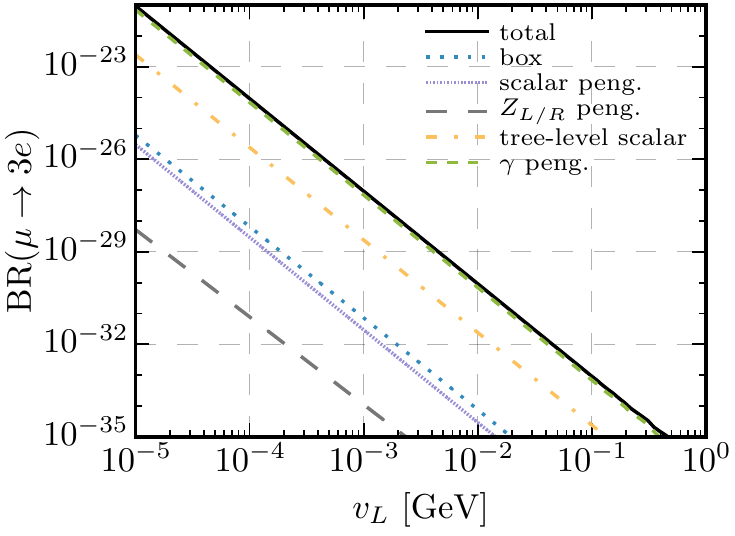}\\
\includegraphics{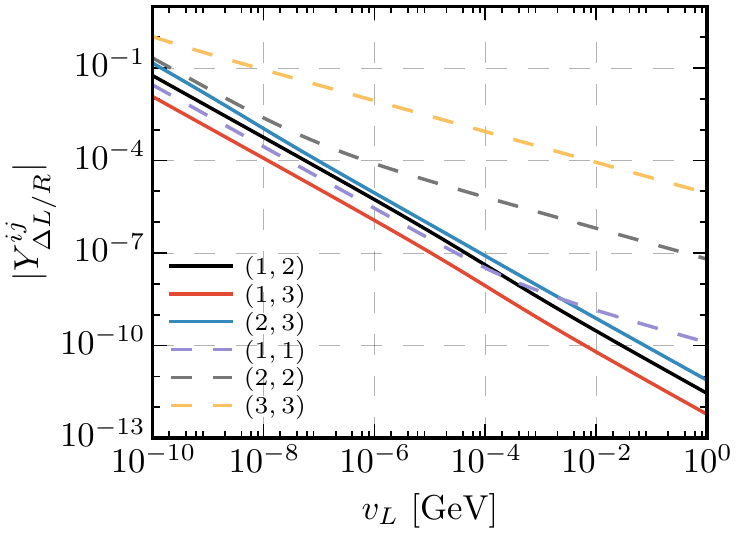}
\includegraphics{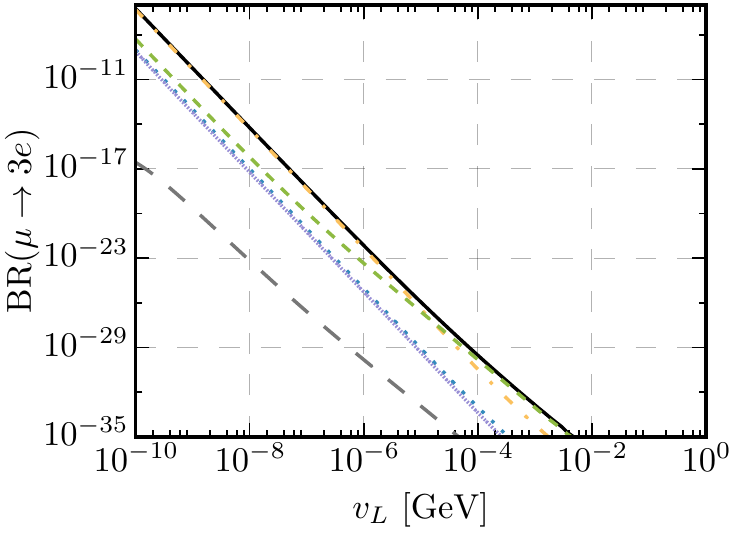}
\caption{Illustration of the different decoupling behaviour resulting from varying $v_L$ for different $x$ values, using $M_D = \MUP$ and the $(+++)$ sign choice. Here $v_L$ is varied between the allowed regions, where the lower bound arises from non-perturbative couplings and the upper bound $v_L=\SI{1}{\GeV}$ from the rho-parameter. The top and bottom rows correspond to $x=10^{-2}$ and $x=10^{-5}$ respectively.}
\label{fig:Mu_Sign_1_vL_variation}
\end{figure}

In \cref{fig:Mu_Sign_1_vL_variation} we then show the decoupling behaviour for two different choices of $x$ as the triplet Yukawa VEV $v_L$ is varied over the allowed domain.\footnote{The $v_L$ domains between the different choices of $x$ differ  due to the triplet Yukawa parametrisation. For $x=10^{-2}$ values of $v_L$ smaller than approximately \SI{E-5}{\GeV} lead to non-perturbative couplings, while for both cases values of $v_L$ greater than $\mathcal O(\SI{1}{\GeV})$ are not permitted due to constraints from the rho-parameter.} The case that $x=10^{-2}$ corresponds to the parameter choice used for the discussion to this point, and the same arguments hold in what concerns the dominance of the $\gamma$ penguins for the entire range of $v_L$, as explicitly depicted for $\mu \to 3e$ in \cref{fig:Mu_Sign_1_vL_variation}. Here we have the situation that $\BDi \ll \sqrt{v_L/v_R}$ for all shown $v_L$ choices. As discussed before for the $M_D \propto \mathbb{1}$ case and illustrated in \cref{sec:TripCoupParameterization}, for the sign choice $(+++)$ all off-diagonal terms vanish at leading order. Subsequently, the numerical calculation yields heavily suppressed off-diagonal entries that scale as $1/v_L$. In the case that $x=10^{-5}$, all off-diagonal $Y_\Delta$ entries still scale with $1/v_L$. The diagonal elements, however, show differences: the 
approximation $\BDi \ll \sqrt{v_L/v_R}$ only holds for $i=3$ over the entire range of $v_L$. For $i=1,2$, $\BDi \simeq \mathcal{O}\left( \sqrt{v_L/v_R} \right)$ for small values of $v_L$. Therefore, just like the off-diagonal terms which are generated by the first non-vanishing order in $\BDi/\sqrt{v_L/v_R}$, also the diagonal $Y^{(i,i)}_\Delta$ elements scale as $1/v_L$ for small $v_L$ values. For increasing $v_L$, first the $(2,2)$ and then also the $(1,1)$ elements  fall into the limit $\BDi \ll \sqrt{v_L/v_R}$, eventually resulting in a decoupling at a rate proportional to $1/\sqrt{v_L}$. 
As a result, the $\gamma$ penguin dominance in $\mu\to 3e$ only kicks in for 
$v_L \gtrsim 10^{-5}$~GeV. Before that, $Y_\Delta^{(1,1)} \gg m_u/m_c \, Y_\Delta^{(2,2)}$,
 giving a boost to the tree-level contribution.

\begin{figure}[htbp]
\includegraphics{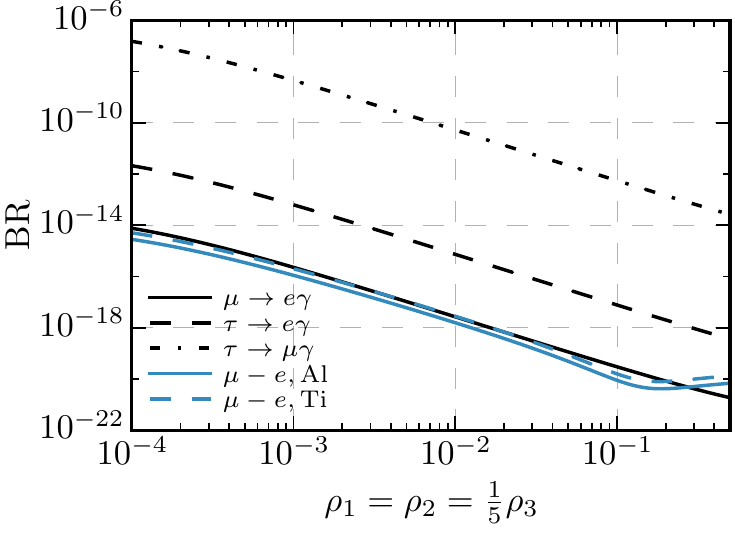}
\includegraphics{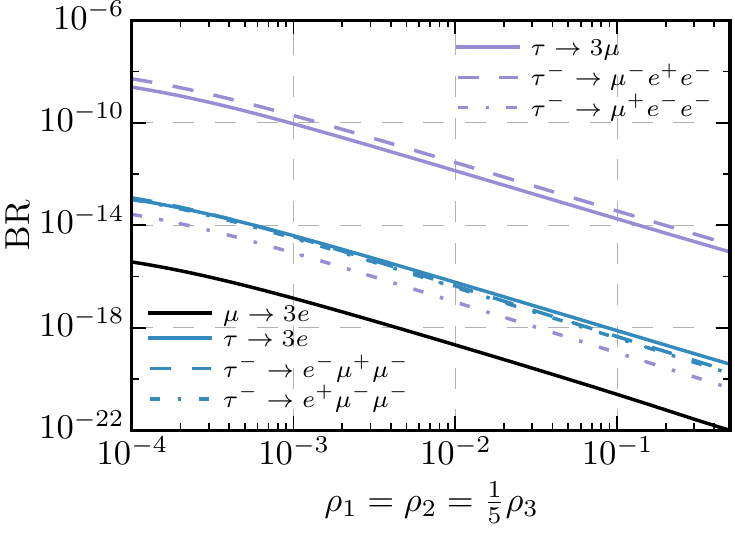}
\caption{A cross-section of different LFV observables for the choice $M_D=x \MUP$, with $x=10^{-2}$, $v_L=\SI{5E-5}{\GeV}$ and the Yukawa solution sign choice $(+-+)$.}
\label{fig:Mu_Sign_3_scalar_variation}
\end{figure}

{\emph{\bf{$(+-+)$ Solution.}}}
As for the $M_D \propto \mathbb 1$ case, we now turn to a different solution to $Y_\Delta$ for the same input parameters. As described in \cref{sec:TripCoupParameterization}, the effect of switching to a $(+-+)$ solution rather than the $(+++)$ solution is not qualitatively different to the case $M_D\propto \mathbb 1$ given that $\BDi \ll \sqrt{v_L/v_R}$. First we consider varying the scalar sector choosing $x=10^{-2}$ and $v_L=\SI{5E-5}{\GeV}$. This results in a triplet Yukawa that reads
\begin{equation}
Y_\Delta^{(+-+)} = \begin{pmatrix}
\SI{1.74E-5}{} & \SI{-7.00E-5}{} & \SI{9.25E-5}{}  \\
\SI{-7.00E-5}{}  & \SI{-8.61E-3}{} & \SI{2.33E-2}{}   \\
\SI{9.25E-5}{}  & \SI{2.33E-2}{} & \SI{1.20}{}
\end{pmatrix}\,.
\label{eq:Ydelta_Mu_pmp}
\end{equation}
The results of these choices are shown in \cref{fig:Mu_Sign_3_scalar_variation} as a function of the triplet-scalar masses. In comparison to \cref{fig:Mu_Sign_1_scalar_variation}, many of the flavour observables are within reach of current or upcoming experiments. In this region of parameter space the change of sign does not modify the relative size of the $Y_\Delta^{(1,1)}$ or $Y_\Delta^{(2,2)}$ entries. Correspondingly, for $\mu \to 3e$ and $\tau \to 3e$ the dominant modes remain the $\gamma$ penguins. These observables are however far larger as the corresponding off-diagonal Yukawas $Y_{\Delta}^{(1,2)}$ and $Y_\Delta^{(1,3)}$  are typically four orders of magnitude larger compared to the $(+++)$ sign choice. Additionally, since $Y_\Delta^{(2,3)}$ has changed by five orders w.r.t.\ to the $(+++)$ choice, the ratio of $\rm{BR}(\tau \to 3\mu) /\rm{BR}( \mu \to 3e)$ is increased by two orders of magnitude. 

\subsubsection{Case III: $M_D \propto \MUPCKM$}
\label{subsec:Mu_CKM}

\begin{figure}
\includegraphics{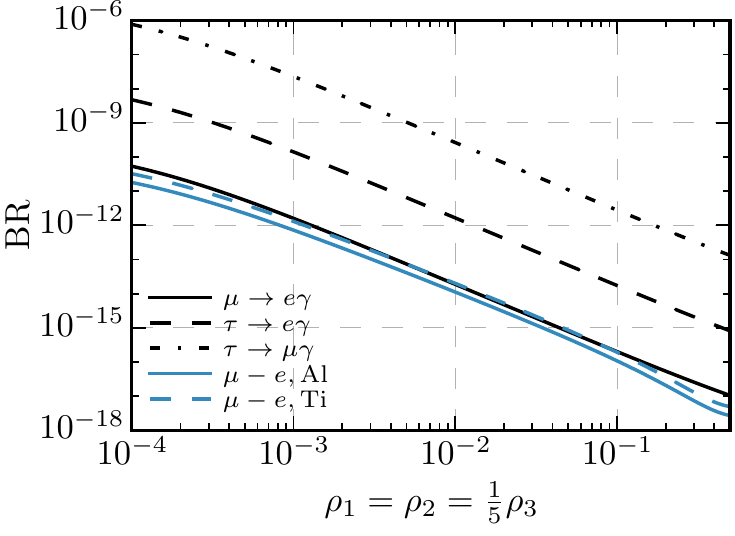}
\includegraphics{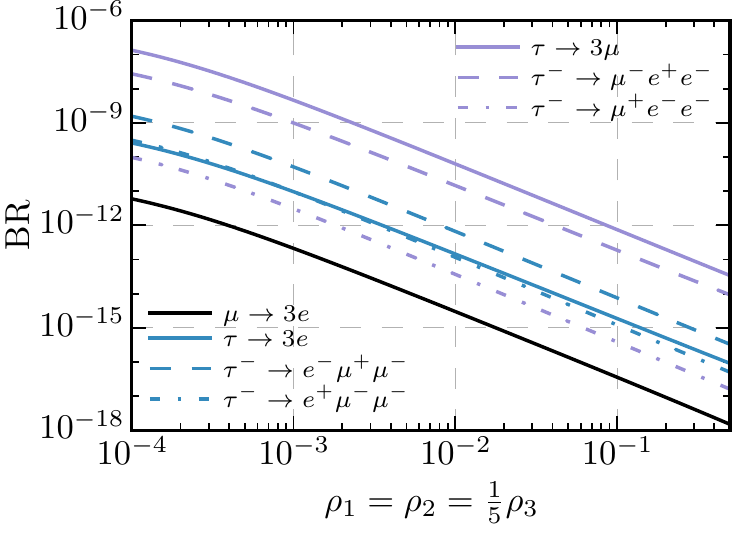}
\caption{A cross-section of different LFV observables for the choice $M_D=x \MUPCKM$, with $x=10^{-2}$, $v_L=\SI{5E-5}{\GeV}$ and the Yukawa solution sign choice $(+++)$.}
\label{fig:Mu_CKM_Sign_1_scalar_variation}
\end{figure}

\begin{figure}
\includegraphics{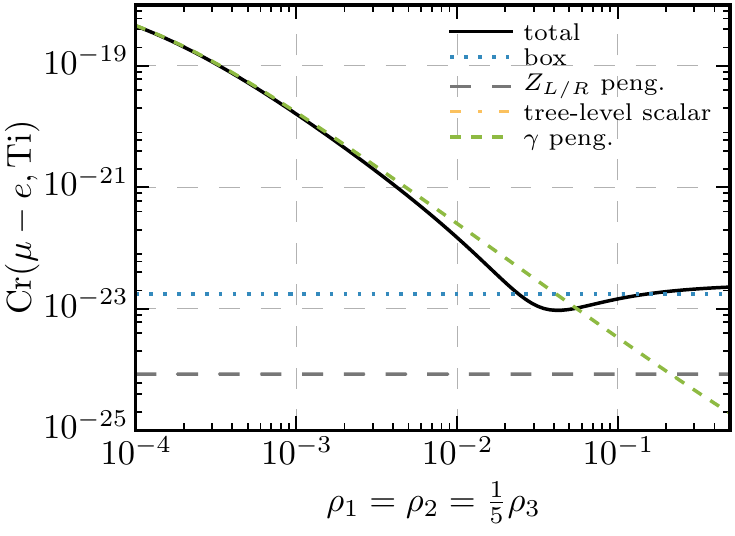}
\includegraphics{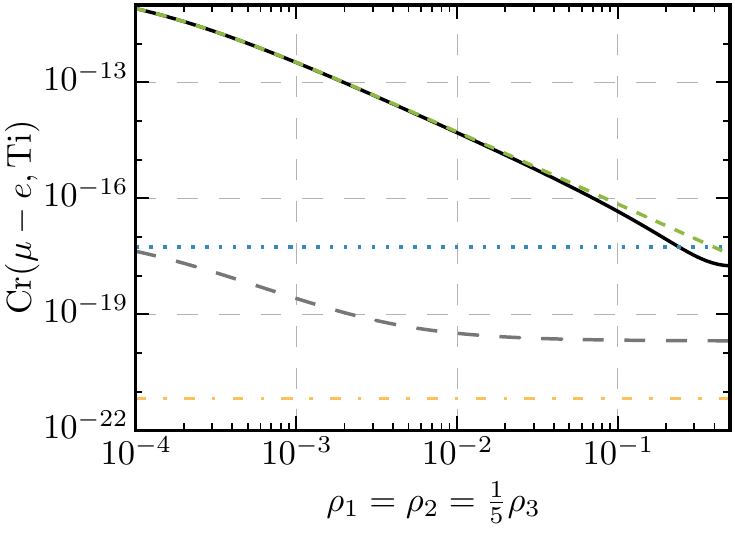}\\
\caption{The $\mu-e$ conversion rates in Ti when varying the triplet scalar sector for the choice
$v_L=10^{-6}$~GeV and the sign choice $(+++)$ for the solution of the triplet-Yukawa. Two different choices of $M_D$
 are made: $M_D=x \MUP$ on the left-hand panel and $M_D = x \MUPCKM$ on the right-hand panel, where in both cases $x=10^{-3}$.}
\label{fig:MuEconv_Mell_vs_Mell_CKM}
\end{figure}

Let us now go ahead and consider $M_D = x \MUPCKM$, which is motivated by Yukawa unification due to the intimate connection between the up- and down-type Yukawas.\footnote{In LR-symmetric theories, the up-type mass matrix can be written as $m_u = V_{\rm CKM}^{L\dagger} m_u^{\rm diag} V_{\rm CKM}^R $, where $m_u^{\rm diag} = \MUP$, $V_{\rm CKM}^L = V_{\rm CKM}$ is the usual CKM matrix and $V_{\rm CKM}^R$ is the according quantity in the $SU(2)_R$ sector. Parity symmetry relates $V_{\rm CKM}^R = V_{\rm CKM}^{L}$ (up to a diagonal matrix of free phases on either side which we choose to set to zero here) so that $m_u = \MUPCKM$. See also the Appendix A in Ref.~\cite{Bertolini:2014sua}.} We once again start by considering the $(+++)$ sign choice, $v_L=\SI{5E-5}{\GeV}$ and $x=10^{-2}$, completely analogous to the previous subsection. This results in a triplet Yukawa of the form 
\begin{equation}
Y_\Delta^{(+++)} = \begin{pmatrix}
\SI{4.87E-4}{} & \SI{2.14E-3}{} & \SI{4.12E-3}{}  \\
\SI{2.14E-3}{} & \SI{1.06E-2}{} & \SI{5.01E-2}{}   \\
\SI{4.12E-3}{} & \SI{5.01E-2}{} & \SI{1.22}{}
\end{pmatrix}\,.
\label{eq:Ydelta_Mu_CKM_ppp}
\end{equation}
Multiplication of the CKM matrix on both sides results in a slight decrease of the hierarchy amongst the diagonal entries and an increase in the size of the off-diagonal entries, similar to the case where $M_D=\MUP$ with the $(+-+)$ sign choice. Shown in \cref{fig:Mu_CKM_Sign_1_scalar_variation} is the effect of varying the triplet scalar sector with this choice of the triplet-Yukawa. 
Note that all of the parameter space shown in this figure could be probed by the proposed PRISM/PRIME experiment for $\mu-e$ conversion \cite{PRIME,Barlow:2011zza}. In \cref{fig:MuEconv_Mell_vs_Mell_CKM}, we further decompose the rate into the different contributions, directly comparing the $M_D=\MUP$ and $M_D=\MUPCKM$ scenarios. Here, the size of the $\gamma$ penguin contribution is determined by the sum of the product of couplings $\sum_{i} Y_\Delta^{(2,i)}Y_\Delta^{(i,1)}$ which increases from the former to the latter $M_D$ choice. Note that, due to the large off-diagonal entries in \cref{eq:Ydelta_Mu_CKM_ppp}, the combination  $Y_\Delta^{(2,3)}Y_\Delta^{(3,1)}$
becomes sizeable, two orders of magnitude larger than $Y_\Delta^{(2,2)}Y_\Delta^{(2,1)} + Y_\Delta^{(2,1)}Y_\Delta^{(1,1)}$ which is the relevant contribution for the $M_D\propto \MUP$ choice. Additionally, multiplication by the CKM matrix also introduces contributions arising from the bidoublet scalar sector. However, under the given constraints that the heavy bidoublet Higgs mass is around \SI{20}{\TeV}, these contributions are extremely sub-dominant. This contribution can nevertheless be seen in the right-hand panel of \cref{fig:MuEconv_Mell_vs_Mell_CKM}.

In the considered case $M_D \propto \MUPCKM$,  the effect of switching to a different sign choice is less drastic than in the diagonal $\MUP$ case. The reason is that with the inherently flavour-violating nature of $M_D$, there is already a direct flavour-violating insertion into $Y_\Delta$. A change from a same-sign to a mixed-sign solution still has an impact here, but it is no longer as pronounced as in the case with diagonal $M_D$. As a result, while all LFV observables are generically two orders of magnitude larger than in the $(+++)$ case, the relative magnitude of the observables remains almost unchanged. A parameter point with a certain value of $v_L$ and the $(+++)$ solution is therefore almost indistinguishable from the same point with larger $v_L$ but the $(+-+)$ solution. 

\begin{figure}[htbp]
\includegraphics{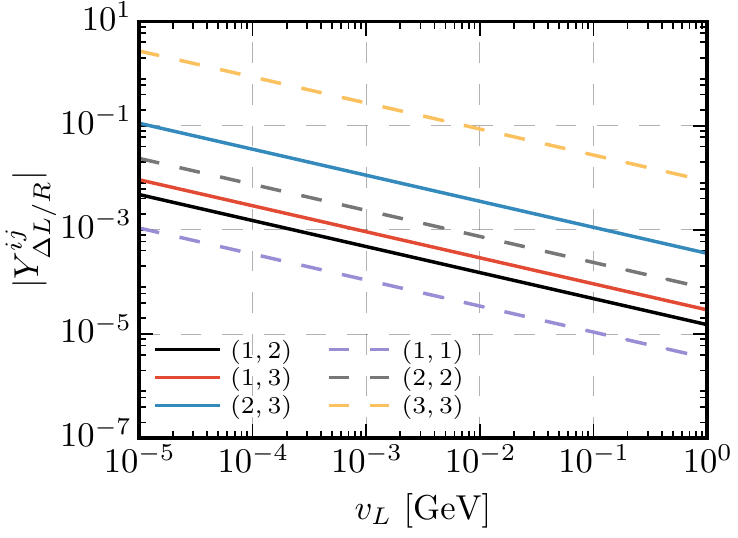}
\includegraphics{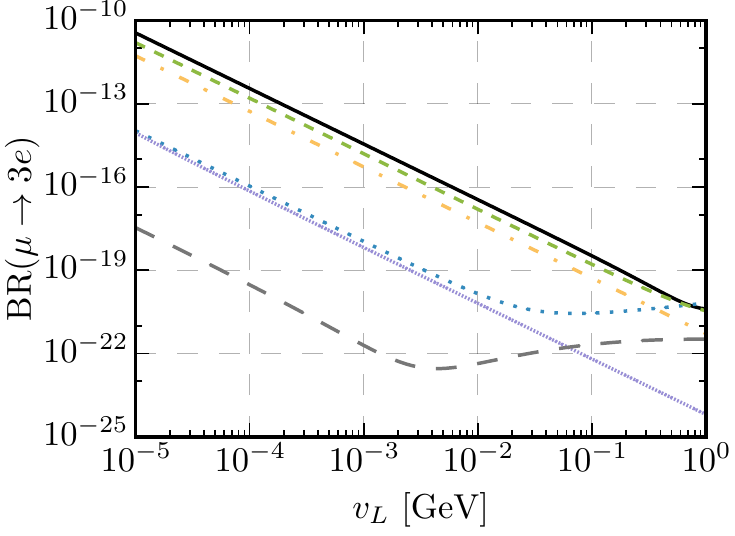}\\
\includegraphics{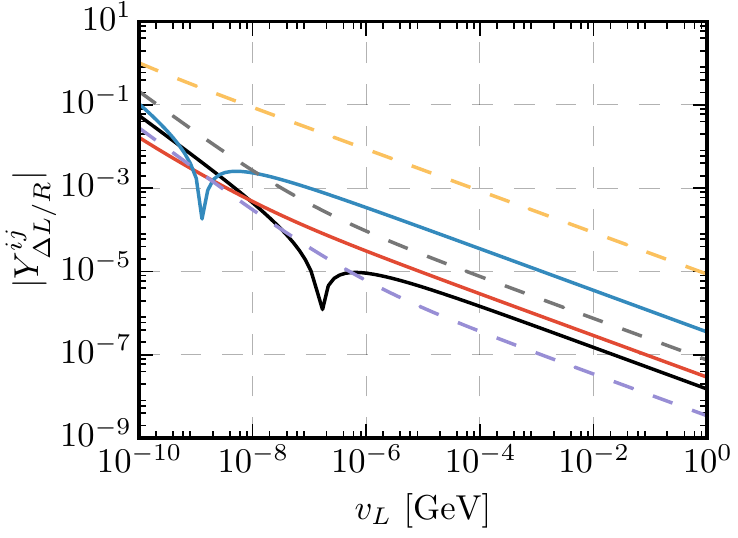}
\includegraphics{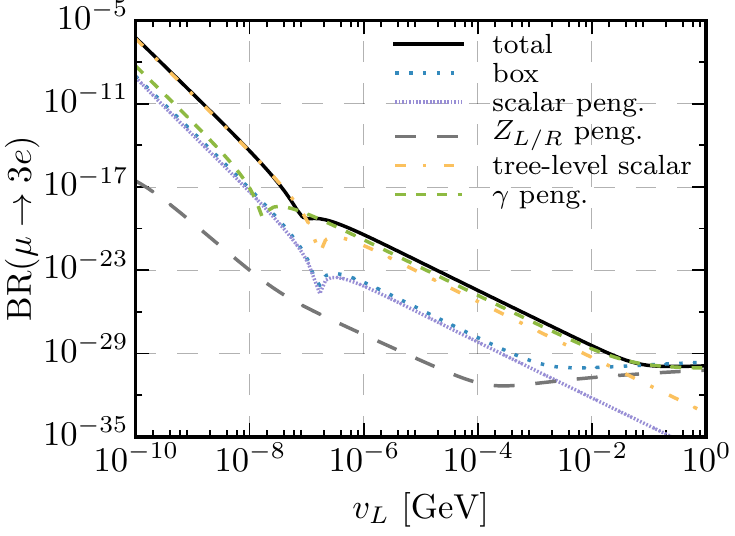}
\caption{Illustration of the different decoupling behaviour resulting from varying $v_L$ with $M_D = \MUPCKM$ using different $x$ values and the $(+++)$ choice. Here $v_L$ is varied between the allowed regions, where the lower bound  arises from non-perturbative couplings and the upper bound $v_L\sim \mathcal O ({\rm GeV})$ from the rho-parameter. The top and bottom rows correspond to $x=10^{-2}$ and $x=10^{-5}$ respectively.}
\label{fig:Mu_CKM_Sign_1_vL_variation}
\end{figure}

To conclude this section we show in \cref{fig:Mu_CKM_Sign_1_vL_variation} the equivalent of \cref{fig:Mu_Sign_1_vL_variation} for $M_D = \MUPCKM$, namely the variation of $v_L$ given two different choices of $x$. For the choice $x=10^{-2}$ we see that all entries of $Y_\Delta$ decrease at the same rate. This is a direct consequence of the multiplication by the CKM matrix. Subsequently we see that the $\gamma$ penguins and tree-level contributions to $\mu \to 3e$ are of comparable size. Additionally we observe that the box and $Z_{L/R}$-penguin diagrams do not completely decouple with increasing $v_L$. This is due to the $W_{L/R}-\nu_R$ loops which are independent of $v_L$. However, the actual rates in this region of parameter space will not be directly probed in upcoming experiments. Lastly we consider the case where $x=10^{-5}$. Here, we observe that we end up in regions where the triplet Yukawa entries change sign (seen as the dips in the figure) in addition to the change in decoupling behaviour due to the relative sizes of $B_D$ and $\sqrt{v_L/v_R}$ as was already observed for the case $M_D=\MUP$.

\subsubsection{Impact of the CP phase}
\label{sec:CPphase}

So far we have always assumed the CP phase to be zero. However, this need not be the case. Actually, recent fits even slightly prefer an angle of $\dCP \simeq 3\pi/2$ \cite{Esteban:2016qun}. Therefore we discuss here the impact of of the CP phase on the LFV observables and consider scenarios which are parity-symmetric. When looking at the parametrization of  $Y_\Delta$ according to \cref{eq:TripletCouplSolution} one readily sees that $B_D$ becomes complex, requiring the rotation matrix $R$ to be a complex unitary matrix. As explained in \cref{sec:TripCoupParameterization}, the effect is similar to switching from a $(\pm\pm\pm)$ solution to a mixed-sign solution namely. This holds even in the case where $M_D$ is diagonal and $ \sqrt{v_L/v_R}\gg \BDi$, off-diagonal $Y_\Delta$ entries are already induced at the zeroth order in $B_D/ \sqrt{v_L/v_R}$. Therefore, when turning on $\dCP$ in the case of diagonal $M_D$ and a $(\pm\pm\pm)$ choice, large differences of the LFV observables are expected w.r.t. the $\dCP=0$ case. In those cases, however, where there is either a mixed-sign choice or a non-diagonal $M_D$ such as in \cref{subsec:Mu_CKM}, the effect is far less pronounced.

\begin{figure}[htbp]
\includegraphics{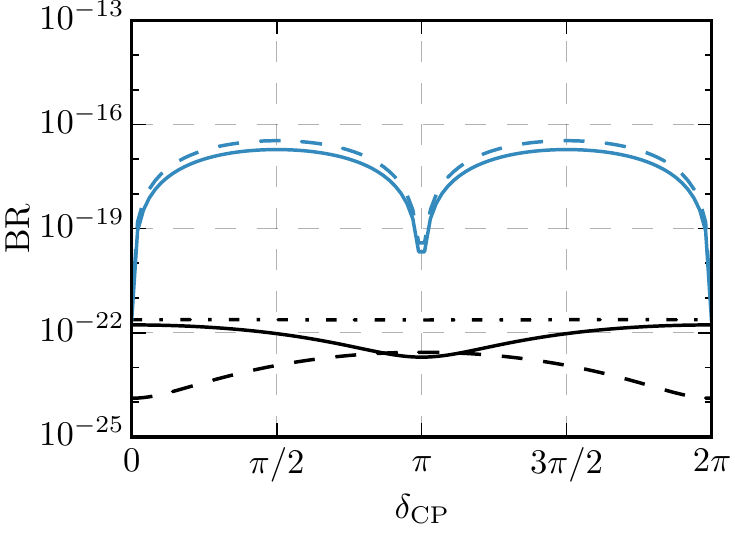}
\includegraphics{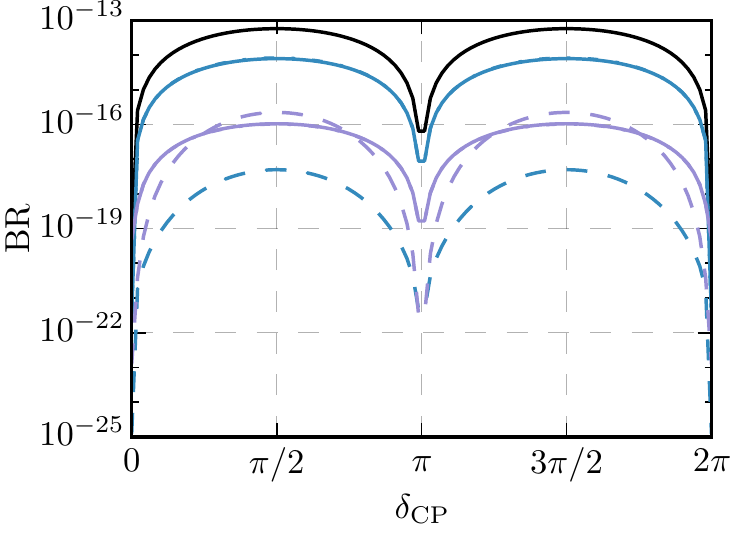}\\
\includegraphics{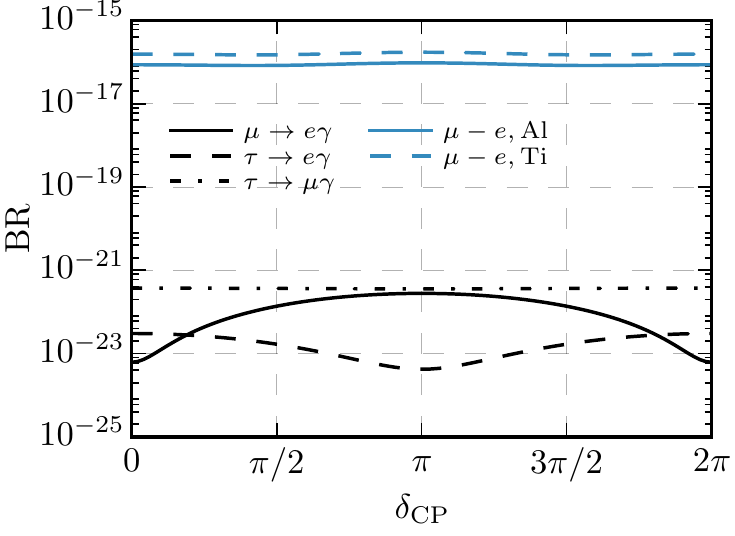}
\includegraphics{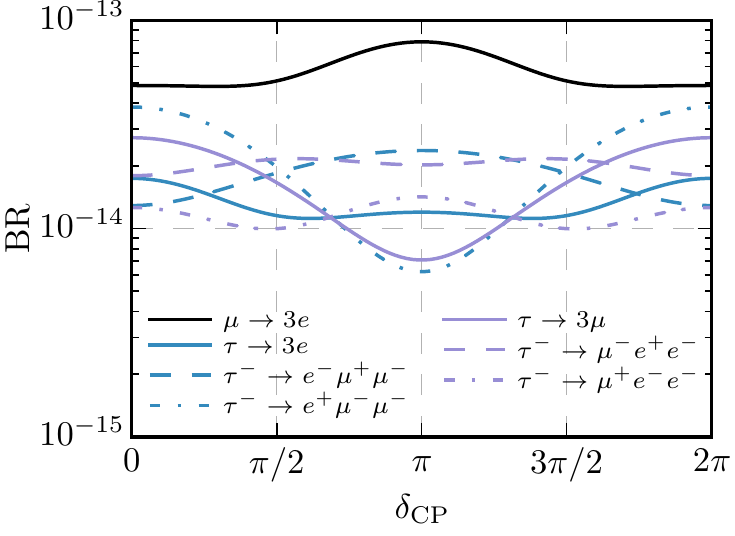}
\caption{The main LFV observables varying the $\dCP$ phase for $M_D=\mathbb{1}\,$MeV, and $v_L=10^{-5}$~GeV. The top row corresponds to the $(+++)$ solution while the bottom row corresponds to the $(+-+)$ solution.}
\label{fig:dCP_MD_Diag_GeV}
\end{figure}

We show this behaviour in \cref{fig:dCP_MD_Diag_GeV} with the parametrisation $M_D \propto \mathbb{1}$, both the $(+++)$ and $(+-+)$ solutions. While there are many orders of magnitude difference between the cases of zero and maximal CP phase when applying the $(+++)$ solution, the differences are only of $\mathcal O(1)$ in case of $(+-+)$. The same arguments hold for the other $M_D$ parametrisations; we observe large differences in LFV rates between different CP phases for the diagonal $M_{\rm up-type}$ and $(+++)$ choice but only comparably small changes in the other cases. This is clearly illustrated in the next subsection where we show our main results in the cases where $\dCP=0$ and $\dCP=3\pi/2$. 

Clearly, allowing for complex phases in the neutrino and, thus, in the Yukawa sector will
give rise to an electric dipole moment (edm) for the leptons. Here in particular
the bound on the electron is rather severe as its edm must be below $8.7\times10^{-29}~e$cm
\cite{Olive:2016xmw}.
In the parameter region of \cref{fig:dCP_MD_Diag_GeV} we find values of up to approximately $10^{-33}$ where
the main contribution is due to the doubly charged Higgs bosons. However, this
contribution is suppressed as one can show that in the limit $m_F/m_B \to 0 $ the contribution
to the edm vanishes \cite{Clavelli:2000ua}, where $m_F$ and $m_B$ are the masses of the
fermion and the boson in the loop. The other potentially dangerous contribution due to
the singly charged Higgs boson is suppressed because the lighter one is essentially
the $\Delta_L$ and, thus, the corresponding fermion is the left-handed neutrino. The contribution
of  the heavier state is suppressed by its mass of around 20 TeV. As a result the electron edm will likely not be testable at the upgraded ACME experiment
\cite{acme}.

\subsubsection{Measurement prospects}
\label{sec:2D}

In this section we ask the question: what are the prospects of measuring a signal of lepton flavour violation given a triplet scalar sector with masses at the TeV scale? Here, we choose the scalar sector and model parameters according to \cref{tab:benchmark}. For each parametrisation of $M_D$ we perform 2D scans in both $v_L$ as well as the overall scale of $M_D$ which we define, as before, by the continuous parameter $x$. While the structure of $Y_\Delta$ is determined by the parametrisation of $M_D$ as well as by the choice of one of the eight possible solutions to \cref{eq:TripletCouplSolution}, the overall $Y_\Delta$ magnitude is governed by the sizes of $v_L$ and $M_D$. Therefore, by scanning these two quantities for the different $M_D$ parametrisations one obtains a robust prediction as to the extent of the parameter space which is probeable by current and future experiments. It should be noted that the choice of the scalar sector maximises the rates of the flavour observables. In this sense these projections are a best case scenario, as the LHC will begin to increase the bounds on the masses of the triplet-scalar sector.

\begin{figure}
\includegraphics{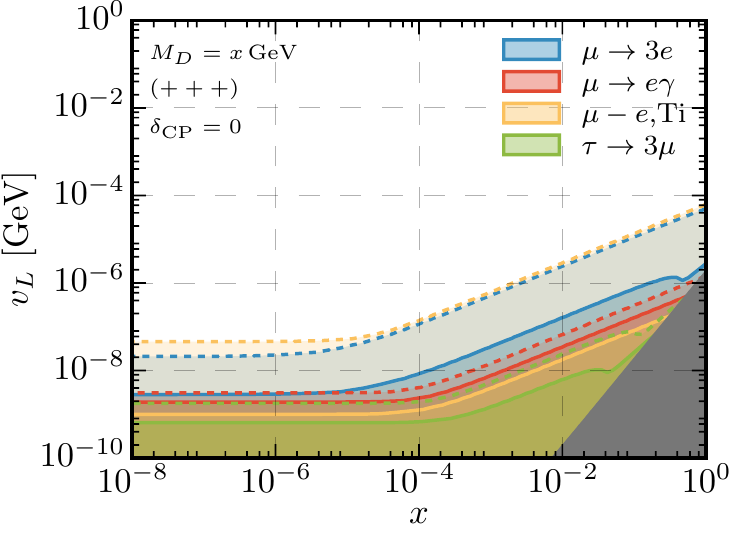}
\includegraphics{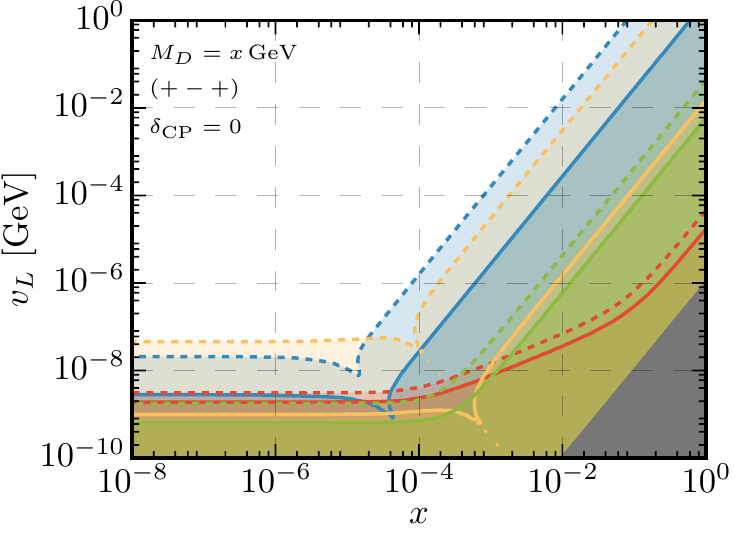}\\
\includegraphics{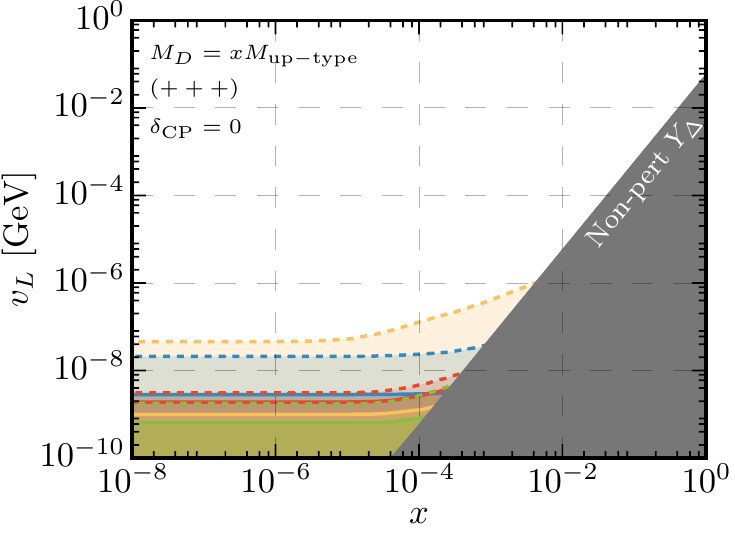}
\includegraphics{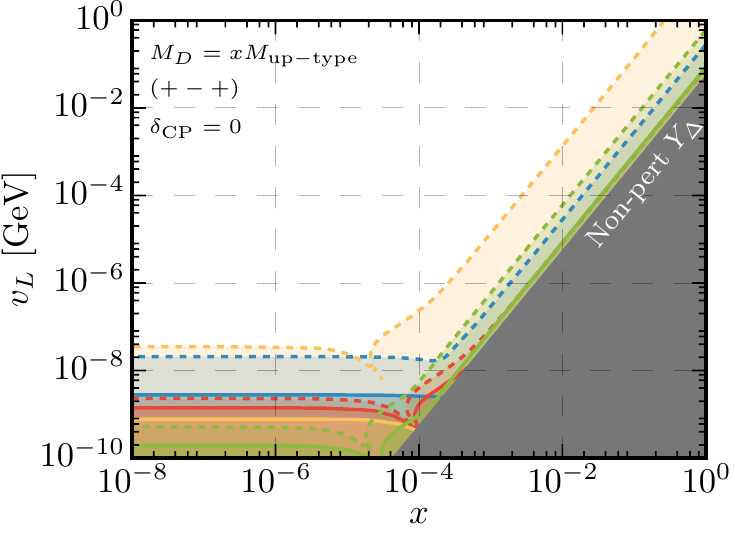}\\
\includegraphics{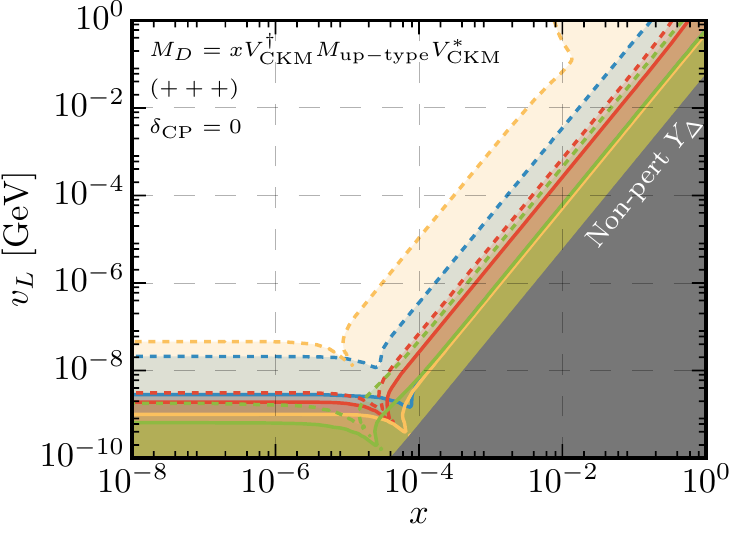}
\includegraphics{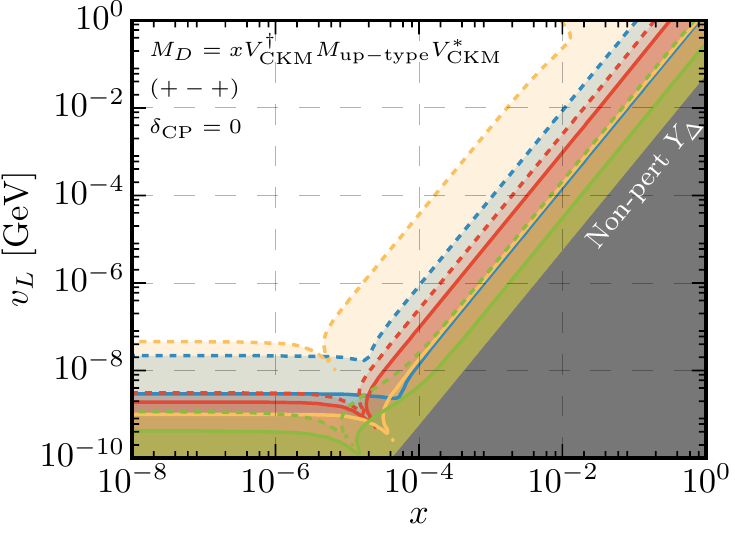}
\caption{Sensitivity of current and future experiments in the $(x,v_L)$ plane. Solid lines are the current bounds, while the dashed lines are the projected sensitivities of upcoming experiments, see \cref{tab:sensi} for the numerical values. The colour scheme for the shaded regions is $\mu \to 3e$ (blue), $\mu\to e\gamma$ (red), $\mu-e$,Ti (yellow) and finally $\tau \to 3\mu$ (green). The non-perturbative regions (grey) correspond to $\rm{Max}(|Y_\Delta|)\geq\sqrt{4\pi}$.}
\label{fig:2D_diag}
\end{figure}

The results for $\dCP=0$ are presented in \cref{fig:2D_diag}: in each panel, we shade the region excluded by current experiments for the most sensitive channels.\footnote{All flavour observables that where shown in \cref{sec:numericalresults} are considered in \cref{fig:2D_diag}, however to improve readability only the four most sensitive channels are shown in subsequent figures.} We also depict the sensitivity for future experiments with the lighter shaded regions with a dashed border. The plots have to be read as follows: in the upper left-hand corner of each figure (shown in white), the LFV rates are too small to be measured in the near future. Going to smaller $v_L$ and larger $x$ values, the rates increase and many of the current or near-future experiments start to become sensitive.

A generic feature of all plots, irrespective of the $M_D$ parametrisation or the sign choice, is that the LFV rates are almost independent of $x$ in the small $x$ regime. However, at a certain $x$-value, depending on both the $M_D$ choice and the particular observable, the LFV rates begin to increase. The reason is as follows. For small $x$, $\BDi$ is of the order of $\sqrt{v_L/v_R}$ or even larger.\footnote{Recall that $B_D = R^\dag M_D^{-1/2} m_\nu^{\rm light} M_D^{(*)\,-1/2} R^*$.} This means that the $M_D$ dependence in  
\begin{align}
Y_\Delta &= \frac{1}{2\sqrt{2} v_L} M_D^{(*)\,1/2} R^* (B_D\pm\sqrt{B_D^2+4\alpha}) R^\dag M_D^{1/2}\,,
\end{align}
cancels to first order and the off-diagonal $Y_D$ structure is determined by the PMNS matrix which enters in the rotation matrix $R$. With increasing $x$, however, we enter the limit $\sqrt{v_L/v_R} \gg \BDi$ and therefore the arguments outlined in \cref{subsec:diag_MD} hold:
\begin{enumerate}[(i)]
\item if $M_D$ is diagonal, then for the $(\pm\pm\pm)$ choices, the $Y_\Delta$ off-diagonal elements scale with $x/v_L$ 
\item for mixed sign choices the entries scale as\footnote{The LFV amplitudes scale quadratically with $Y_\Delta$. However, this is typically the product of a diagonal with an off-diagonal entry of $Y_\Delta$. As mentioned above the diagonal entries scale with $x/\sqrt{v_L}$ in the limit $\sqrt{v_L/v_R}\gg \BDi$, meaning the LFV rates scale with either $x^4/v_L^3$ or $x^4/v_L^2$ in the majority of the parameter space.} $x/\sqrt{v_L}$
\item If, in turn, $M_D$ contains non-diagonal elements, then the same-sign choices also scale like $x/\sqrt{v_L}$. The only difference with respect to the mixed-sign choice being an overall smaller LFV rate.
\end{enumerate}

\begin{figure}
\includegraphics{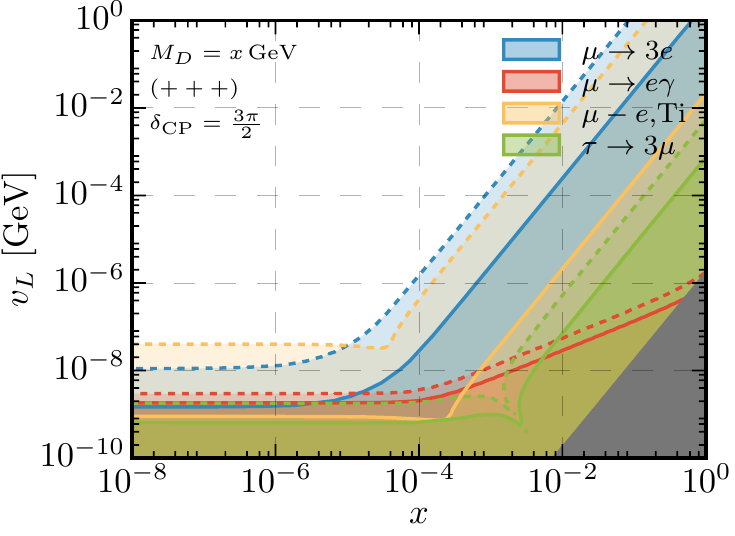}
\includegraphics{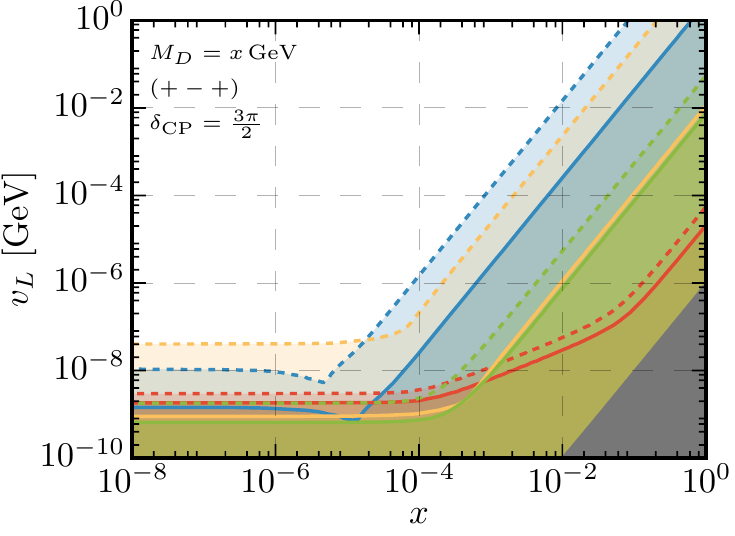}\\
\includegraphics{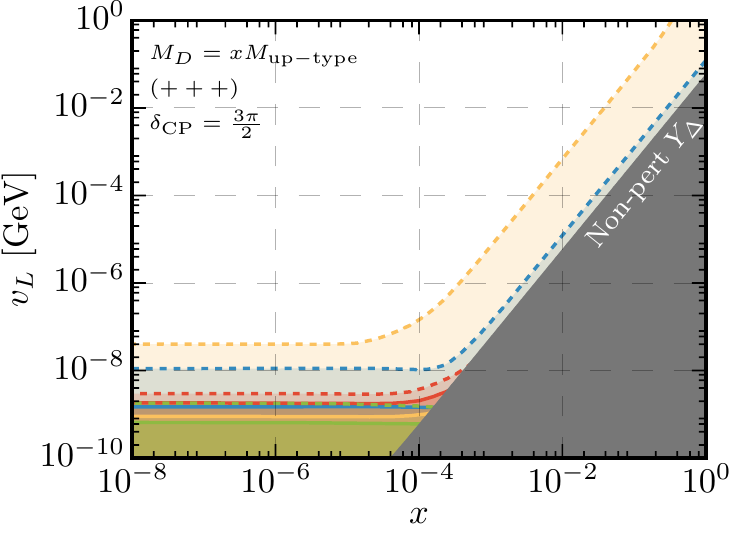}
\includegraphics{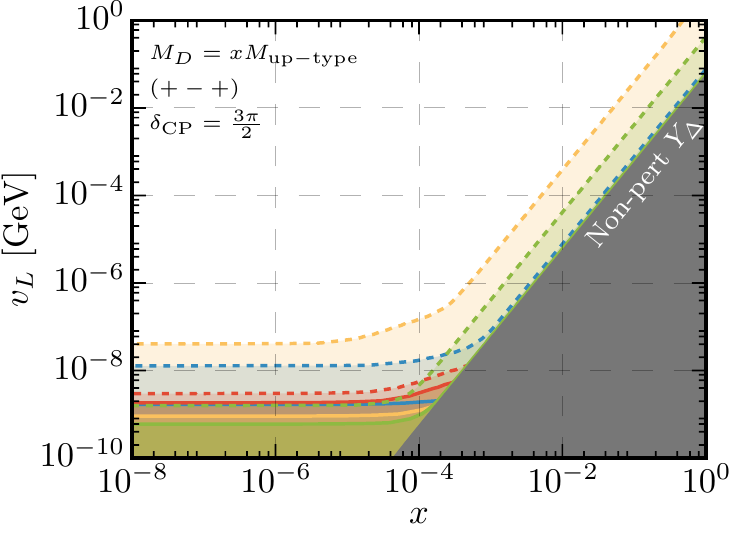}\\
\includegraphics{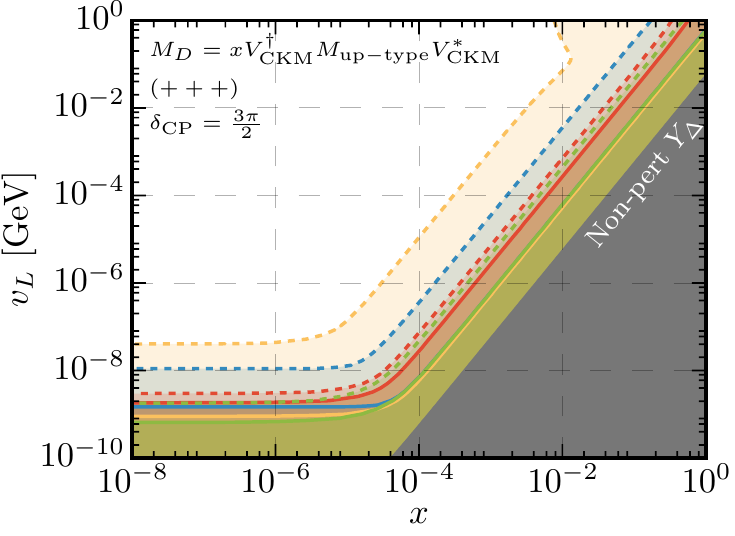}
\includegraphics{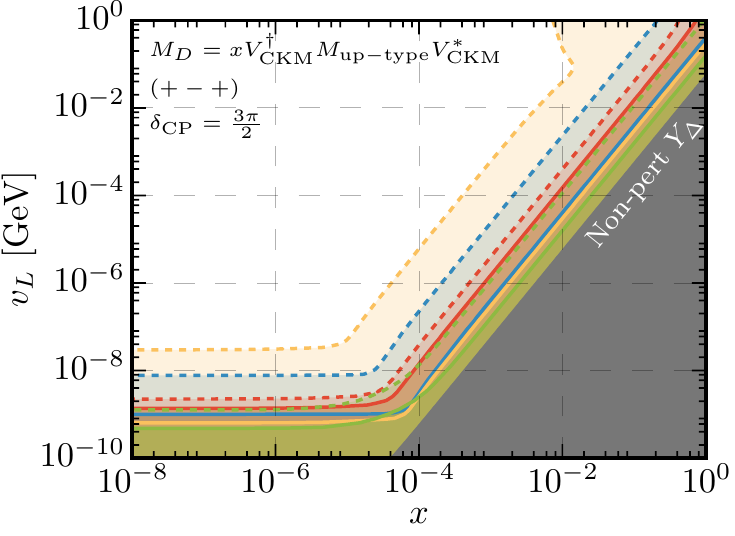}
\caption{Sensitivity of current and future experiments in the $(x,v_L)$ plane for $\dCP=3\pi/2$. Solid lines are the current bounds, while the dashed lines are the projected sensitivities of upcoming experiments, see \cref{tab:sensi} for the numerical values. The colour scheme for the shaded regions is $\mu \to 3e$ (blue), $\mu\to e\gamma$ (red), $\mu-e$,Ti (yellow) and finally $\tau \to 3\mu$ (green).}
\label{fig:2D_diag_dCP_3pion2}
\end{figure}

Let us begin with the parametrisation $M_D = x \mathbb{1}~$GeV. In the top row of \cref{fig:2D_diag} 
we show the respective planes for both the $(+++)$ and $(+-+)$ solutions. As discussed in some 
detail in \cref{subsec:diag_MD}, $\mu \to 3e$ is the observable with the best prospects of being 
measured in the near future, as there exists no real hierarchy between the $Y_\Delta$ entries. 
However, if the PRISM/PRIME experiment reaches the expected sensitivity of $10^{-18}$ for $\mu-e$ 
conversion in Ti, then the future reach will be comparable with the projected sensitivity of the 
Mu3e experiment \cite{Blondel:2013ia} for the $(+++)$ sign choice. Nevertheless, for very small $x$-values, $\mu-e$ conversion 
is more sensitive for both sign choices. The case $M_D \propto \mathbb{1}$ also leads to the most 
drastic change in the region which is experimentally probeable when changing between the sign 
choices. Here we see that the change in sign choice drastically increases the rate of the 
observables in the regime where $x \gtrsim 10^{-3}$.

For the case that $M_D=\MUP$, the coverage of both current and upcoming experiments is limited. The vast majority of the sensitive region occurs in the small $x$ and $v_L$ regime. For the sign choice $(+++)$, there is no prospect of future experiments probing perturbative parameter regions where $x \geq \SI{2E-3}{}$ irrespective of the $v_L$ choice. Whereas, for the mixed sign choice, future and current experiments have some sensitivity in the regimes where $Y_\Delta^{(3,3)}$ is close to becoming non-perturbative. Interestingly, due to the increased rate of $\tau\to 3\mu$ decays, see the discussion in \cref{subsec:Mu}, the corresponding measurement prospects for BELLE~II \cite{Aushev:2010bq,BelleII2015} are a little higher than for $\mu\to 3e$ despite the unprecedented sensitivity of the Mu3e experiment.
The sensitivity for small $x$ regions is largely unchanged between the sign choices. The best future prospects in this case is through the measurement of $\mu-e$ conversion.

The last remaining choice studied is $M_D= \MUPCKM$, shown in the bottom row of \cref{fig:2D_diag}. 
There is an increase of the LFV observables w.r.t. the $\MUP$ case in the region $\sqrt{v_L/v_R}\gg \BDi$ due to the CKM multiplication, which boosts sensitivities for the large-$x$ region. 
With upcoming experiments even regions where $x\simeq 10^{-2}$ and $v_L \simeq \SI{1}{\GeV}$ will be detectable through these observables, in particular $\mu-e$ conversion in Titanium. The change in shape of the $\mu-e$ conversion projections for large $v_L$ are due to the $W_{L/R}-\nu_R$ boxes which become important in this region of parameter space, see also \cref{fig:MuEconv_Mell_vs_Mell_CKM}.

Finally we repeat the same procedure for the case $\dCP=3\pi/2$ in \cref{fig:2D_diag_dCP_3pion2}, motivated by recent global fits \cite{Esteban:2016qun}. As discussed in \cref{sec:CPphase}, the differences w.r.t. the $\dCP=0$ case are most drastic for the same-sign solution and a flavour-diagonal $M_D$ as the LFV rates obtain a significant boost in the regions with large $x$ due to the non-orthogonality of the rotation matrix $R$ in the complex case. Therefore, all six cases shown in \cref{fig:2D_diag_dCP_3pion2} also feature measurable LFV rates in the large-$x$ regions. Interestingly, due to different cancellations in the different LFV observables due to the complex phase, see also  \cref{fig:dCP_MD_Diag_GeV}, the relative magnitude of some LFV observables is altered.
In particular, all the parameter region above $x\simeq 10^{-4}$ for $M_D=x\,\MUP$ and $(+-+)$ probeable by the Mu3e experiment is already excluded by $\tau\to 3\mu$. Here, BELLE~II has the best measurement prospects for the near future. However, also for this maximal CP phase, the best prospects in the long run are found in the $\mu-e$ conversion rate should the PRISM/PRIME experiment reach its expected sensitivity. The $M_D \propto \mathbb{1}$ case, however, is best probed by Mu3e.

\section{Conclusions and outlook}
\label{sec:conclusion}

We have investigated left-right symmetric models containing scalar triplets, paying particular attention to a consistent
treatment of the lepton and Higgs sectors. Furthermore, we have advanced a 
 method to consistently calculate the triplet-Yukawa
 couplings taking into account both the experimental data and the underlying
 symmetries without any approximations. 
For a given parameter point in the model there exists an eightfold degeneracy in the solution of the triplet-Yukawas due to different sign choices in the quadratic equations for each fermion generation. We find that these eight cases can be
divided into two sub-classes. 

 The model is completely left-right symmetric
in view of its particle content and the differences between the bilinear terms of the
scalar potential. 
We have considered several different realisations of the neutrino Dirac mass term, namely, a flavour diagonal case
with either degenerate entries or a hierarchy similar to the up-quark sector as well
as a scenario where there is CKM-like mixing. For each case we have studied
in detail the consequences for lepton flavour violating observables, considering both classes of sign choice for the triplet-Yukawa solution.
 Using this knowledge
we have surveyed which parts of the parameter space can be probed by upcoming lepton flavour violation
experiments. This entailed a calculation of the rates
for $\mu\to e \gamma$, $\mu \to 3 e$, their counterparts in the $\tau$-sector
as well as $\mu-e$ conversion in heavy nuclei, studying in particular their dependence on
the Yukawa couplings as well as on various parameters of the Higgs potential.

Naively one would expect that
flavour-violating three-body decays of the leptons, most importantly
$\mu \to 3 e$, will give the best sensitivity
and discovery potential, due to the tree-level contributions via the 
doubly charged Higgs bosons. 
While this is correct for some regions of parameter
space, we find that there is also a large part where upcoming $\mu-e$ conversion experiments
will be more sensitive. 
This occurs over the majority of the parameter space due to $\gamma$-penguins with charged scalars running in the loops, however for regions
where the triplet Yukawas are small, the  $W_R$-$\nu_R$ loop contributions can dominate.
These conclusions hold despite the fact that existing electroweak precision data implies that
the additional vector bosons are too heavy to be discovered at the 14 TeV LHC.  

Given the case that all signs in the solution to the triplet-Yukawa are equal, there are significant differences between the different parametrisations of the Dirac mass term. In particular, the case with a CKM-like flavour mixing in the Dirac mass matrix exhibits LFV rates which are, in most of the parameter space, several orders of magnitude larger than for the other parametrisations. When switching to the other class of sign choices or allowing a non-zero CP phase in the neutrino mixing matrix, the respective differences are reduced.

For completeness we note, that in some parts of the parameter space investigated
the doubly charged Higgs bosons are light enough that they might be discovered in the
next years at the LHC. However, 
some  are sufficiently heavy that they could only be studied
at a 100 TeV $p$-$p$ collider.

\section*{Acknowledgements}
The authors would like to thank Florian Staub for valuable assistance on {\tt SARAH}-related questions as well as Martin Hirsch for collaboration in the early stages of this project. M.E.K.\ acknowledges support from the DFG Research Unit 2239 ``New Physics
at the LHC'' while both M.E.K and T.O are supported by the SFB-Transregio TR33 ``The Dark Universe''. W.P.\ is supported by the
DFG, project nr.\ PO-1337/3-1. C.B. thanks the Institut f\"{u}r Theoretische Physik und Astronomie in W\"urzburg for 
its hospitality while part of this work was carried out. C.B. is supported by the Spanish grants FPA2011-22975, FPA2014-58183-P, Multidark 
CSD2009-00064 and SEV-2014-0398 (MINECO), and PROMETEOII/2014/084 (Generalitat Valenciana).

\addtocontents{toc}{\protect\setcounter{tocdepth}{1}}

\appendix 
\section{Determination of the triplet-Yukawa couplings}
\label{sec:TripCoupParameterization}

As discussed in \cref{sec:ParameterisationYuks}, one can find a suitable parametrisation of the triplet-Yukawa couplings for either one or both discrete LR symmetries, depending on whether or not there is a CP phase present in the PMNS matrix. To reiterate: 
\begin{itemize}
\item $\dCP = 0$: Charge-conjugation or parity symmetric
\item $\dCP \neq 0$: Only parity symmetric
\end{itemize}
In the following we describe in more detail the method used to determine the triplet-Yukawa couplings as well as the requirement of invoking different symmetries in the presence of CP phases.

We begin with the expressions for the light neutrino mass matrices that are re-written using invariance under charge-conjugation and parity, namely \cref{eq:lightnumassmatrix,eq:lightnumassmatrixParity}
\begin{align}
\mnul &\overset{\mathcal{C}}{=}\left(\frac{v_L}{v_R} \,M_R -M_D\,M_{R}^{-1}\,M_D\right)\,,\\
 \mnul&\overset{\mathcal{P}}{=}\left(\frac{v_L}{v_R} \,M_R^* -M_D\,M_{R}^{-1}\,M_D^*\right)\,.
\end{align}
For the charge conjugation symmetric case we multiply the left- and right-hand side by $M_D^{-1/2}$, while for the parity symmetric case multiplication from the right-hand side requires an additional conjugation, yielding
\begin{align}
M_D^{-1/2} \mnul M_{D}^{-1/2}&\overset{\mathcal{C}}{=} \frac{v_L}{v_R} M_D^{-1/2} M_R M_D^{-1/2}-M_D^{1/2} M_R^{-1} M_D^{1/2}\,, \\
M_D^{-1/2} \mnul M_{D}^{*\,-1/2}&\overset{\mathcal{P}}{=} \frac{v_L}{v_R} M_D^{-1/2} M_R^* M_D^{*\,-1/2}-M_D^{1/2} M_R^{-1} M_D^{*1/2}\,,
\end{align}
which if we make the following definitions
\begin{align}\label{eq:ABalpha_defs}
\alpha &\equiv \frac{v_L}{v_R}\,, & A &\equiv M_D^{(*)\,-1/2} M_R M_D^{-1/2}\,, & B &\equiv M_D^{-1/2} \mnul M_D^{(*)\,-1/2}\,,
\end{align}
where $(*)$ refers to the additional conjugation required for the parity symmetric scenario, allows one to write 
\begin{align}
\label{eq:ABalpha_relation}
 B &\overset{\mathcal{C}}{=} \alpha A - A^{-1}\,,\\
  B &\overset{\mathcal{P}}{=} \alpha A^* - A^{-1}\,. \label{eq:ABalpha_relation1}
\end{align}
However, in what follows we exploit the fact that the matrices $A$ and $B$ are either: $(i)$ real symmetric ($\dCP =0$), or $(ii)$ complex symmetric ($\dCP \neq 0$). As a result, $B$ and subsequently $A$ are diagonalised by $R$ which is either: $(i)$ a real orthogonal matrix, or $(ii)$ a complex unitary matrix.\footnote{For more details regarding the various diagonalisation procedures see appendix D of Ref.~\cite{Dreiner:2008tw}.}
For case $(i)$, if the matrix $R$ diagonalizes $A$ then this same matrix also diagonalizes the inverse matrix $A^{-1}$. As a result \cref{eq:ABalpha_relation} can be written as
\begin{align}\label{eq:simulBandAdiag}
B \overset{\mathcal{C}\text{ or }\mathcal{P}}{=}\alpha A - A^{-1}= R \left(\alpha A_D-A_D^{-1}\right)R^T \,,
\end{align}
where the subscript $D$ indicates the matrix is in a real diagonal form. Here we observe that both charge-conjugation and parity invariance are equivalent if $A$ is real. As requiring real $A_D$ necessitates a unitary $R$, we cannot simmultaneously diagonalise both $A$ and $A^{-1}$ for case $(ii)$ as
\begin{align}
B \overset{\mathcal{C}}{=}\alpha A - A^{-1}= R^*\alpha A_D R^\dag-RA_D^{-1}R^T \neq R \left(\alpha A_D-A_D^{-1}\right)R^T \,.
\end{align}
However, here this procedure indeed applies for $A^*$ and $A^{-1}$ namely
\begin{align}
B \overset{\mathcal{P}}{=}\alpha A^* - A^{-1}= R \left(\alpha A_D-A_D^{-1}\right)R^T \,,
\end{align}
so that one can find a suitable parametrisation of the triplet-Yukawa in the $\mathcal P$-symmetric case also with $\delta_{\rm CP}\neq 0$, as we shall see in what follows.
%

\cref{eq:ABalpha_relation,eq:ABalpha_relation1} are identical in their respective real diagonal forms, namely
\begin{equation}
 \BDi=\alpha \ADi-\left(\ADi\right)^{-1}\,,
\end{equation}
where $i=1,2,3$. Solving the decoupled quadratic equations for $\ADi$ yields
\begin{align}
\label{eq:polysol}
\ADi &= \frac{\BDi \pm \sqrt{\left(\BDi\right)^2 + 4 \alpha}}{2\alpha}\,.
\end{align}
Using the definitions in \cref{eq:ABalpha_defs,eq:ML_MR_mD_defs} we arrive at expressions for the triplet Yukawas
\begin{align}
Y_{\Delta}^{(\pm\pm\pm)} \equiv Y_{\Delta_{L/R}}^{(\pm\pm\pm)} &= \frac{1}{{2\sqrt{2} v_L}} M_D^{(*)1/2}R^* \text{diag}\left(\BDi \pm \sqrt{\left(\BDi\right)^2 + 4 \alpha}\right)
R^\dag M_D^{1/2}\,,
\label{eq:Ydelta_appendix}
\end{align}

As explained above, this expression holds for any $\delta_{\rm CP}$ in the case of discrete $\mathcal P$ symmetry whereas it can be applied to both $\mathcal C$ and $\mathcal P$ symmetries in the absence of a CP phase. 

\cref{eq:polysol} leads to an eightfold degeneracy in the solutions due to the choice of sign for each diagonal entry of $A_D$, as first noted in Ref. \cite{Akhmedov:2005np}. However, these eight solutions can be categorized into two distinct cases. The differences between these two cases is best illustrated through an example where we choose $M_D$ to be diagonal and real. In this case $B \propto \mnul$ which, for realistic choices of the neutrino oscillation parameters and large enough $v_L$, leads to the hierarchy $\alpha \gg (\BDi)^2 $. Subsequently, expanding \cref{eq:polysol} for small $\BDi$ yields
\begin{align}
\ADi &= \pm \alpha^{-1/2} + \mathcal{O}\left(\BDi\right)\,.
\label{eq:ad_appendix}
\end{align}
Therefore the principle difference between the degenerate solutions is simply a sign choice. But, this sign choice has large ramifications on the resulting triplet Yukawa matrices. To demonstrate this consider the two neutrino generation case, where we examine both mixed and same-sign choices for the cases $\dCP =0$ and $\dCP \neq 0$.

{\emph{\bf{$\dCP = 0$:}}} For the same-sign scenario we have
\begin{align}
A^{(++)} &= R A_D R^T =\begin{pmatrix} \cos\theta & \sin\theta \\ -\sin\theta & \cos\theta \end{pmatrix} \begin{pmatrix} \alpha^{-1/2} & 0 \\ 0 & \alpha^{-1/2} \end{pmatrix} \begin{pmatrix} \cos\theta & -\sin\theta \\ \sin\theta & \cos\theta \end{pmatrix} = A_D\,.
\end{align}
whereas for the mixed sign case we obtain
\begin{align}
A^{(+-)}&= R \begin{pmatrix} \alpha^{-1/2} & 0 \\ 0 & -\alpha^{-1/2} \end{pmatrix}R^T =  \alpha^{-1/2} \begin{pmatrix} \cos2\theta & \sin2\theta \\ \sin2\theta & - \cos2\theta \end{pmatrix}\,.
\end{align}
We therefore see that in this example the choice of sign dictates whether or not there are flavour violating off-diagonal entries at leading order. Note that the above argumentation generalizes to the realistic scenario of three neutrino generations. 

This argumentation is, of course, still valid if $M_D$ is non-diagonal as it relies on already-diagonalised quantities. However, when plugging \cref{eq:ad_appendix} into the full expression, \cref{eq:Ydelta_appendix}, one sees that the impact of the above-mentioned effect is weakened comparatively when $M_D$ itself contains a flavour-violating structure. Therefore, in this situation the solution for $Y_\Delta$ with different sign choices in general contains larger off-diagonal elements than the solution with equal signs, the relative difference of these off-diagonals is small compared to the case in which $M_D$ is diagonal.

Shown in \cref{fig:Mu3e_sign_dep} are numerical results of the branching ratio for $\mu\to3e$ as a function of the triplet VEV $v_L$. Here, all possible sign choices are considered in two extreme scenarios, namely diagonal $M_D$ and $M_D = \MUPCKM$. As illustrated in the toy two-generation example, same-sign choices for the diagonal $M_D$ lead to highly suppressed off-diagonals in the resulting triplet Yukawas in comparison to the mixed-sign case. 
However, in the scenario that $M_D$ is no longer diagonal then the effect between same or mixed-sign solutions is comparatively smaller. 

{\emph{\bf{$\dCP \neq 0$:}}} Here we demonstrate that there is a significant difference in the same-sign scenario when introducing this phase. As $B$ is now a complex symmetric matrix $R$ must necessarily be a unitary matrix. Therefore for the same-sign case we obtain 
\begin{align}
A^{(++)} &= R^* A_D R^\dagger \,,\\
&= e^{-2i\phi_3}\begin{pmatrix} e^{-i \phi_1}\cos\theta & e^{-i \phi_2}\sin\theta \\ -e^{-i \phi_2}\sin\theta & e^{i \phi_1} \cos\theta \end{pmatrix} \begin{pmatrix} \alpha^{-1/2} & 0 \\ 0 & \alpha^{-1/2} \end{pmatrix}
\begin{pmatrix} e^{-i \phi_1}\cos\theta & e^{-i \phi_2}\sin\theta \\ e^{-i \phi_2}\sin\theta & e^{i \phi_1} \cos\theta \end{pmatrix} \,,\notag\\
&=\alpha^{-1/2}e^{2i\phi_3} \begin{pmatrix}
e^{-2i\phi_1}\left(\cos^2 \theta + e^{-2i(\phi_2-\phi_1)}\sin^2 \theta\right) & i \sin2\theta \sin(\phi_1-\phi_2) \\
i \sin2\theta \sin(\phi_1-\phi_2) & e^{2i\phi_1}\left(\cos^2 \theta + e^{2i(\phi_2-\phi_1)}\sin^2 \theta\right)
\end{pmatrix}\,, \notag
\end{align}
where $\phi_i$ are the three phases of a generic unitary $2\times2$ matrix. We observe, in contrast to the case with the same-sign solution and $\dCP =0$, that there is a complex off-diagonal generated at leading order even in the case that $M_D$ is proportional to the unit matrix. This off-diagonal is in general non-zero as the three phases $\phi_1$, $\phi_2$ and $\phi_3$ must be chosen such that the matrix $A$ is brought into its real diagonal form. The resulting structure shares similarities to the case with mixed sign and $\dCP =0$. Namely, we see an off-diagonal entry, which in this case is complex, proportional to $\sin2\theta$.

\section{Alternative neutrino parameters}
\label{sec:alt_neutrino_params}
\begin{figure}
\includegraphics{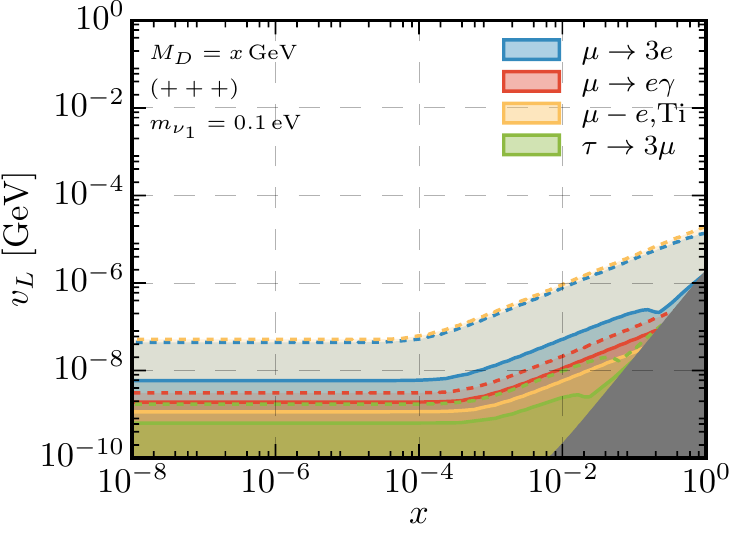}
\includegraphics{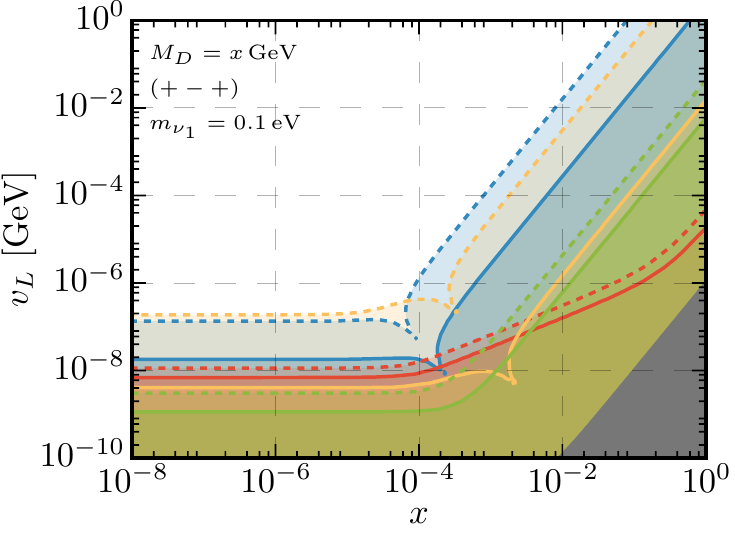}\\
\includegraphics{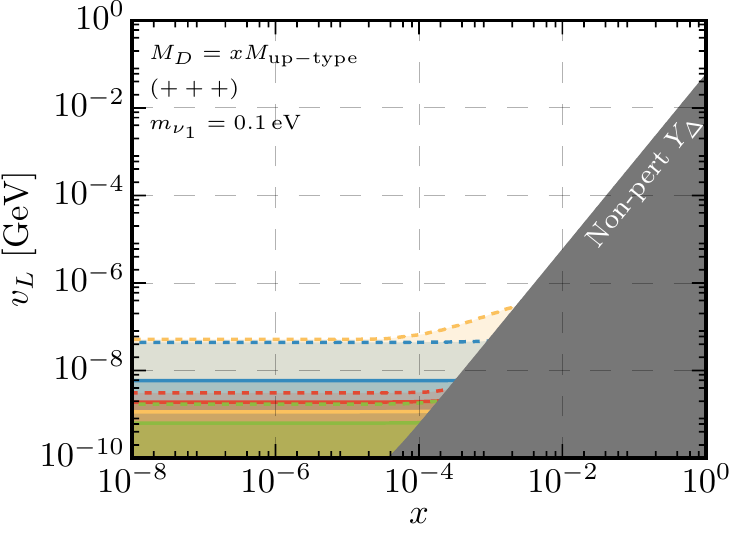}
\includegraphics{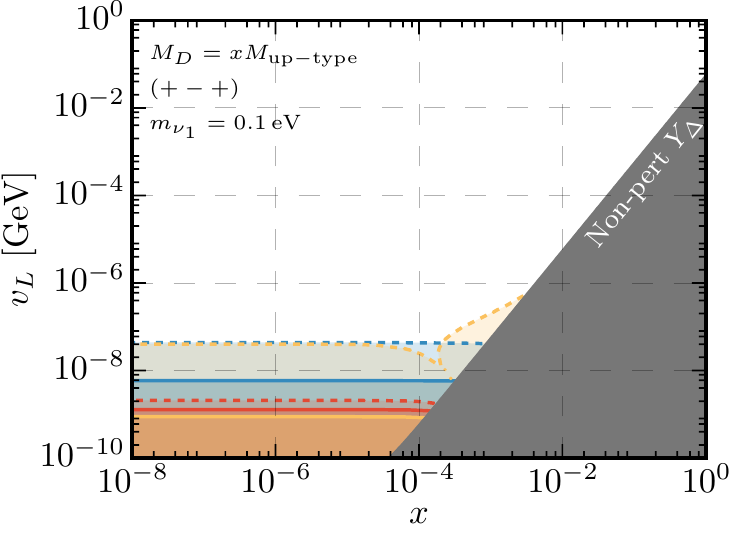}\\
\includegraphics{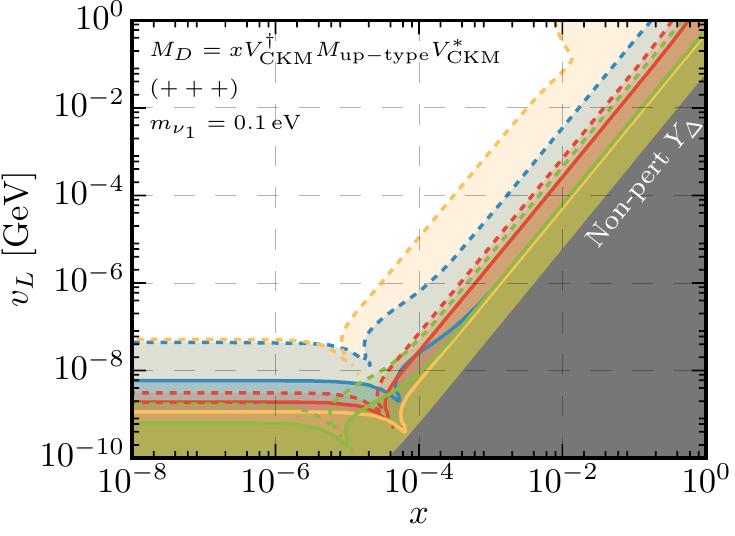}
\includegraphics{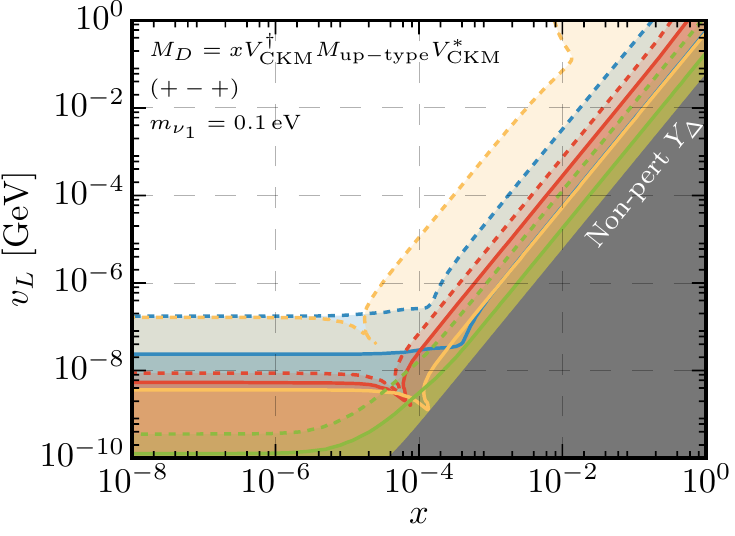}
\caption{Sensitivity of current and future experiments in the $(x,v_L)$ plane. Here we take $\dCP=0$, and $m_{\nu_1}=\SI{0.1}{\eV}$ where once again all other model parameters are given in \cref{tab:benchmark}. See \cref{fig:2D_diag} for a description of the colours and contours.}
\label{fig:2D_diag_mnu1_eV}
\end{figure}

\begin{figure}
\includegraphics{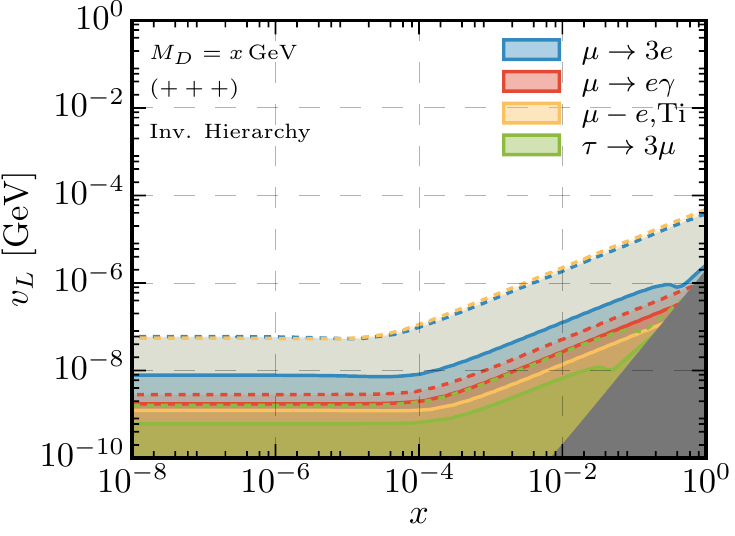}
\includegraphics{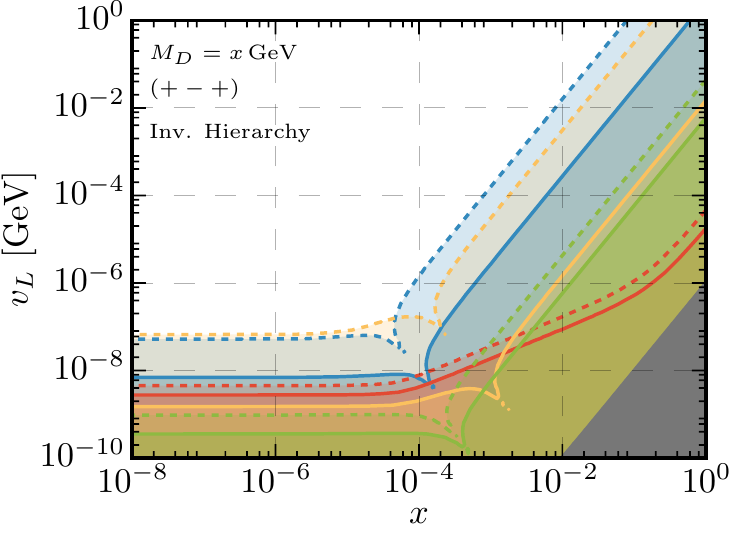}\\
\includegraphics{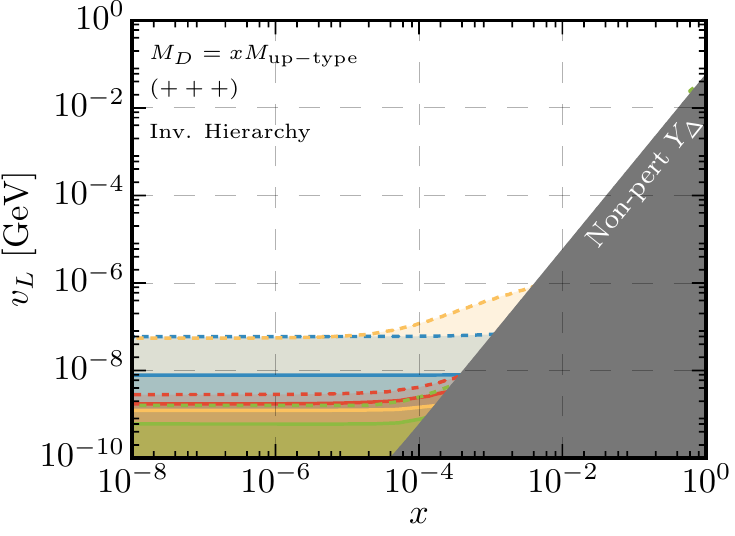}
\includegraphics{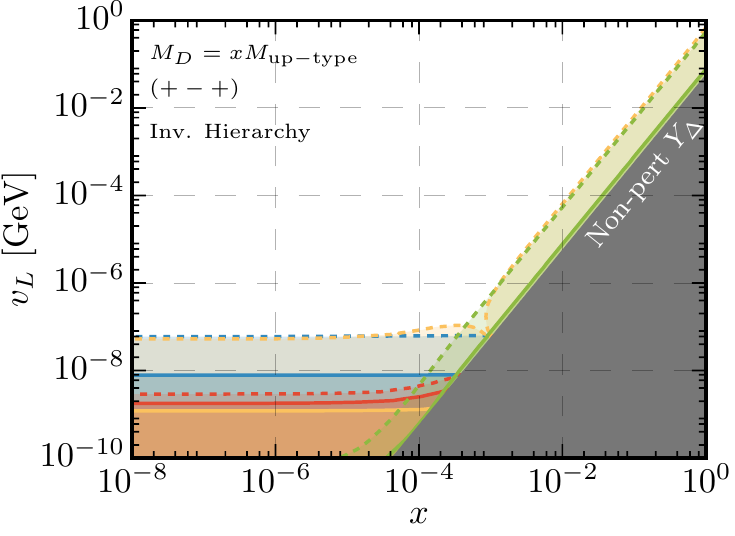}\\
\includegraphics{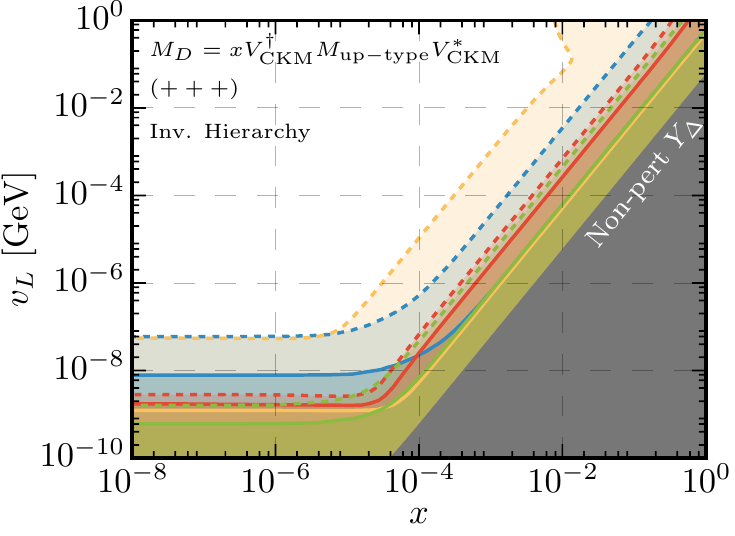}
\includegraphics{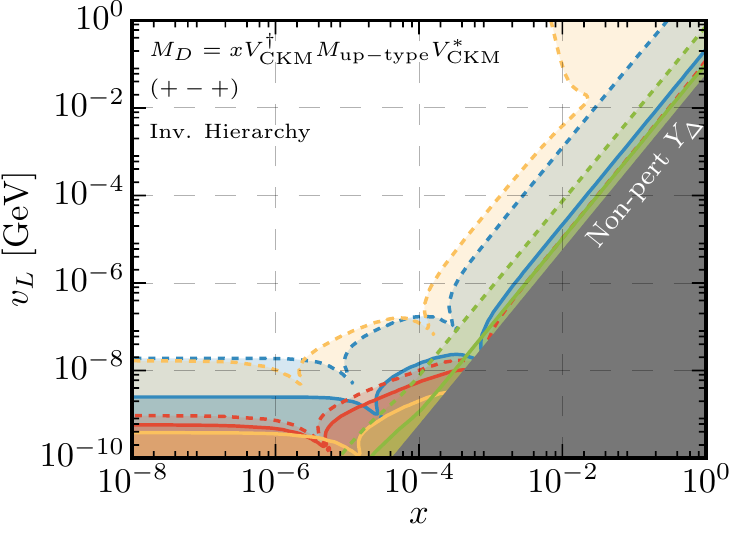}
\caption{Sensitivity of current and future experiments in the $(x,v_L)$ plane where we assume an inverted neutrino mass hierarchy. Here we take $\dCP=0$, and $m_{\nu_1}=\SI{E-4}{\eV}$ where once again all other model parameters are given in \cref{tab:benchmark}. See \cref{fig:2D_diag} for a description of the colours and contours.}
\label{fig:2D_IH}
\end{figure}

In this appendix we show the results of current bounds and future sensitivities using the values of \cref{tab:sensi} while varying the neutrino masses and hierarchies. Firstly, we show the effect of altering the lightest neutrino mass to $m_{\nu_1} = \SI{0.1}{\eV}$ resulting in a quasi-degenerate light neutrino mass spectrum. The results of which are shown in \cref{fig:2D_diag_mnu1_eV}. We then also consider the case of an inverse hierarchy of the neutrino masses. In this scenario the best-fit values used for the neutrino oscillation parameters are those from \cite{Forero:2014bxa}. Setting $\dCP = 0$ and assuming $m_{\nu_3}=\SI{E-4}{\eV}$ we obtain the results shown in \cref{fig:2D_IH}.

\section{Scalar mass matrices}
\label{sec:scalarmasses}
\subsection{Doubly charged}
The mass matrix is written in the basis $\{ \delta_R^{--}, \delta_L^{--}\}$
\begin{align}
\mathcal{M}_{H^{\pm\pm}}^2 &=\frac12\begin{pmatrix}
 m_{RR} & m_{RL}\\
\dotso &  m_{LL}
\end{pmatrix}\,,
\end{align}
\begin{subequations}
with entries
\begin{align}
m_{RR}&=(\rho_3-2\rho_1)v_L^2+\rho_2 v_R^2-\alpha_3 v^2 \frac{t_\beta^2-1}{t_\beta^2+1} \,,\\
m_{LL}&=(\rho_3-2\rho_1)v_R^2+\rho_2 v_L^2-\alpha_3 v^2 \frac{t_\beta^2-1}{t_\beta^2+1}\,,\\
m_{RL}&= v^2\left(\beta_3(t_\beta^2-1)-\beta_1\frac{t_\beta}{t_\beta^2+1} \right) +v_L v_R \left(4 \rho_4 +(2\rho_1-\rho_3)t_\beta^2\right)\,.
\end{align}
\end{subequations}
Expanding in two expansion parameters $x=v_L/v$ and $y=v/v_R$ we obtain for the masses to leading order 
\begin{subequations}
\begin{align}
m_{H^{\pm\pm}_1}&= 2 \rho_2 v_R^2 +\frac{1}{2} \alpha_3 v^2 \cos2\beta +\mathcal{O}\left(x,y^2\right)\,,\\
m_{H^{\pm\pm}_2}&= \frac{1}{2} \left((\rho_3-2\rho_1)v_R^2 + \alpha_3 v^2 \cos 2\beta\right)+\mathcal{O}\left(x,y^2\right)\,.
\end{align}
\end{subequations}
\subsection{Singly charged}
Here the basis is defined as $\{\phi^-, (\phi^+)^c, (\delta_R^+)^c, (\delta_L^+)^c \}$
\begin{align}
\mathcal{M}_{H^\pm}^2 &=\frac12\begin{pmatrix}
 m_{\phi^-\phi^-} & m_{\phi^-\phi^+} & m_{\phi^- R} & m_{\phi^- L}\\
 \dotso & m_{\phi^+\phi^+} & m_{\phi^+ R} & m_{\phi^+ L} \\
 \dotso & \dotso & m_{R R} & m_{R L} \\
    \dotso & \dotso & \dotso & m_{L L} \\
\end{pmatrix}\,,
\end{align}
with entries
\begin{subequations}
\begin{align}
m_{\phi^-\phi^-} &=\frac{1}{t_\beta^2-1} \left(2v_L v_R t_\beta(2\beta_3-\beta_1 t_\beta)+ \frac{v_L^2 v_R^2}{v^2}(2 \rho_1 - \rho_3)(t_\beta^2+1) -\alpha_3 \left(v_L^2 + t_\beta^2 v_R^2\right)\right),\\
m_{\phi^-\phi^+} &=\frac{t_\beta^2+1}{t_\beta^2-1} \left(v_L v_R (2\beta_3t_\beta-\beta_1 )+ \frac{2v_L^2 v_R^2}{v^2}t_\beta(2 \rho_1 - \rho_3) -\alpha_3 \frac{t_\beta}{t_\beta^2+1}\left(v_L^2 +  v_R^2\right)\right)\,,\\
m_{\phi^+\phi^+} &=\frac{1}{t_\beta^2-1} \left(2v_L v_R t_\beta(2\beta_3t_\beta-\beta_1 )+ \frac{v_L^2 v_R^2}{v^2}(2 \rho_1 - \rho_3)(t_\beta^2+1) -\alpha_3 \left(t_\beta^2 v_L^2 +v_R^2\right)\right),\\
m_{\phi^- R} &= \frac{\sqrt{2} v}{2\sqrt{t_\beta^2+1}}\left(v_L(\beta_1- 2\beta_3 t_\beta)+ v_R \alpha_3 t_\beta\right)\,,\\
m_{\phi^+ L} &= \frac{\sqrt{2} v}{2\sqrt{t_\beta^2+1}}\left(v_R(\beta_1- 2\beta_3 t_\beta)+ v_L \alpha_3 t_\beta\right)\,,\\
 m_{\phi^- L} &= \frac{\sqrt{2}}{2\sqrt{t_\beta^2+1}} \left( v v_L \alpha_3 + v v_R t_\beta(\beta_1-2\beta_3 t_\beta)-2 \frac{v_L v_R^2}{v} (2 \rho_1 - \rho_3)(t_\beta^2+1)\right)\,,\\
 m_{\phi^+ R} &= \frac{\sqrt{2}}{2\sqrt{t_\beta^2+1}}\left( v v_R \alpha_3 + v v_L t_\beta(\beta_1-2 \beta_3 t_\beta) -2 \frac{v_L^2 v_R}{v} (2 \rho_1 - \rho_3) (t_\beta^2 +1)\right)\,,\\
 m_{R R} &=  v_L^2(\rho_3 - 2\rho_1) - \frac{1}{2} v^2 \alpha_3 \frac{t_\beta^2 -1}{t_\beta^2+1}\,,\\
  m_{L L} &=  v_R^2(\rho_3 - 2\rho_1) - \frac{1}{2} v^2 \alpha_3 \frac{t_\beta^2 -1}{t_\beta^2+1}\,,\\
  m_{R L} &= \frac{1}{2}\left(2v_L v_R t_\beta(2\rho_1 - \rho_3) - v^2(\beta_1-2\beta_3 t_\beta) \frac{t_\beta^2-1}{t_\beta^2+1}\right)\,.
\end{align}
\end{subequations}
Expanding once again in two expansion parameters $x$ and $y$ as well as working in the limit $\tan\beta\to 0$ we obtain for the masses to leading order 
\begin{subequations}
\begin{align}
m_{H^{\pm}_L}&= \frac{1}{4}\left(v^2\left[\alpha_3 \frac{\beta_1^2}{\rho_3-2\rho_1-\alpha_3}\right]+2v_R^2(\rho_3 -2 \rho_1)\right)+ \mathcal{O}\left(x,y^2\right)\,,\\
m_{H^{\pm}}&= \frac{1}{4}\left(v^2\left[ \frac{\beta_1^2}{2\rho_1+\alpha_3-\rho_3}\right]+\alpha_3(v^2+2v_R^2)\right)+ \mathcal{O}\left(x,y^2\right)\,.
\end{align}
\end{subequations}
\subsection{Neutral CP-odd}
Here the basis is defined as $\{\varphi_1, \varphi_2, \varphi_R, \varphi_L \}$
\begin{align}
\mathcal{M}^2_A &= \frac12\begin{pmatrix}
 m_{\varphi_1\varphi_1} & m_{\varphi_1\varphi_2} & m_{\varphi_1 R} & m_{\varphi_1 L}\\
 \dotso & m_{\varphi_2\varphi_2} & m_{\varphi_2 R} & m_{\varphi_2 L} \\
 \dotso & \dotso & m_{R R} & m_{R L} \\
    \dotso & \dotso & \dotso & m_{L L} \\
\end{pmatrix}\,,
\end{align}
with entries
\begin{subequations}
\begin{align}
m_{\varphi_1\varphi_1} &= 4v^2\frac{t_\beta^2}{t_\beta^2+1}\left[2 \lambda_2- \lambda_3 \right] + 2v_L v_R\frac{1}{t_\beta^2-1} \left[\beta_3 t_\beta^2 (t_\beta^2-3) - \beta_1 t_\beta (t_\beta^2-2)\right] \notag \\
&\qquad+ \frac{2v_L^2 v_R^2}{v^2}\frac{1}{t_\beta^2-1} \left[(2 \rho_1 - \rho_3) (1+t_\beta^2)((t_\beta^2-2)\right] + \alpha_3 \frac{t_\beta^2}{t_\beta^2-1} (v_L^2+ v_R^2)\,, \\
m_{\varphi_1\varphi_2} &= 4v^2\frac{t_\beta^2}{t_\beta^2+1}\left[\lambda_3 - 2 \lambda_2 \right] + 2v_L v_R \frac{1}{t_\beta^2-1}\left[\beta_3  (t_\beta+t_\beta^3) - \beta_1 )\right] \notag \\
&\qquad+ \frac{2v_L^2 v_R^2}{v^2}\frac{(t_\beta+t_\beta^3)}{t_\beta^2-1} \left[(2 \rho_1 - \rho_3) (1+t_\beta^2)((t_\beta^2-2)\right] - \alpha_3 \frac{t_\beta}{t_\beta^2-1} (v_L^2+ v_R^2)\,, \\
m_{\varphi_2\varphi_2} &= 4v^2\frac{1}{t_\beta^2+1}\left[\lambda_3 - 2 \lambda_2 \right] + 2 v_L v_R \frac{1}{t_\beta^2-1} \left[(3 t_\beta^2 - 1) \beta_3 - \beta_1 t_\beta \right] \notag\\
&\qquad + \frac{2v_L^2v_R^2}{v^2}\frac{t_\beta^2+1}{t_\beta^2-1}\left[2 \rho_1 -\rho_3 \right]-\alpha_3 \frac{1}{t_\beta^2-1} (v_L^2+v_R^2)\,, \\
m_{\varphi_1 R} &= \frac{1}{\sqrt{t_\beta^2+1}}\left(v_L v t_\beta \left(2\beta_3 t_\beta - \beta_1\right)+\frac{2v_L^2 v_R}{v} (t_\beta^2+1)\left(2\rho_1-\rho_3\right)\right)\,,\\
m_{\varphi_1 L} &= \frac{1}{\sqrt{t_\beta^2+1}}\left(v_R v t_\beta \left(\beta_1- 2\beta_3 t_\beta \right)-\frac{2v_L v_R^2}{v} (t_\beta^2+1)\left(2\rho_1-\rho_3\right)\right)\,, \\
m_{\varphi_2 R} &= \frac{v_L v}{\sqrt{t_\beta^2+1}} \left(2 \beta_3 t_\beta - \beta_1\right)\,, \qquad
m_{\varphi_2 L} = \frac{v_R v}{\sqrt{t_\beta^2+1}} \left(\beta_1 - 2 \beta_3 t_\beta\right)\,,\\
m_{R R} &= v_L^2(\rho_3 - 2 \rho_1)\,, \quad
m_{L L} = v_R^2(\rho_3 - 2 \rho_1)\,, \quad
m_{R L} = v_L v_R \left(2 \rho_1 - \rho_3\right)\,. 
\end{align}
\end{subequations}
Expanding once again in two expansion parameters $x$ and $y$ as well as working in the limit $\tan\beta\to 0$ we obtain for the masses to leading order 
\begin{subequations}
\begin{align}
m_{A_L}^2 &= \frac{1}{2}\left((\rho_3 - 2\rho_1) v_R^2 - \frac{\beta_1^2}{\alpha_3 + 2\rho_1 - \rho_3}v^2\right)+ \mathcal{O}\left(x,y^2\right)\,, \\
m_{A}^2 &= \frac{1}{2}\left(4 \alpha_3 v_R^2 + \left[4(\lambda_3 - 2 \lambda_2)   + \frac{\beta_1^2}{\alpha_3 + 2\rho_1 - \rho_3}\right] v^2 \right)+ \mathcal{O}\left(x,y^2\right)\,.
\end{align}
\end{subequations}
\subsubsection{Neutral CP-even}
Here the basis is defined as $\{\sigma_1, \sigma_2, \sigma_R, \sigma_L \}$
\begin{align}
\mathcal{M}^2_A &= \frac12\begin{pmatrix}
 m_{\sigma_1\sigma_1} & m_{\sigma_1\sigma_2} & m_{\sigma_1 R} & m_{\sigma_1 L}\\
 \dotso & m_{\sigma_2\sigma_2} & m_{\sigma_2 R} & m_{\sigma_2 L} \\
 \dotso & \dotso & m_{R R} & m_{R L} \\
    \dotso & \dotso & \dotso & m_{L L} \\
\end{pmatrix}\,,
\end{align}
with entries
\begin{subequations}
\begin{align}
m_{\sigma_1\sigma_1} &= 4v^2 \frac{1}{t_\beta^2+1} \left[\lambda_1 + t_\beta((2\lambda_2 + \lambda_3) t_\beta - 2\lambda_4)\right] + 2 v_L v_R \frac{t_\beta^2}{t_\beta^2-1} \left[\beta_3 (t_\beta^2+1)- \beta_1 t_\beta \right] \notag \\
&\qquad +\frac{2 v_L^2 v_R^2}{v^2} \frac{t_\beta^2}{t_\beta^2-1}\left[(2\rho_1-\rho_3)(t_\beta^2+1)\right] -\alpha_3 \frac{t_\beta^2}{t_\beta^2-1}(v_L^2 + v_R^2)\,,\\
m_{\sigma_2\sigma_2} &= 4 v^2 \frac{1}{t_\beta^2+1} \left[2\lambda_2 + \lambda_3 + t_\beta (\lambda_1 t_\beta - 2 \lambda_4))\right] + 2 v_L v_R \frac{1}{t_\beta^2-1}\left[\beta_3 (t_\beta^2+1)- \beta_1 t_\beta \right] \notag \\
&\qquad +\frac{2 v_L^2 v_R^2}{v^2} \frac{1}{t_\beta^2-1}\left[(2\rho_1-\rho_3)(t_\beta^2+1)\right] -\alpha_3 \frac{1}{t_\beta^2-1}(v_L^2 + v_R^2)\,,\\
m_{\sigma_1\sigma_2} &= 4 v^2 \frac{1}{t_\beta^2+1} \left[ t_\beta (\lambda_1  - 2 \lambda_2 + \lambda_3)- \lambda_4 (1+ t_\beta^2)\right] + 2 v_L v_R \frac{t_\beta}{t_\beta^2-1}\left[\beta_1 t_\beta - \beta_3 (t_\beta^2+1) \right] \notag \\
&\qquad -\frac{2 v_L^2 v_R^2}{v^2} \frac{t_\beta}{t_\beta^2-1}\left[(2\rho_1-\rho_3)(t_\beta^2+1)\right] +\alpha_3 \frac{t_\beta}{t_\beta^2-1}(v_L^2 + v_R^2)\,,\\
m_{\sigma_1 R} &= \frac{t_\beta}{\sqrt{1+t_\beta^2}}\left(v_L v(2 \beta_3 t_\beta - \beta_1) + 2 v_R v(\alpha_1 - 2 \alpha_2 t_\beta) +  \frac{2 v_L^2 v_R}{v}(t_\beta^2+1)(2 \rho_1 - \rho_3)\right)\,,\\
m_{\sigma_1 L} &= \frac{1}{\sqrt{1+t_\beta^2}}\left(v_R v t_\beta(2 \beta_3 t_\beta - \beta_1) + 2 v_L v(\alpha_1 - 2 \alpha_2 t_\beta) +  \frac{2 v_L v_R^2}{v}(t_\beta^2+1)(2 \rho_1 - \rho_3)\right)\,,\\
m_{\sigma_2 R} &= \frac{v}{\sqrt{t_\beta^2+1}} \left(v_L [\beta_1 - 2\beta_3 t_\beta] + 2 v_R \left[(\alpha_1+ \alpha_2) t_\beta -2 \alpha_2\right]\right)\,,\\
m_{\sigma_2 L} &= \frac{v}{\sqrt{t_\beta^2+1}} \left(v_R [\beta_1 - 2\beta_3 t_\beta] + 2 v_L \left[(\alpha_1+ \alpha_2) t_\beta -2 \alpha_2\right]\right)\,,\\
m_{R R} &= (\rho_3 - 2 \rho_1) v_L^2 + 4\rho_1 v_R^2\,, \quad
m_{L L} = (\rho_3 - 2 \rho_1) v_R^2 + 4\rho_1 v_L^2\,, \quad
m_{R L} = (2\rho_1 + \rho_3) v_L v_R\,.
\end{align}
\end{subequations}
In order to obtain analytic results for the masses one must specify to a region of parameter space where the triplet and bi-doublet scalars do not mix. This corresponds to the limit $v_L,\alpha_1,\alpha_2,\beta_1 \to 0$. Additionally we also once again perform an expansion in the two parameters $x$ and $y$ as well as working in the limit $\tan\beta\to 0$. This yields the results
\begin{subequations}
\begin{align}
m_h^2 &\simeq 2 \lambda_1 v^2-\frac{8 \lambda_4^2 v^4}{\alpha_3 v_R^2} \,, & m_{H_R}^2 &= 2 \rho_1 v_R^2\,,\\
m_H^2 &= 2(2\lambda_2+\lambda_3)v^2 + \frac{\alpha_3}{2} v_R^2 \,, & m_{H_L}^2 &= \frac{1}{2} \left(\rho_3 - 2 \rho_1\right) v_R^2\,.
\end{align}
\end{subequations}

\section{{\tt SARAH} model file}
\label{sec:app:model_file}
Here we present the {\tt SARAH} model definitions which we have used for the study above and which we have made public on the {\SARAH} website. Alongside the more exotic left-right-symmetric models presented together with Ref.~\cite{Staub:2016dxq}, we present here for the first time a public code featuring the minimal left-right-symmetric model, including the full scalar potential. 

{\bf Gauge groups}
\begin{lstlisting}
Gauge[[1]]={B,   U[1], bminl,       gBL,False};
Gauge[[2]]={WL, SU[2], left,        g2,True};
Gauge[[3]]={WR, SU[2], right,       gR,True};
Gauge[[4]]={G,  SU[3], color,       g3,False};
\end{lstlisting}
{\bf Matter fields.} Here we use the $B-L$ charge normalization such that $Q_{B-L} = \frac{B-L}{2}$, i.e. $Q_{em} = T_{3L}+T_{3R}+Q_{B-L}$
\begin{lstlisting}
FermionFields[[1]] = {QLbar, 3, {conj[uL], conj[dL]},         -1/6, -2,  1, -3};  
FermionFields[[2]] = {LLbar, 3, {conj[nuL], conj[eL]},         1/2, -2,  1,  1};
FermionFields[[3]] = {QR,    3, {uR,  dR},                     1/6,  1,  2,  3};
FermionFields[[4]] = {LR,    3, {nuR, eR},                    -1/2,  1,  2,  1};

ScalarFields[[1]]  = {Phi, 1, {{H0, Hp},{Hm, HPrime0}},          0,  2, -2,  1}; 
ScalarFields[[2]]  = {deltaR,1, {{deltaRp/Sqrt[2],deltaRpp},
                                  {deltaR0, - deltaRp/Sqrt[2]}}, 1,  1,  3,  1};
ScalarFields[[3]]  = {deltaL,1, {{deltaLp/Sqrt[2],deltaLpp},
                                  {deltaL0, - deltaLp/Sqrt[2]}}, 1,  3,  1,  1};
\end{lstlisting}
{\bf Definition of the scalar potential and the Yukawa interactions}
\begin{lstlisting}
DEFINITION[GaugeES][LagrangianInput]= {
	{LagHC, {AddHC->True}},
	{LagNoHC,{AddHC->False}}};
\end{lstlisting}

 Definitions of index contractions for the scalar potential
\begin{lstlisting}
contractionMu12=Delta[rig1,rig2] Delta[lef2,lef1];

contractionLam1=Delta[rig1,rig2] Delta[lef2,lef1] Delta[rig3,rig4] Delta[lef4,lef3];
contractionLam2a=epsTensor[lef2,lef1] epsTensor[rig2,rig1] epsTensor[lef4,lef3] epsTensor[rig4,rig3];
contractionLam2b=epsTensor[rig2,rig1] epsTensor[lef2,lef1] epsTensor[rig4,rig3] epsTensor[lef4,lef3];
contractionLam3=epsTensor[lef2,lef1] epsTensor[rig2,rig1] epsTensor[rig4,rig3] epsTensor[lef4,lef3];
contractionLam4a=- Delta[rig1,rig2] Delta[lef2,lef1] epsTensor[lef4,lef3] epsTensor[rig4,rig3];
contractionLam4b=- Delta[rig1,rig2] Delta[lef2,lef1] epsTensor[rig4,rig3] epsTensor[lef4,lef3];

contractionRho1a=Delta[rig1b,rig2b] Delta[rig2,rig1] Delta[rig3b,rig4b] Delta[rig4,rig3];
contractionRho1b=Delta[lef1b,lef2b] Delta[lef2,lef1] Delta[lef3b,lef4b] Delta[lef4,lef3];
contractionRho2a=Delta[rig1b,rig2] Delta[rig2b,rig1] Delta[rig3,rig4b] Delta[rig4,rig3b];
contractionRho2b=Delta[lef1b,lef2] Delta[lef2b,lef1] Delta[lef3,lef4b] Delta[lef4,lef3b];
contractionRho3=Delta[lef1b,lef2b] Delta[lef2,lef1] Delta[rig3b,rig4b] Delta[rig4,rig3];
contractionRho4a=Delta[rig1b,rig2] Delta[rig2b,rig1] Delta[lef3,lef4b] Delta[lef4,lef3b];
contractionRho4b=Delta[lef1b,lef2] Delta[lef2b,lef1] Delta[rig3,rig4b] Delta[rig4,rig3b];

contractionAlp1a=Delta[rig1,rig2] Delta[lef2,lef1] Delta[lef3b,lef4b] Delta[lef4,lef3];
contractionAlp1b=Delta[rig1,rig2] Delta[lef2,lef1] Delta[rig3,rig4] Delta[rig4b,rig3b];
contractionAlp2a=- epsTensor[rig2,rig1] epsTensor[lef2,lef1] Delta[rig3,rig4] Delta[rig4b,rig3b];
contractionAlp2b=- epsTensor[rig2,rig1] epsTensor[lef2,lef1] Delta[lef3,lef4] Delta[lef4b,lef3b];
contractionAlp2c=- epsTensor[lef2,lef1] epsTensor[rig2,rig1] Delta[rig3,rig4] Delta[rig4b,rig3b];
contractionAlp2d=- epsTensor[lef2,lef1] epsTensor[rig2,rig1] Delta[lef3,lef4] Delta[lef4b,lef3b];
contractionAlp3a=Delta[rig1,rig2] Delta[lef2,lef3] Delta[lef3b,lef4b] Delta[lef4,lef1];
contractionAlp3b=Delta[lef1,lef2] Delta[rig2,rig3] Delta[rig3b,rig4b] Delta[rig4,rig1];

contractionBeta1a=Delta[rig1,rig2] Delta[rig2b,rig3] Delta[lef3,lef4b] Delta[lef4,lef1];
contractionBeta1b=Delta[lef1,lef2] Delta[lef2b,lef3] Delta[rig3,rig4b] Delta[rig4,rig1];
contractionBeta2a= epsTensor[rig1,rig2] Delta[rig2b,rig3] Delta[lef3,lef4b] epsTensor[lef4,lef1];
contractionBeta2b= epsTensor[lef2,lef1] Delta[lef2b,lef3] Delta[rig3,rig4b] epsTensor[rig1,rig4];
contractionBeta3a= Delta[rig1,rig2] epsTensor[rig3,rig2b] epsTensor[lef4b,lef3] Delta[lef4,lef1];
contractionBeta3b= Delta[lef1,lef2] epsTensor[lef2b,lef3] epsTensor[rig3,rig4b] Delta[rig4,rig1];
\end{lstlisting}

 Scalar potential
\begin{lstlisting}
LagNoHC = ( mu12 contractionMu12 Phi.conj[Phi]           
           - mu22 ( conj[Phi].conj[Phi]
           +  Phi.Phi )
           + muLR2 ( deltaR.conj[deltaR]  
           + deltaL.conj[deltaL]  )
           - lam1 contractionLam1 Phi.conj[Phi].Phi.conj[Phi] 
          - lam2 ( contractionLam2a conj[Phi].conj[Phi].conj[Phi].conj[Phi] 
          + contractionLam2b Phi.Phi.Phi.Phi )
          - lam3 contractionLam3 conj[Phi].conj[Phi].Phi.Phi
          - lam4 ( contractionLam4a Phi.conj[Phi].conj[Phi].conj[Phi] 
          + contractionLam4b Phi.conj[Phi].Phi.Phi )                     
          - rho1 ( contractionRho1a deltaR.conj[deltaR].deltaR.conj[deltaR] 
          + contractionRho1b deltaL.conj[deltaL].deltaL.conj[deltaL] )
          - rho2 ( contractionRho2a deltaR.deltaR.conj[deltaR].conj[deltaR] 
          + contractionRho2b deltaL.deltaL.conj[deltaL].conj[deltaL] )
          - rho3 contractionRho3 deltaL.conj[deltaL].deltaR.conj[deltaR] 
          - rho4 ( contractionRho4a deltaR.deltaR.conj[deltaL].conj[deltaL] 
          + contractionRho4b deltaL.deltaL.conj[deltaR].conj[deltaR]   )         
          - alp1 ( contractionAlp1a Phi.conj[Phi].deltaL.conj[deltaL] 
          + contractionAlp1b Phi.conj[Phi].deltaR.conj[deltaR] )
          - alp2 ( contractionAlp2a Phi.Phi.deltaR.conj[deltaR] 
          + contractionAlp2b Phi.Phi.deltaL.conj[deltaL] 
          + contractionAlp2c conj[Phi].conj[Phi].deltaR.conj[deltaR] 
          + contractionAlp2d conj[Phi].conj[Phi].deltaL.conj[deltaL]  )     
          - alp3 ( contractionAlp3a Phi.conj[Phi].deltaL.conj[deltaL] 
          + contractionAlp3b conj[Phi].Phi.deltaR.conj[deltaR]  )
          - beta1 ( contractionBeta1a Phi.deltaR.conj[Phi].conj[deltaL] 
          + contractionBeta1b conj[Phi].deltaL.Phi.conj[deltaR] )
          - beta2 ( contractionBeta2a conj[Phi].deltaR.conj[Phi].conj[deltaL] 
          + contractionBeta2b Phi.deltaL.Phi.conj[deltaR] )
          - beta3 ( contractionBeta3a Phi.deltaR.Phi.conj[deltaL] 
          + contractionBeta3b conj[Phi].deltaL.conj[Phi].conj[deltaR] ) );
\end{lstlisting}

 Yukawa interactions
\begin{lstlisting}
LagHC = - ( YL1 Phi.LLbar.LR 
          + YL2 conj[Phi].LLbar.LR 
          + YQ1 QLbar.Phi.QR 
          - YQ2 QLbar.conj[Phi].QR 
          + YDR LR.deltaR.LR 
          + YDL conj[LLbar].deltaL.conj[LLbar] );
\end{lstlisting}
{\bf Rotations in gauge sector}
\begin{lstlisting}
DEFINITION[EWSB][GaugeSector] =
{ {{VB,VWL[3],VWR[3]},{VP,VZ,VZR},ZZ},
  {{VWL[1],VWL[2],VWR[1],VWR[2]},{VWLm,conj[VWLm],VWRm,conj[VWRm]},ZW} };     
\end{lstlisting}
{\bf VEVs}
\begin{lstlisting}
DEFINITION[EWSB][VEVs]={
{H0,       {vHd, 1/Sqrt[2]}, 
                     {sigmaH10, I/Sqrt[2]},{phiH10, 1/Sqrt[2]}},
{HPrime0, {vHu, 1/Sqrt[2]}, 
                     {sigmaH20, I/Sqrt[2]},{phiH20,1/Sqrt[2]}},
{deltaR0, {vR, 1/Sqrt[2]}, 
                     {sigmaR0, I/Sqrt[2]},{phiR0,1/Sqrt[2]}},
{deltaL0, {vL, 1/Sqrt[2]}, 
                     {sigmaL0, I/Sqrt[2]},{phiL0,1/Sqrt[2]}} };
\end{lstlisting}
{\bf Rotations in the matter sector}
 \begin{lstlisting}
DEFINITION[EWSB][MatterSector]=   
    { (*Neutral scalars*)
     {{phiH10,phiH20,phiR0,phiL0},{hh,ZH}},
      (*Pseudoscalars*)
     {{sigmaH10,sigmaH20,sigmaR0,sigmaL0},{Ah,UP}},
      (*Singly charged scalars*)
     {{Hm,conj[Hp],conj[deltaRp],conj[deltaLp]},{Hpm,UC}},
      (*Doubly charged scalars*)
     {{conj[deltaRpp],conj[deltaLpp]},{Hppmm,UCC}},
      (*Fermions*)
     {{{dL}, {conj[dR]}}, {{DL,Vd}, {DR,Ud}}},
     {{{uL}, {conj[uR]}}, {{UL,Vu}, {UR,Uu}}},
     {{{eL}, {conj[eR]}}, {{EL,Ve}, {ER,Ue}}},
     {{nuL, conj[nuR]},{Fv0,PMNS}} };  
\end{lstlisting}
{\bf Dirac spinors}
 \begin{lstlisting}
DEFINITION[GaugeES][DiracSpinors]={
  Fd1 -> {dL, 0},
  Fd2 -> {0, dR},
  Fu1 -> {uL, 0},
  Fu2 -> {0, uR},
  Fe1 -> {eL, 0},
  Fe2 -> {0, eR},
  Fv1  -> {nuL,0},
  Fv2 -> {0,nuR} };

DEFINITION[EWSB][DiracSpinors]={
 Fd ->{  DL, conj[DR]},
 Fe ->{  EL, conj[ER]},
 Fu ->{  UL, conj[UR]},
 Fv ->{Fv0, conj[Fv0]}};
\end{lstlisting}


\providecommand{\href}[2]{#2}\begingroup\raggedright\endgroup

\end{document}